\pdfoutput=1
\documentclass{article}

\usepackage[preprint]{preprint}

\usepackage[utf8]{inputenc}
\usepackage[T1]{fontenc}

\usepackage{hyperref}
\usepackage{url}
\usepackage{nicefrac}
\usepackage{microtype}

\usepackage{amsmath}
\usepackage{amssymb}
\usepackage{amsfonts}

\usepackage{booktabs}
\usepackage{graphicx}

\usepackage{enumitem}
\usepackage{subcaption}
\usepackage{multirow}
\usepackage{makecell}
\usepackage{wrapfig}
\usepackage[table]{xcolor}
\usepackage[final]{changes} %
\usepackage{placeins} %
\usepackage{algorithm}
\usepackage{algpseudocode}

\title{Querying Counterfactuals on Tissue Graphs with Supervised Disentanglement}

\author{%
  \setcounter{footnote}{1}%
  \parbox{\dimexpr\textwidth-2\tabcolsep\relax}{\centering\bfseries
    Abdul Moeed\textsuperscript{1,2} \quad
    Stefan Schrod\textsuperscript{1,3} \quad
    Martin Rohbeck\textsuperscript{1} \quad
    Marc Jan Bonder\textsuperscript{4,5} \\
    Pavlo Lutsik\textsuperscript{6} \quad
    Oliver Stegle\textsuperscript{1,3}\thanks{Joint senior and corresponding authors: \texttt{daniel.dimitrov@embl.de}, \texttt{o.stegle@dkfz-heidelberg.de}} \quad
    Daniel Dimitrov\textsuperscript{1,3}\footnotemark[2]
  }\\[1em]
  \parbox{\dimexpr\textwidth-2\tabcolsep\relax}{\centering\small
    \textsuperscript{1}DKFZ, Heidelberg \quad
    \textsuperscript{2}HIDSS4Health \quad
    \textsuperscript{3}EMBL, Heidelberg \quad
    \textsuperscript{4}UMCG, Groningen \\
    \textsuperscript{5}Oncode Institute, Utrecht \quad
    \textsuperscript{6}KU Leuven \quad
  }
}
\begin{document}

\maketitle

\begin{abstract}
\textit{Tissue graph counterfactuals} ask what a cell would express if its spatial neighborhood were altered, a question central to predicting cell behavior in tissues. Existing methods, however, present only specific instances of such queries.
In this work, we first formalize \textit{tissue graph counterfactuals} as a class of spatial interventions that either rewire connections between cells (\textit{edge perturbation}) or modify the expression of their neighbors (\textit{node perturbation}). We then introduce \textit{Cellina} {\renewcommand{\thefootnote}{\ddag}\footnote{\url{https://cellina.readthedocs.io}}\addtocounter{footnote}{-1}}, a framework that uses supervised disentanglement to decompose a cell's intrinsic state from its spatial context, using the latter as a conditioning input for in-silico perturbation predictions. 
Across benchmarks spanning over 2.5 million spatially-resolved cells in observational and interventional screens, \textit{Cellina} outperforms both spatially-informed and non-spatial competitors in in-silico perturbations, while training as fast as non-spatial methods. Moreover, we show that \textit{Cellina} reveals cancer subdomains with distinct signaling programs in an unsupervised manner and enables targeted neighbor perturbation simulations.

\end{abstract}

\section{Introduction}
\label{sec:introduction}

A central goal of computational single-cell biology is to predict how cells respond to perturbations, including in cell types or other contexts that have not been measured directly \citep{bunne_how_2024, roohani2025virtual, dimitrov2026interpretation}. Existing single-cell perturbation methods typically rely on at least one of two assumptions: (i) that perturbations act as shared stimuli applied uniformly across cells, and (ii) that cells are conditionally independent, giving rise to an effectively i.i.d.\ learning problem. Tissues violate both assumptions. Within a tissue, a cell's transcriptional state is shaped by its local neighborhood, that is, the composition of nearby cells and the signals they emit \citep{armingol2021deciphering}. Consequently, modeling tissues requires methods that reason about neighbor-driven stimuli, which are unique to every cell.
This motivates two complementary prediction tasks: what would a cell express if it were surrounded by a different set of neighbors, and what would it express if its existing neighbors changed the activity of specific genes or pathways? To address these tasks, a first generation of spatially-informed perturbation models has recently emerged, ranging from label-swapping flow models~\citep{akbarnejad2025mapping} and coordinate-aware covariate models~\citep{lin2025concert} to neighbor-propagating graph neural networks~\citep{sun2025spatialprop}. Yet, each method targets a specific intervention type and is commonly evaluated in-domain. 
Here, we formalize \textit{tissue graph counterfactuals}: interventions on either the neighborhood connections of a focal cell (\textit{edge perturbation}) or the expression of its neighbors (\textit{node perturbation}), corresponding to the two mutable components of the tissue graph (see Section~\ref{sec:counterfactuals}). Each operator corresponds to a distinct class of biological intervention. An edge perturbation asks about composition and location: the infiltration of an immune population into a tumor, the depletion of a cell type that sustains its growth, or, prospectively, how an administered therapeutic cell would behave once it lodges in one region rather than another. A node perturbation asks about signaling in a fixed tissue, as when an agonist, an inhibitor, or a knockdown shifts a coherent gene program across a cell's existing neighbors. We evaluate an instance of each: transplanting cells between perturbed and healthy regions, and shifting pathway activity across their neighbors.

\begin{figure}[t]
    \centering
    \includegraphics[width=\textwidth]{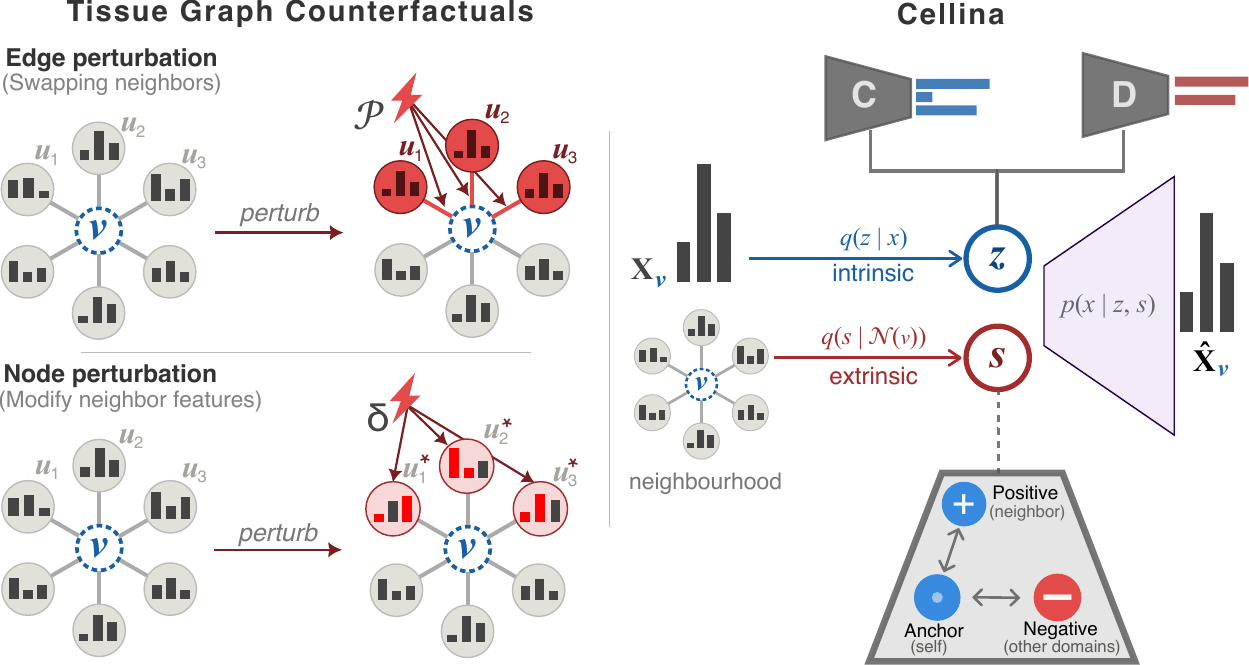}
    \caption{\textbf{Tissue graph counterfactuals and \textit{Cellina} overview.}
    (\emph{Left}) Two interventions on a focal cell $v$ with neighbors $u$:
    \textbf{edge perturbation} rewires $v$'s neighborhood $\mathcal{N}(v)$ to a counterfactual neighbor pool $\mathcal{P}$, and \textbf{node perturbation} alters neighbor
    expression on a feature (gene) subset $S$.
    (\emph{Right}) \textit{Cellina} encodes intrinsic identity $z \sim q(z \mid x_v)$
    and spatial representation $s$ from $v$'s local neighborhood, and decodes
    $p(x \mid z, s)$. Supervision anchors $z$ to cell type and removes
    spatial-domain information adversarially; \textit{Cellina-GAT}
    additionally applies a contrastive loss on $s$.}
    \label{fig:method}
\end{figure}

We present \textit{Cellina} (Figure~\ref{fig:method}), a (graph) variational autoencoder (VAE) that addresses tissue graph counterfactuals by decomposing each cell's gene expression into an intrinsic representation $z$ encoding cell identity and an extrinsic (spatial) representation $s$ encoding the effect of its microenvironment. Because such disentangled representations cannot be reliably recovered without supervision~\citep{locatello2019challenging}, we use cell-type and spatial-domain labels as a soft inductive bias without formal identifiability guarantees~\citep{khemakhem2020variational}, anchoring $z$ to cell-type identity and adversarially removing spatial-domain information, which routes microenvironmental variation to $s$.
We validate this separation under out-of-distribution regimes via in-silico neighborhood alterations, tasks at which a model that conflates the drivers of variation cannot succeed~\citep{scholkopf2021toward}.

\textbf{Contributions:}
\begin{enumerate}
\def\labelenumi{\arabic{enumi}.}
    \item We formalize \textit{tissue graph counterfactuals} as a class of spatial interventions with two operators, edge and node perturbations, and use them to jointly define and evaluate neighborhood-driven cell responses.
    \item We introduce \textit{Cellina}, a dual-encoder graph VAE with supervised disentanglement. On in-silico perturbations derived from observational data, it outperforms spatial and non-spatial baselines on colorectal cancer ($+0.18$ Pearson over the closest baseline) and across held-out mouse brain domains. On \textit{in vivo} spatial interventional data, where it predicts how T cells respond to specific CRISPR knockouts in  neighboring cancer cells, it leads the closest competitor by $+0.17$ ($+0.24$ with \textit{Cellina-GAT}) Pearson.
    \item We use \textit{Cellina}'s spatial latents to identify biologically distinct cancer subdomains without supervision, and to simulate pathway-targeted neighbor perturbations using existing priors.
\end{enumerate}

\section{Related Work}
\label{sec:related}

\textbf{Perturbations and context transfer.} scGen~\citep{lotfollahi2019scGen} and CPA~\citep{hetzel2022predicting, lotfollahi2023predicting} are standard methods for predicting cellular responses to perturbations. Both models assume i.i.d.\ data and neither represents continuous neighbor composition or cell-specific spatial contexts. 
More recent methods based on optimal transport \citep{bunne2023learning} and flow matching \citep{klein2025cellflow} model individual cell trajectories, yet still work on the same shared-stimulus intervention assumption.
Extending this paradigm to tissue perturbations requires disentangling intrinsic cellular state from extrinsic influence, and reasoning about continuous variation in individual neighborhoods rather than shared or discretized stimuli.

\textbf{Spatially-informed disentanglement.}
A related line of work leverages spatial information to separate intrinsic cell
states from extrinsic tissue influences. For example, MISTy~\citep{tanevski2022explainable} and
NCEM~\citep{fischer2023modeling} model neighborhood effects through multi-view
regression and graph neural networks, respectively, while SIMVI~\citep{dong2025simvi}
uses a graph VAE with unsupervised disentanglement to isolate
spatially-induced variation. These approaches yield interpretable decompositions of
spatial scales, but do not support in-silico perturbations.

\textbf{Tissue graph perturbations.} The state-of-the-art methods most directly related to simulating tissue graph counterfactuals in spatial omics are MintFlow \citep{akbarnejad2025mapping}, Concert \citep{lin2025concert}, Celcomen \citep{megas2025estimation}, and SpatialProp \citep{sun2025spatialprop}. Celcomen models spatial in-silico perturbations through learned gene-gene interactions, but learns a global interaction matrix shared across the tissue, perturbing gene values rather than nodes or edges. SpatialProp proposed modeling the downstream effects of neighbor perturbations, making it directly related to our \textit{node perturbation} task. MintFlow and Concert both perform in-silico perturbations via label conditioning, but MintFlow approaches it via an operation on the graph itself, enabling adaptation to our \textit{edge perturbation} task. \textit{Cellina} supports both edge and node perturbations, disentangles intrinsic from spatial factors under label supervision, and focuses on predicting cell type $\times$ domain pairs unseen during training (Table~\ref{tab:method_comparison}).

\textbf{Graph counterfactuals.} The broader graph ML literature offers a complementary perspective: counterfactual reasoning over graphs has been explored via instance-level adjacency perturbations for explainability \citep{lucic2022cf, bajaj2021robust}, and in generative graph VAEs with input-conditioned priors \citep{ma2022clear}. \textit{Cellina} draws on this line of work but reorients it from model explanation to biological intervention, replacing label-swap objectives with simulating counterfactual neighborhood queries over tissue graphs. See Appendix~\ref{app:baselines} for full descriptions of competitors.

\section{Method}
\label{sec:method}

\subsection{Notation and Problem Definition}
\label{sec:notation}

Let $\mathcal{V} = \{v_1, \ldots, v_N\}$ denote $N$ cells on a spatial tissue slide. Each cell $v \in \mathcal{V}$ is associated with:
\begin{itemize}[noitemsep, topsep=0pt]
  \item $x_v \in \mathbb{Z}_{\geq 0}^G$: gene expression counts across $G$ genes
  \item $y_v \in \{1, \ldots, C\}$: cell-type label (e.g., T cell, epithelial, fibroblast)
  \item $d_v \in \{1, \ldots, D\}$: spatial domain label -- a discrete partition of a tissue according to pre-defined regions or niches (e.g., tumor vs.\ healthy regions)
\end{itemize}

Spatial domain and cell-type labels capture complementary aspects of a cell's identity: cell-type label $y_v$ encodes intrinsic cell identity (\emph{what} kind of cell it is), while spatial domain $d_v$ encodes which tissue region it inhabits (\emph{where} it is located).

The spatial proximity between cells is encoded by a weighted graph $\mathcal{G} = (\mathcal{V}, \mathcal{E}, W)$ whose nodes $\mathcal{V}$ are cells and whose edges $\{u,v\} \in \mathcal{E}$ connect a focal cell $v$ to its neighbors $u$, with edge weights $W_{uv} \geq 0$. $W$ is computed from the two-dimensional spatial coordinates of each cell $v$ relative to all others using a Gaussian proximity kernel (Appendix~\ref{app:data}).
For each cell $v$, we denote the spatial neighborhood as:
\[
\mathcal{N}(v) = \{u \in \mathcal{V} \mid \{v,u\} \in \mathcal{E}\}
\]

On this graph we formalize two intervention operations: \emph{edge} and \emph{node} perturbations (Section~\ref{sec:counterfactuals}). Because matched pre- and post-intervention measurements of the same cells are generally unavailable \emph{in situ}~\citep{bunne2023learning, dimitrov2026interpretation}, we evaluate them as \emph{in-silico perturbations}, mirroring existing single-cell context transfer benchmarks~\citep{roohani2025virtual, wu2024perturbench, NEURIPS2024_24c4d51f}; throughout, we report \emph{in-silico perturbation predictions} for empirical results and reserve \emph{tissue graph counterfactual} for the formal task definition.

\subsection{Model Architecture}
\label{sec:model_overview}

\textbf{Generative model.} \textit{Cellina} is a (graph) variational autoencoder (VAE)
\citep{kingma2013auto, kipf2016variational} that decomposes each cell's expression into two latent variables: an intrinsic representation $z$ encoding cell identity and a spatial representation $s$ encoding microenvironmental influence. Both have standard normal priors, $\text{Normal}(0, I)$.
Following standard practice for single-cell counts \citep{lopez2018deep, gayoso2022python}, the likelihood $p(x \mid z, s)$ is a negative binomial distribution parametrized by a decoder $\mathrm{Dec}([z;\, s])$ ($[\cdot\,;\,\cdot]$ denotes concatenation; Appendix~\ref{app:generative}).

\textbf{Inference.}
We propose two variants of the model. Both use an MLP encoder $\mathrm{Enc}_z(x_v)$ for the intrinsic posterior $q(z \mid x_v) = \text{Normal}(\mu_z, \sigma^2_z)$, and differ in how the spatial posterior is parametrized:
\begin{itemize}[noitemsep]
    \item \textbf{Cellina} uses a degree-normalized aggregation of neighbor expression
    $\varphi(v) = \bigl(\sum_{u \in \mathcal{N}(v)} W_{uv}\,\tilde{x}_u\bigr) \big/ \bigl(\sum_{u \in \mathcal{N}(v)} W_{uv}\bigr)$ where $\tilde{x}_u \in \mathbb{R}^G$ denotes log-normalized $x_u$; encoded via an MLP encoder $\mathrm{Enc}_s(\varphi)$ that outputs $\mu_s, \sigma_s$.
    \item \textbf{Cellina-GAT} replaces the fixed aggregator with a
    multi-layer GATv2 \citep{brody2021attentive} operating on $v$'s local subgraph $\mathcal{G}_v = (x_v,\, \{x_u\}_{u \in \mathcal{N}(v)},\,
    \mathcal{E}_v, W_{uv})$ (where $\mathcal{E}_v \subseteq \mathcal{E}$ is an edge set of $\mathcal{N}(v)$); self-loops are excluded, so $s$ does not directly aggregate $v$'s own expression, which is
    primarily captured by $z$; $\mu_s, \sigma_s$ are linear heads on the
    focal-node representation.
    For this variant, we add a modified contrastive loss $\mathcal{L}_\mathrm{spatial}$ (Appendix \ref{app:model_details}).
\end{itemize}

These two variants trade off efficiency and expressivity: \textit{Cellina}'s linear aggregator decouples neighborhood construction from training and exhibits training-time scaling similar to non-spatial baselines (Figure~\ref{fig:scalability}), whereas \textit{Cellina-GAT} learns attention over each subgraph at additional computational cost per step (Appendix~\ref{app:inference}). All approximate posteriors are diagonal Gaussians whose parameters are output by the encoders, and samples are drawn using the reparameterization trick \citep{kingma2013auto}.

\textbf{Supervised disentanglement.}
Because optimizing the ELBO of the VAE alone does not prevent $z$ from absorbing spatially-driven variation, we introduce two auxiliary objectives that route spatial signal to $s$ by removing it from $z$.
Specifically, (i) we anchor $z$ to cell-type identity using a cell type classifier with loss $\mathcal{L}_\mathrm{clf}$, and (ii) adversarially strip spatial-domain information from $z$ through a two-part adversarial objective. First, $\mathcal{L}_\mathrm{disc}$ is optimized to train the discriminator to predict the domain label from $z$; second, $\mathcal{L}_\mathrm{adv}$ is optimized to encourage the encoder to render $z$ domain-invariant.
For \textit{Cellina-GAT}, we further add a modified graph-supervised contrastive loss $\mathcal{L}_\mathrm{spatial}$ on $s$ as a biologically grounded inductive bias. We use a cell's spatial neighbors as positives, rather than cells sharing a domain label, to discourage $s$ from collapsing onto domain labels and erasing within-domain neighborhood variation; domain labels enter only through the negatives, which separate distinct domains. Appendix~\ref{app:ablation_disc} ablates each auxiliary objective independently, probing the cell-type and domain information retained in the frozen latents; the contrastive term in particular improves latent informativeness and fit quality.
The combined training objective (minimized in step 2 over encoder and decoder parameters) follows as:
\[\mathcal{L} = \mathcal{L}_\mathrm{VAE}
  + \alpha_\mathrm{clf}\,\mathcal{L}_\mathrm{clf}
  + \alpha_\mathrm{spatial}\,\mathcal{L}_\mathrm{spatial}
  - \alpha_\mathrm{adv}\,\mathcal{L}_\mathrm{adv},
  \]
where $\alpha$ are data-adaptive normalization scales fixed after the first training epoch (Appendix~\ref{app:loss_norm}).

\textbf{Training procedure.} We optimize this objective via alternating updates: (1) a discriminator step that updates the discriminator on detached $z$ (encoder parameters frozen), and (2) a VAE step that updates encoder and decoder with the discriminator frozen, using the combined objective above. Alternating updates ensure the discriminator provides informative gradients to the encoder \citep{goodfellow2014generative}; for details and hyperparameters, see Appendices~\ref{app:hyperparams} and~\ref{app:model_details}.

\subsection{Tissue Graph Counterfactuals}
\label{sec:counterfactuals}

A tissue graph counterfactual asks: \emph{what would cell $v$ express if its spatial neighborhood were altered while its intrinsic identity remained fixed?} That alteration acts on the graph itself (the cell's microenvironment), not on an intrinsic per-cell covariate.
The tissue graph $\mathcal{G}$ has two mutable
components: its edges $\mathcal{E}$ with associated weights $W$, encoding the neighborhood topology, and the neighbor node feature matrix $X$, encoding cell expression. This naturally gives rise to two counterfactual queries: 

\textbf{Definition 1 (Edge Perturbation).} An edge perturbation intervenes on a cell's local neighborhood $\mathcal{N}(v)$, replacing it with an alternative neighborhood $\mathcal{N}'$:
\[
\mathcal{N}(v) := \mathcal{N}'.
\]
This is an intervention on the topology alone: it admits arbitrary modifications to the neighborhood, including the addition, removal, or substitution of neighbors, with replacements drawn from cells typically observed in the data, without altering the expression of neighboring cells, i.e., the neighbor node features $X$. Biologically, it asks about composition and location: the infiltration of a cell population into a region, the depletion of one that sustains it, or the arrival of a cell type that was previously absent. In this work, we evaluate \textbf{edge perturbation} as an in-silico domain edge rewiring: focal cells of type $y$ in a source domain are assigned new cell-type-matched neighborhoods from the target domain; type $y$ itself is excluded from the donor pool and from $W$ in training and inference, so that none of its target domain expression enters the model (Appendix~\ref{app:donor_sampling}).

\textbf{Definition 2 (Node Perturbation).} A node perturbation modifies the feature vectors of $v$'s neighbors while preserving the graph topology. Let $\mathcal{S}$ denote the set of intervened genes. For each neighbor $u \in \mathcal{N}(v)$, the perturbed gene feature vector is defined as:
\[
x_{u,g}^{\mathrm{cf}} =
\begin{cases}
  T_g(x_{u,g}) & g \in \mathcal{S} \\
  x_{u,g}      & g \notin \mathcal{S}
\end{cases}
\]
where $T_g : \mathbb{R}_{\geq 0} \to \mathbb{R}_{\geq 0}$ is a gene-specific transformation encoding the counterfactual expression under the intervention. The approach is agnostic to the choice of $T_g$, which may correspond to additive shifts, knockouts, overexpression, or learned counterfactual values. In our setting, we instantiate $T_g$ as an in-silico perturbation, specifically a multiplicative scaling $T_g(x_{u,g}) = x_{u,g} \cdot e^{\delta_g}$, where $\delta_g \in \mathbb{R}$ is a gene-specific log fold-change applied to each cell $x_u$ (approximately, an additive shift in log-normalized space; Appendix~\ref{app:convergence}). Biologically, it asks about signaling in a fixed tissue: an agonist, an inhibitor, or a knockdown that shifts a coherent program across a cell's existing neighbors. Operating on a subset $\mathcal{S}$ rather than the full feature vector follows from this: genes act in co-regulated programs, so $\mathcal{S}$ can be chosen to target a specific biological process such as a pathway or regulon.

In sum, an \textbf{edge perturbation} alters the topology of the tissue graph (which and how many cells are a focal cell's neighbors), but not neighbor expression. In contrast, a \textbf{node perturbation} alters the expression of any number of genes but not the neighbor topology; Appendix~\ref{app:operators} details the operations each perturbation admits, their applications (Table~\ref{tab:operator_mapping}), and the narrow case in which they coincide.

\section{Experiments}
\label{sec:experiments}

\subsection{Evaluation Setting}
\label{sec:evaluation}
\begin{wrapfigure}[23]{r}{0.4\textwidth}
    \centering
    \includegraphics[width=0.4\textwidth]{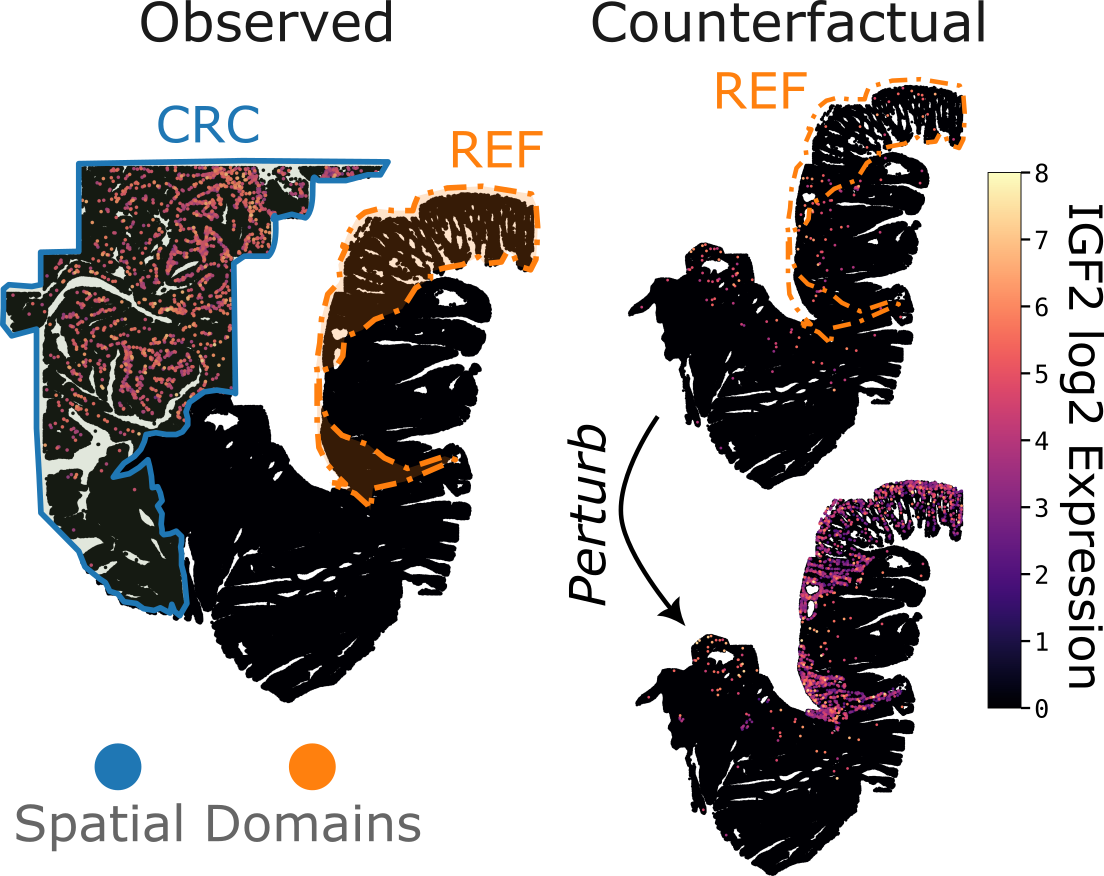}
    \caption{\textbf{In-silico expression prediction with \textit{Cellina}.}
    Observed IGF2 expression in Fibroblasts with pre-annotated spatial domains. Rewiring $\text{Fibroblasts}_{\text{REF}}$ neighborhoods to those of $\text{Fibroblasts}_{\text{CRC}}$ yields predicted IGF2 expression consistent with observed CRC Fibroblasts.}
    \label{fig:example}
\end{wrapfigure}

Interventional spatial data remain scarce, as current spatial perturbation screens are limited by their assays, which often provide limited cellular resolution or incomplete transcript readouts, and by their designs, which generally lack matched controls~\citep{dhainaut2022spatial, breinig2026integrated, binanSimultaneousCRISPRScreening2025}.
We therefore evaluate the models under held-out prediction tasks. During training, cells of type $y$ are observed in a source condition $d$ but held out in a target condition $d'$, where they are also excluded from the neighborhood graph $W$ at training and inference. At test time, source cells of type $y$ are assigned the neighborhood they would have in $d'$, either neighbors drawn from $d'$ (edge perturbation) or their own neighbors with expression shifted towards $d'$ (node perturbation; Section~\ref{sec:counterfactuals}), and the predicted expression is compared with that of the held-out cells of type $y$ in $d'$. Since assays are destructive, this reference is an observed population rather than the same cells. This mirrors, in a spatial setting, the \textit{context transfer} protocol established for perturbation benchmarks~\citep{NEURIPS2024_24c4d51f, wu2024perturbench, roohani2025virtual}, in which a perturbation's effect is observed in some cell types and must be predicted for a cell type in which it was never observed; here, the perturbation is the change of neighborhood from $d$ to $d'$. In the CRC and mouse brain cohorts, $d$ and $d'$ correspond to spatial domains (e.g., healthy colon and pathologist-annotated cancer areas; Figure~\ref{fig:example}). In Perturb-FISH interventional data, we predict how control T cells respond to tumor cell neighborhoods $d'$ defined by distinct \textit{in situ} CRISPR knockouts (Table~\ref{tab:cellina-invivo}).

This approach allows for assessing both types of graph perturbation defined in Section~\ref{sec:counterfactuals}. In \emph{edge perturbation}, the neighbors of seed cells from the source domain are swapped with cells from the target domain; in \emph{node perturbation}, seed cells retain their original neighbors; their neighbors' feature vectors are then altered using a cell-type-specific factor $\delta_{y,g}$;
for the held-out type this is the global shift $\delta_{\setminus y}$, so that the perturbed input does not provide the held-out population's shift (Appendix~\ref{app:convergence}).

\subsection{Benchmarking Approach}
\label{sec:benchmark}

\textbf{Metrics.} Following recent benchmarking practices~\citep{wu2024perturbench, roohani2025virtual, heidari2026evaluating}, we evaluate in-silico perturbation predictions along four complementary axes: (i) whether methods recover the correct direction and relative magnitude of gene-level effects (Pearson~$r$ on log fold-changes of top differentially expressed genes), (ii) whether they correctly identify the most differentially expressed genes (Signed Precision), (iii) whether predicted and observed expression shifts agree in absolute magnitude (RMSE of log fold-change vectors), and (iv) overall distributional fit (E-distance). All metrics are assessed for each held-out cell type; formal definitions
and an expanded metrics set with per-cell-type breakdown appear in Appendices~\ref{app:metrics} and~\ref{app:extended-results}.

\textbf{Alternative methods.} We consider the following methods, spanning spatially-informed models with a native neighborhood-perturbation interface and popular non-spatial baselines (see Appendix~\ref{app:baselines} for full descriptions; all methods, including \textit{Cellina}, are run at their default settings, Appendix~\ref{app:hyperparams}):
\emph{Mean shift}, an average expression shift between domains, applied to each cell type, and motivated by similar baselines \citep{ahlmann2025deep, vinas2025systema}; \emph{scGen, CPA}, spatially-uninformed perturbation-focused autoencoders, which treat the spatial domain as a label embedding or shift \citep{lotfollahi2019scGen, lotfollahi2023predicting}; \emph{Concert}, a spatially-informed model that propagates perturbations across cells through a Gaussian process over their coordinates, but simulates them by swapping categorical attribute embeddings rather than through graph operations \citep{lin2025concert}; \emph{SpatialProp}, an efficient GNN-based method tailored for in-domain prediction of neighbor perturbations \citep{sun2025spatialprop}; and \emph{MintFlow}, a spatially-informed flow-matching model for in-domain tissue perturbations via neighbor cell-type label swapping, adapted to our edge perturbation task \citep{akbarnejad2025mapping}. MintFlow is trained on all cells, including held-out cell types. SIMVI~\citep{dong2025simvi} was excluded from our in-silico perturbation evaluations due to a lack of native support for such queries and prohibitive GPU memory requirements (Appendix~\ref{app:scalability}, Figure~\ref{fig:scalability}); a disentanglement comparison with it favors \textit{Cellina} (Appendix~\ref{app:disentanglement}).

We evaluate three variants of our model (Section~\ref{sec:method}): (1) \emph{Cellina}, the base model with pre-aggregated neighborhoods and both supervision objectives on $z$; (2) \emph{Cellina (ablated)}, which removes both supervision objectives and retains only the dual-encoder architecture (per-component and architecture ablations in Appendix~\ref{app:ablation_disc}); and (3) \emph{Cellina-GAT}, which replaces the pre-aggregated encoder with a graph attention network and a contrastive loss on $s$.

\subsection{Results}
\label{sec:results}

\textbf{Cellina accurately predicts in-silico perturbations in colorectal cancer data.}

To evaluate in-silico perturbation predictions in a clinically relevant setting, we consider a spatial transcriptomics dataset of approximately 2.4 million cells across six colorectal cancer tissue slides (an 18{,}000-gene panel, see Appendix~\ref{app:data})~\citep{crowell2025tracing}.
We use pathologist-annotated healthy colonic mucosa (REF) and colorectal cancer (CRC) regions to define the prediction task: each REF cell of type $y$ receives new neighbors sampled from CRC cells that neighbor type $y$, and the predicted expression is evaluated against held-out CRC cells of the same type.

Both supervised \textit{Cellina} variants outperform every competitor on Pearson, Signed Precision and RMSE\textsubscript{LFC}. Even \textit{Cellina (ablated)}, trained without cell-type or domain supervision, leads or matches all non-\textit{Cellina} methods on correlation and precision, a result we attribute to encoding continuous, cell-resolved neighborhoods. Supervised disentanglement adds a further $+0.15$ Pearson in the full model (Table~\ref{tab:loo_summary_crc}; component ablations in Appendix~\ref{app:ablation_disc}). Notably, none of the spatially-informed competitors surpasses the non-spatial CPA on Pearson or RMSE. The mean shift baseline, which applies a group-average expression change without knowledge of individual neighborhoods, is competitive with scGen and spatially-informed Concert on Pearson $r$. scGen, nevertheless, attains the lowest E-distance ($4.65$), followed by \textit{Cellina (ablated)}, but this metric is computed in the log-normalized PCA space that scGen is trained to reconstruct; when scored in count space by negative-binomial deviance, scGen falls behind \textit{Cellina} and CPA (Table~\ref{tab:nb_deviance}). We therefore prioritize Pearson $r$ and Signed Precision, as they capture which genes respond to a perturbation, i.e.\ the biologically relevant signal that would guide follow-up experiments.

We also confirm that these results are robust to random model initialization across seeds (Appendix~\ref{app:seed-stability}). Moreover, when an entire patient is held out from training together with the target cell type, \textit{Cellina} widens its gain over CPA from $+0.18$ to $+0.27$ Pearson, leading on all four metrics (Appendix~\ref{app:cross_patient}), indicating that it generalizes to unseen donors.

\begin{table}[h]
\centering
\caption{Leave-one-cell-type-out performance (top 50 DEGs) for predicting the in-silico perturbed state of healthy colon cells placed in tumor region (REF $\rightarrow$ CRC). Mean $\pm$ std across cell types and patient samples. \textbf{node-pert} refers to node perturbation task. Best per metric within each block (edge- vs.\ node-perturbation) in \textbf{bold}; \textit{Cellina} variants in gray, spatially-informed competitors in orange.}
\label{tab:loo_summary_crc}
\begin{tabular}{lcccc}
\toprule
Method & Pearson $\uparrow$ & Precision\textsubscript{signed} $\uparrow$ & E-distance $\downarrow$ & RMSE\textsubscript{LFC} $\downarrow$ \\
\midrule
Mean shift & 0.51 $\pm$ 0.26 & 0.18 $\pm$ 0.15 & 29.89 $\pm$ 10.49 & 5.08 $\pm$ 2.51 \\
CPA & 0.68 $\pm$ 0.19 & 0.22 $\pm$ 0.19 & 6.44 $\pm$ 2.27 & 1.50 $\pm$ 0.56 \\
scGen & 0.50 $\pm$ 0.37 & 0.19 $\pm$ 0.20 & \textbf{4.65 $\pm$ 3.49} & 1.99 $\pm$ 0.84 \\
\rowcolor{orange!10}
Concert & 0.49 $\pm$ 0.43 & 0.21 $\pm$ 0.22 & 29.98 $\pm$ 5.22 & 1.81 $\pm$ 0.67 \\
\rowcolor{orange!10}
MintFlow & 0.65 $\pm$ 0.31 & 0.23 $\pm$ 0.21 & 13.01 $\pm$ 1.76 & 1.51 $\pm$ 0.43 \\
\rowcolor{gray!10}
Cellina (ablated) & 0.71 $\pm$ 0.24 & 0.31 $\pm$ 0.22 & 4.93 $\pm$ 1.66 & 1.51 $\pm$ 0.88 \\
\rowcolor{gray!10}
Cellina & 0.86 $\pm$ 0.17 & \textbf{0.45 $\pm$ 0.19} & 7.00 $\pm$ 1.05 & 1.14 $\pm$ 0.56 \\
\rowcolor{gray!10}
Cellina-GAT & \textbf{0.89 $\pm$ 0.09} & 0.42 $\pm$ 0.20 & 8.59 $\pm$ 1.90 & \textbf{1.09 $\pm$ 0.61} \\
\midrule
\rowcolor{gray!10}
Cellina\textsubscript{node-pert} & 0.80 $\pm$ 0.18 & 0.36 $\pm$ 0.19 & \textbf{7.20 $\pm$ 1.45} & 1.39 $\pm$ 0.76 \\
\rowcolor{gray!10}
Cellina-GAT\textsubscript{node-pert} & \textbf{0.88 $\pm$ 0.09} & \textbf{0.40 $\pm$ 0.18} & 11.35 $\pm$ 3.26 & \textbf{1.08 $\pm$ 0.49} \\
\rowcolor{orange!10}
SpatialProp\textsubscript{node-pert} & 0.46 $\pm$ 0.29 & 0.06 $\pm$ 0.11 & 27.12 $\pm$ 8.51 & 7.39 $\pm$ 1.23 \\
\bottomrule
\end{tabular}
\end{table}

\textbf{Node perturbation evaluation.}
Next, we evaluate \textit{Cellina} in the same (REF $\rightarrow$ CRC) regime, but under a node-perturbation setting: rather than changing the neighborhood context entirely, we retain each focal cell's original neighbors and shift their expression by cell-type-specific log-fold vectors $\delta_{y,g}$ ($\delta_{\setminus y}$ for the held-out type; top 200 genes) that reflect the expression difference between REF and CRC (Appendix~\ref{app:convergence}). By design, SpatialProp is the direct competitor in this evaluation; \textit{Cellina} leads it on Pearson, Signed Precision, E-distance, RMSE and NB deviance (Table~\ref{tab:nb_deviance}).
For \textit{Cellina-GAT}, \textbf{edge} and \textbf{node perturbation} are on par on correlation and precision, while base \textit{Cellina}'s node perturbation trails its edge counterpart (CRC $0.80$ vs.\ $0.86$ Pearson). We also see that node perturbation peaks at ${\approx}100$--$200$ perturbed genes and degrades beyond (Appendix~\ref{app:convergence}). Node perturbations are, however, more targeted, enabling scenarios where only specific gene programs in the microenvironment are modified, which we explore in Section~\ref{sec:subtypes}.

\textbf{Cellina performs competitively across domains in the mouse brain.}

To further assess multi-domain generalization, we repeat the same evaluation on a whole-adult-mouse MERFISH cohort~\citep{zhang2023molecularly} (1{,}122-gene panel, \textasciitilde146K cells, three slides), holding out two independently annotated spatial domains (Fiber-tracts and Isocortex). Averaged over held-out domains, \textit{Cellina-GAT} leads all baselines on Pearson ($0.84$, tied with \textit{Cellina}), Signed Precision ($0.53$) and RMSE\textsubscript{LFC} ($5.97$), with CPA as its closest competitor ($0.82$, $0.44$, $6.36$) and scGen with the lowest E-distance ($5.73$; Table~\ref{tab:loo_summary_merfish}). Concert and MintFlow are below CPA on all four metrics, and the node-perturbation ranking likewise holds, with \textit{Cellina-GAT} leading SpatialProp by a safe margin.

\begin{table}[h]
\centering
\caption{Leave-one-cell-type-out performance (top 50 DEGs) for the in-silico perturbations Thalamus $\rightarrow$ Isocortex, Fiber-tracts. Mean $\pm$ std across cell types and three slides, averaged over both held-out domains; per-domain results in Appendix Table~\ref{tab:loo_summary_merfish_full}. Best per metric within each block (edge- vs.\ node-perturbation) in \textbf{bold}; \textit{Cellina} variants in gray, spatially-informed competitors in orange.}
\label{tab:loo_summary_merfish}
\begin{tabular}{lcccc}
\toprule
Method & Pearson $\uparrow$ & Precision\textsubscript{signed} $\uparrow$ & E-distance $\downarrow$ & RMSE\textsubscript{LFC} $\downarrow$ \\
\midrule
Mean shift & 0.43 $\pm$ 0.23 & 0.12 $\pm$ 0.09 & 25.18 $\pm$ 4.75 & 10.98 $\pm$ 2.78 \\
CPA & 0.82 $\pm$ 0.16 & 0.44 $\pm$ 0.13 & 6.53 $\pm$ 1.99 & 6.36 $\pm$ 4.71 \\
scGen & 0.77 $\pm$ 0.16 & 0.20 $\pm$ 0.14 & \textbf{5.73 $\pm$ 4.84} & 6.48 $\pm$ 3.97 \\
\rowcolor{orange!10}
Concert & 0.77 $\pm$ 0.16 & 0.18 $\pm$ 0.15 & 24.91 $\pm$ 2.08 & 7.45 $\pm$ 4.99 \\
\rowcolor{orange!10}
MintFlow & 0.81 $\pm$ 0.17 & 0.29 $\pm$ 0.16 & 19.52 $\pm$ 1.81 & 7.27 $\pm$ 5.43 \\
\rowcolor{gray!10}
Cellina (ablated) & 0.81 $\pm$ 0.15 & 0.37 $\pm$ 0.15 & 8.33 $\pm$ 4.76 & 6.78 $\pm$ 4.54 \\
\rowcolor{gray!10}
Cellina & \textbf{0.84 $\pm$ 0.16} & 0.51 $\pm$ 0.17 & 6.93 $\pm$ 0.83 & 6.08 $\pm$ 5.06 \\
\rowcolor{gray!10}
Cellina-GAT & \textbf{0.84 $\pm$ 0.16} & \textbf{0.53 $\pm$ 0.17} & 7.67 $\pm$ 0.83 & \textbf{5.97 $\pm$ 5.07} \\
\midrule
\rowcolor{gray!10}
Cellina\textsubscript{node-pert} & 0.82 $\pm$ 0.15 & 0.42 $\pm$ 0.15 & 8.96 $\pm$ 2.92 & 6.65 $\pm$ 4.63 \\
\rowcolor{gray!10}
Cellina-GAT\textsubscript{node-pert} & \textbf{0.83 $\pm$ 0.16} & \textbf{0.51 $\pm$ 0.16} & \textbf{8.61 $\pm$ 1.15} & \textbf{5.98 $\pm$ 4.98} \\
\rowcolor{orange!10}
SpatialProp\textsubscript{node-pert} & 0.66 $\pm$ 0.14 & 0.04 $\pm$ 0.04 & 30.79 $\pm$ 5.18 & 8.18 $\pm$ 2.05 \\
\bottomrule
\end{tabular}
\end{table}

\newpage
\textbf{Cellina predicts \textit{in vivo} spatial neighbor perturbation effects.}

Next, we evaluate \textit{Cellina} using \textit{in vivo} interventional Perturb-FISH data from a human melanoma xenograft \citep{binanSimultaneousCRISPRScreening2025}. The data measure CRISPR knockouts (KOs) of NF$\kappa$B-pathway genes in cancer cells and infiltrating T cells. The task is to predict how control T cells (far away from perturbed cancer cells) respond if their neighbors expressed the profile of perturbed cancer cells. During training we hold out T cells that have perturbed cancer cells as neighbors. At inference, we set the expression of cells neighboring the control T cells to the per-gene mean of the KO-specific cancer population and compare the predicted expression against held-out T cells (Appendix \ref{app:data}). 

Predicting the full T cell response is dominated by a shared transcriptional shift that varies little across perturbations, which even the mean shift captures ($0.51$ vs.\ $0.55$ Pearson for \textit{Cellina-GAT}; Appendix Table~\ref{tab:invivo_full_results}); we, therefore, isolate the KO-specific component by removing the average perturbation shift from the predicted and observed logFC~\citep{vinas2025systema}. On this task, \textit{Cellina} variants lead SpatialProp, and a \textit{Cellina} variant is best on every metric except E-distance (\textit{Cellina-GAT} on Pearson, $0.40$, and RMSE\textsubscript{LFC}; base \textit{Cellina} on Signed Precision, $0.68$; Table~\ref{tab:cellina-invivo}). The mean shift has the lowest E-distance yet no KO-specific signal (Pearson $0.06$). To further verify that the perturbation response is due to the knockout effects, rather than a shared effect of having cancer neighbors, we perform an additional experiment in which the neighbors instead express the profile of randomly drawn cancer cells. All methods drop in performance under this random control, and \textit{Cellina}'s KO-specific Pearson falls to zero ($0.33 \rightarrow 0.00$; Appendix Table~\ref{tab:invivo_full_results}), confirming that its predictions are knockout-specific.

\begin{table}[ht]
\centering
\caption{Node perturbation performance for the Perturb-FISH \textit{in vivo} CRISPR KO dataset. KO-specific effect; mean over 9 KOs, $\pm$ std over 5 seeds. A \textit{Cellina} variant (gray) is best on Pearson, Signed Precision and RMSE\textsubscript{LFC}. Spatial competitors in orange.}
\label{tab:cellina-invivo}
\begin{tabular}{lcccc}
\toprule
Method & Pearson $\uparrow$ & Precision\textsubscript{signed} $\uparrow$ & E-distance $\downarrow$ & RMSE\textsubscript{LFC} $\downarrow$ \\
\midrule
Mean shift & 0.06 $\pm$ 0.06 & 0.51 $\pm$ 0.03 & \textbf{0.13 $\pm$ 0.01} & 0.46 $\pm$ 0.02 \\
\rowcolor{gray!10}
Cellina & 0.33 $\pm$ 0.08 & \textbf{0.68 $\pm$ 0.03} & 4.13 $\pm$ 0.15 & 0.43 $\pm$ 0.01 \\
\rowcolor{gray!10}
Cellina-GAT & \textbf{0.40 $\pm$ 0.07} & 0.66 $\pm$ 0.03 & 4.89 $\pm$ 0.19 & \textbf{0.41 $\pm$ 0.01} \\
\rowcolor{orange!10}
SpatialProp & 0.16 $\pm$ 0.06 & 0.55 $\pm$ 0.06 & 7.44 $\pm$ 0.04 & 0.64 $\pm$ 0.04 \\
\bottomrule
\end{tabular}
\end{table}

\subsection{\textit{Cellina} Captures Biologically Meaningful Within-Domain Heterogeneity}
\label{sec:subtypes}

\begin{figure}
    \centering
    \includegraphics[width=0.8\textwidth]{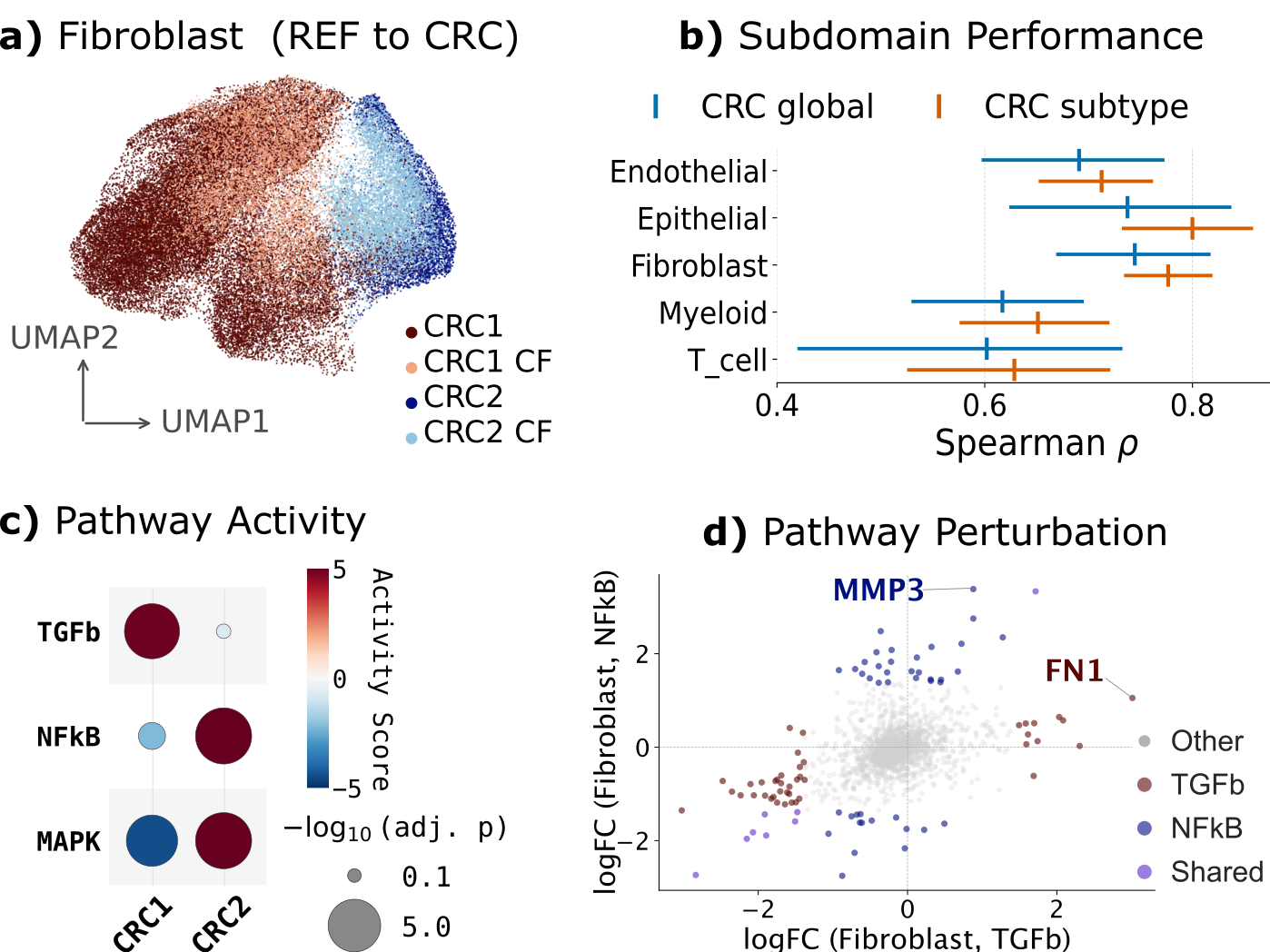}
    \caption{\textbf{Cellina identifies spatial subdomains and
    enables pathway-specific perturbations in CRC.}
    (\textbf{a}) UMAP of \textit{Cellina}-generated Fibroblast counts; in-silico perturbed
    cells (CRC1 CF, CRC2 CF) integrate with their respective observed CRC
    subdomain populations.
    (\textbf{b}) Spearman $\rho$ across cell types (mean, 95\% CI), comparing global (pre-annotated) CRC neighborhoods versus \textit{Cellina}-identified
    subdomains. (\textbf{c}) Inferred pathway activity per CRC
    subdomain. (\textbf{d}) Top-50 most affected genes under
    domain-matched pathway perturbations ($x$-axis: CRC1; $y$-axis: CRC2).}
    \label{fig:application}
\end{figure}

To test whether \textit{Cellina}'s spatial latent $s$ captures microenvironmental variation finer than the domain labels used during training, we ask whether it separates cells within a single annotated domain, i.e., the tumor region of the colorectal cancer cohort \citep{crowell2025tracing}.
We cluster the latent $s$ over CRC cells to discover spatially autocorrelated subdomains \citep{detomaso2021hotspot}, which recovers three modules, of which we analyze the two with the most distinct pathway profiles (denoted CRC1 and CRC2; Appendix~\ref{app:application}). In UMAP space, reconstructed Fibroblast counts from CRC1 and CRC2 form clearly separated clusters, and in-silico perturbed Fibroblasts (REF$\rightarrow$CRC1/2) integrate well with each respective subdomain (Figure~\ref{fig:application}(a)), suggesting that $s$ effectively captures local variation.

\paragraph{\textit{Cellina}'s predictions are specific to the local neighborhood.}
To support that $s$ captures context-specific variation, first, we compare predictions conditioned on neighborhoods sampled from the matched subdomain against those conditioned on the global CRC domain, scored against the same observed reference, and find the former better on average across the five cell types (Figure~\ref{fig:application}(b)). Further, when we rewire every neighbor to a single cell type (Table~\ref{tab:homotypic}), we see that identical neighbors yield different responses across focal cell types, and that predictions do not merely propagate neighbor expression, since a neighborhood of the held-out type itself is rarely the best case. We also observe that edge-perturbation predictions depend on the neighborhood size in a cell-type-specific manner and degrade once the neighborhoods span a large part of the slide (Table~\ref{tab:k_sweep}). Similarly, under node perturbation, the agreement with the observed response increases with the fraction of perturbed neighbors (Table~\ref{tab:dose}).

\paragraph{Subdomains recover interpretable signaling programs.} 
To interpret CRC1 and CRC2 biologically, we score their gene modules against PROGENy pathway gene sets~\citep{schubert2018perturbation, badia2022decoupler}, recovering distinct signaling profiles (Figure~\ref{fig:application}(c)): TGF$\beta$-dominant for CRC1 and NF$\kappa$B/MAPK-dominant for CRC2. These signatures arise without direct supervision and are consistent with the signaling heterogeneity reported by \citet{crowell2025tracing}.

\paragraph{Pathway-specific neighbor perturbations are consistent with biologically grounded responses.}
We next ask whether prior pathway knowledge alone suffices to reproduce these subdomain responses (Figure~\ref{fig:application}(d)). Encouragingly, performing node perturbations with PROGENy pathway weights as alteration vectors $\delta_g$ recovers the subdomain effects to a large extent (Appendix~\ref{app:application}; Pearson $r = 0.77$ (TGF$\beta$$\to$CRC1) and $r = 0.76$ (NF$\kappa$B$\to$CRC2)). Notably, the most up-regulated gene predicted under each pathway (FN1 for TGF$\beta$, MMP3 for NF$\kappa$B) is consistent with established fibroblast programs:
TGF$\beta$ signaling is a canonical driver of extracellular-matrix production (FN1), while NF$\kappa$B activation induces matrix-remodeling metalloproteinases (MMP3)---observed hallmarks of cancer-associated fibroblast and immune microenvironments \citep{crowell2025tracing}.

\section{Discussion and Conclusion}
\label{sec:conclusion}

Foundation models of single cells are scaling rapidly \citep{bunne_how_2024}, yet most still model cells as independent samples \citep{rood2024toward}. 
For such models to serve as virtual cells in tissue context, they must also predict how a cell's state changes with its neighborhood. Recent work has demonstrated compelling instances of modeling tissue perturbations: in-silico cell-type depletion and swapping via label manipulation, including regulatory T cell modulation relevant to cellular therapy \citep{akbarnejad2025mapping}; combinatorial queries on perturbations and covariates in interventional spatial data \citep{lin2025concert}; and gradual microenvironment steering, paired with causal analyses linking perturbation predictions to downstream response \citep{sun2025spatialprop}. 
We hope that our formalization of tissue graph counterfactuals can serve as a common language for such instances, and for building and evaluating models that predict how cells respond to altered neighborhoods. Although such perturbations remain in-silico hypotheses that require wet-lab validation, they offer a way to query virtual cell responses in spatial context and could help guide future experimental design.

Spatial neighborhoods are continuous, compositionally heterogeneous contexts, and averaging over coarse labels or shifts ignores the variation within them.
Two design choices follow directly from this observation and underpin \textit{Cellina}'s performance. First, encoding neighborhoods continuously yields a strong prior even without any supervision: \textit{Cellina (ablated)}, trained without cell-type or domain labels, already matches all non-\textit{Cellina} baselines on correlation and precision in colorectal cancer. Second, supervised disentanglement adds substantial further gains. \textit{Cellina}'s explicit disentanglement additionally enables the discovery of the subdomain structure that can be used to refine pathology annotations, and simulate pathway-specific neighbor perturbations. Together, these components allow a \textit{Cellina} variant to outperform non-spatial baselines on the logFC-based metrics in every dataset we evaluated, whereas none of the other spatially-informed methods consistently outperforms the strongest non-spatial baseline. \textit{Cellina} itself is a relatively simple model, and the more expressive but costlier \textit{Cellina-GAT} leads on node perturbation, but performs largely on par in edge perturbations, indicating that the pre-aggregated neighborhoods already capture the relevant microenvironmental structure; this is
consistent with a broader pattern in single-cell perturbation modeling, where simpler approaches
often match more complex ones \citep{ahlmann2025deep, vinas2025systema}.

\textbf{Limitations and future directions.}
\label{sec:limitations}

\textit{Cellina}'s supervised disentanglement objectives rely on cell-type and spatial-domain annotations, themselves simplifications of continuous biology. While the annotations are required only during training, and not used during inference, they can impact how well the spatial latent $s$ reflects neighborhood variation.
Furthermore, the underlying imaging-based spatial transcriptomics data also depend on cell segmentation, which remains error-prone: transcripts are frequently misassigned across cell boundaries and can dominate downstream niche and neighbor-influence analysis \citep{mitchel2026impact}. A natural extension is to account for segmentation uncertainty within \textit{Cellina}, for instance by jointly modeling raw imaging and transcript-level data.

Beyond these data-related limitations, our predictions are not causal effect estimates, even where they are evaluated against interventional data. Instead, we use \textit{counterfactual} operationally to denote simulated outcomes under altered neighborhoods, rather than in the strict Pearlian sense. Similarly, our evaluation is a context transfer: because spatial assays are destructive, predictions are scored against the observed target-domain population rather than against the same cells after an intervention.
Further, \textit{Cellina} does not impose the
structural assumptions required for identifiability~\citep{khemakhem2020variational,
lachapelle2022disentanglement}, though it could be extended with, e.g., sparse mechanism
priors~\citep{lachapelle2022disentanglement, lopez2023learning}, as a path toward causal representation learning with formal guarantees~\citep{scholkopf2021toward}. Moreover, although \textit{Cellina} is a probabilistic model, we report only mean predictions and leave calibration of its predictive uncertainty to future work.
Finally, emerging spatial perturbation screens~\citep{dhainaut2022spatial, breinig2026integrated, binanSimultaneousCRISPRScreening2025}, albeit limited in scale and resolution, increasingly provide spatially resolved interventional readouts that could ground these predictions empirically. Our task on spatially-resolved knockouts profiled \textit{in vivo} is an early attempt at this, and cell-depletion experiments with single-cell spatial readouts would provide a natural interventional test for future depletion simulations.

\newpage
\section*{Acknowledgments and Disclosure of Funding}
The authors’ work is supported through state funds approved by the State Parliament of Baden-Württemberg for the Innovation Campus Health + Life Science alliance Heidelberg Mannheim, the Data Science Collaborative Research Programme 2022 by the Novo Nordisk Foundation (grant NNF22OC0076414), the Priority Program Translational Oncology of the Deutsche Krebshilfe (grant 70115167), the Helmholtz Association under the joint research school “HIDSS4Health – Helmholtz Information and Data Science School for Health”, the European Research Council (Synergy Grant DECODE 810296), and KU Leuven (BOFZAP starting grant STG/22/024 and C1 grant C14/24/151). We also thank Philipp Sven Lars Schaeffer, Zeynep Edizcan, Ahmet Rifaioglu, Elyas Heidari, Rama Abdulhamid, Ricardo Ramirez Flores, and Julio Saez-Rodriguez for their feedback. The authors declare that they have no competing interests or financial relationships outside the submitted work.

\bibliographystyle{plainnat}
\bibliography{bibliography}

\appendix
\setcounter{figure}{0}
\setcounter{table}{0}
\renewcommand{\thefigure}{A\arabic{figure}}
\renewcommand{\thetable}{A\arabic{table}}
\let\appendixsectionorig\section
\renewcommand{\section}{\FloatBarrier\appendixsectionorig}

\newpage
\section{Evaluation Metrics}
\label{app:metrics}

Unless stated otherwise, all metrics are computed using library-size–normalized gene expression, with fixed library size $\ell_0 = 10^{4}$. Let $c^{(\mathrm{obs})}_{g,v}$ denote the observed raw count of gene $g$ in cell $v$, and $c^{(\mathrm{pred})}_{g,v}$ the corresponding model-predicted raw count. We convert both to normalized expression as
\[
p_{g,v} = \frac{c_{g,v}}{\sum_{g'} c_{g',v}} \cdot \ell_0
\]

We consider perturbation settings where each gene is evaluated between a control and a perturbed condition. The log-fold change (logFC) for gene $g$ is defined as
\[
\mathrm{logFC}_g = \log(p^{(\mathrm{pert})}_g + 1) - \log(p^{(\mathrm{ctrl})}_g + 1),
\]
computed either from observed data or from model predictions.

For observed data, $p^{(\cdot)}_g$ denotes the empirical mean of normalized expression across cells in the corresponding condition. For model predictions, the model outputs raw counts $\hat{c}^{(\mathrm{pert})}_{g,v}$, which are first normalized and then averaged:
\[
p^{(\mathrm{pert})}_g = \mathbb{E}_{v \in \mathrm{pert}} \left[ \frac{\hat{c}^{(\mathrm{pert})}_{g,v}}{\sum_{g'} \hat{c}^{(\mathrm{pert})}_{g',v}} \cdot \ell_0 \right],
\]
and analogously for the control condition.

In all cases, $p$ denotes library-size–normalized gene expression, either computed from observed counts, predicted counts, or model-specific mean parameters (e.g., NB mean for generative models or inverse-transformed outputs for log-normalized models such as scGen). The observed and predicted log-fold change vectors are thus defined as:
    \[
    \mathbf{real} = \left(\mathrm{logFC}^{(\mathrm{obs})}_g\right)_{g \in \mathcal{T}}, \quad
    \mathbf{pred} = \left(\mathrm{logFC}^{(\mathrm{pred})}_g\right)_{g \in \mathcal{T}}
    \]
    We restrict evaluation to the top differentially expressed genes defined as
    \[
    \mathcal{T} = \operatorname*{arg\,top}_n \left( \left| \mathrm{logFC}^{(\mathrm{obs})}_g \right| \right)
    \]

    \textbf{Pearson $r$}: Pearson correlation between observed and predicted log-fold changes across genes, i.e., between $\mathbf{real}$ and $\mathbf{pred}$.

    \textbf{Spearman $\rho$}: Spearman rank correlation between $\mathbf{real}$ and $\mathbf{pred}$.

    \textbf{RMSE (counts)}: Root mean squared error between predicted and observed
    counts on ground-truth differentially expressed (DE) genes
    \[
    \mathrm{RMSE}(\mathbf{p}, \hat{\mathbf{p}}; \mathcal{T}) = \sqrt{\frac{1}{|\mathcal{T}|} \sum_{i \in \mathcal{T}} \big(\hat{p}_i - p_i\big)^2}
    \]
    For better readability, we report $\text{log}_{10}(\mathrm{RMSE}_{\mathrm{counts}})$.

    \textbf{Signed Precision}: the fraction of the top-$n$ observed DE genes $\mathcal{T}$ that are also among the top-$n$ genes by predicted $|\mathrm{logFC}|$ and have the same sign:
    \[ \mathrm{Precision}^{\mathrm{signed}}_n(\mathbf{real}, \mathbf{pred}) = \frac{1}{n} \sum_{i \in \mathcal{T} \cap \mathcal{T}^{\mathrm{pred}}} \mathbf{1}\big(\operatorname{sign}(\mathbf{real}_i) = \operatorname{sign}(\mathbf{pred}_i)\big) \]

    \textbf{RMSE (log-fold change)}: This metric evaluates magnitude differences between predicted and ground truth logFC on a selected set of differentially expressed genes.
    It computes the root mean squared error (RMSE) between the ground-truth and predicted values restricted to that gene set.
    \[
    \mathrm{RMSE}_{\mathrm{LFC}}(\mathbf{real}, \mathbf{pred}; \mathcal{T}) = \sqrt{ \frac{1}{|\mathcal{T}|} \sum_{i \in \mathcal{T}} \big( \mathbf{real}_i - \mathbf{pred}_i \big)^2 }
    \]

    \textbf{E-distance (local)}: Energy distance is a widely used distribution-level metric for single-cell data \citep{peidli2024scperturb}, measuring the overall difference between two cell populations; here, the observed ($X$) and predicted ($Y$) populations. We use the local variant proposed by \citet{heidari2026evaluating}, which restricts pairwise comparisons to $k$-nearest neighborhoods, making it more sensitive to disruptions of gene-gene co-expression that the global E-distance may miss. Here, $\mathcal{N}_k^X(\cdot)$ and $\mathcal{N}_k^Y(\cdot)$ denote the $k=10$ nearest cells in $X$ and $Y$ by Euclidean distance, excluding the query cell itself in the within-population terms:
        \begin{multline*}
        E_{\text{local}}(X,Y)
        =
        \frac{1}{nk}\sum_{i=1}^{n}\Bigg[
        \sum_{y'\in \mathcal{N}_k^Y\!(x_i)}\!\|x_i - y'\|
        +\sum_{x'\in \mathcal{N}_k^X\!(y_i)}\!\|y_i - x'\|\\
        -\sum_{x'\in \mathcal{N}_k^X\!(x_i)}\!\|x_i - x'\|
        -\sum_{y'\in \mathcal{N}_k^Y\!(y_i)}\!\|y_i - y'\|
        \Bigg]
        \end{multline*}
    Unlike the standard E-distance, which captures global distributional discrepancy, this variant measures local manifold differences. It approaches zero whenever the predicted cells lie on the local manifold of the observed ones. Following \citet{fu2026strand}, we fit a 50-component PCA map $T(\cdot)$ on the training split and evaluate $E_{\text{local}}\big(T(X), T(Y)\big)$ on log-normalized expression. For robustness and speed, we subsample $n=200$ cells from each population and average over 10 subsamples.

    \textbf{NB deviance}: E-distance above is computed on log-normalized expression, which is the space
    log-normalized models are trained to reconstruct, whereas \textit{Cellina} and CPA are fit to a
    negative-binomial likelihood on raw counts. To assess distributional fit in count space as well, we also
    report the negative-binomial unit deviance of the predicted mean against the observed mean counts. To keep
    the comparison independent of any single model's fitted dispersion, we estimate one dispersion $\theta_g$
    per gene from the observed counts of the target population and score every method under the same $\theta_g$:
    \[
    D = \frac{1}{G} \sum_{g=1}^{G}
\max\!\left(0,\; 2\left[
  \bar{c}^{(\mathrm{obs})}_{g} \log \frac{\bar{c}^{(\mathrm{obs})}_{g}}{\bar{\mu}_{g}}
  - \bigl(\bar{c}^{(\mathrm{obs})}_{g} + \theta_g\bigr)
    \log \frac{\bar{c}^{(\mathrm{obs})}_{g} + \theta_g}{\bar{\mu}_{g} + \theta_g}
\right]\right)
    \]
    where $\bar{c}^{(\mathrm{obs})}_{g} = \frac{1}{|\mathcal{V}_{\mathrm{obs}}|}\sum_{v \in \mathcal{V}_{\mathrm{obs}}} c^{(\mathrm{obs})}_{g,v}$ is the observed mean, $\bar{\mu}_{g} = \frac{1}{|\mathcal{V}_{\mathrm{pred}}|}\sum_{v \in \mathcal{V}_{\mathrm{pred}}} \hat{\mu}_{g,v}$ is the method's predicted mean, $\mathcal{V}$ the evaluated cells, and the first term is taken as 0 where $\bar{c}^{(\mathrm{obs})}_{g} = 0$. Because scGen and the mean shift
    define no dispersion of their own, and their output is inverse-transformed log-normalized expression, this is the only variant of the metric under which they are not scored
    against a dispersion borrowed from a competing model.

\section{Code and Data Availability, and Pre-Processing}
\label{app:data}

\subsection{Code Availability}
\textit{Cellina} is available at \url{https://github.com/PMBio/cellina} under the BSD 3-Clause License. Scripts and configurations required to download all data and reproduce the experiments presented in this work are available at \url{https://github.com/PMBio/cellina-reproducibility}. Tutorials for \textit{Cellina} and \textit{Cellina-GAT} are at \url{https://cellina.readthedocs.io/}.
All experiments were re-run for this version of the manuscript with \textit{Cellina} v1.2.0, which fixes an issue with log-transformation of spatial features and improves graph sampling; the shared evaluation pipeline was updated accordingly, so some baseline values also differ slightly from the submitted version. Rankings against competitors and main conclusions are unchanged.

\subsection{Colorectal Cancer Patient Cohort}
We downloaded the processed CRC data \citep{crowell2025tracing} as AnnData files from Zenodo (https://zenodo.org/records/15574384). On each slide, we apply a standard feature selection procedure (Seurat-flavor) to subset each slide to the 2000 highly variable genes implemented in \texttt{scanpy} \citep{wolf2018scanpy}. The original CRC dataset contains eight slides, two of which (slide IDs 110, 222) were not used in our analyses for the following reasons: slide 110 contained sequencing artifacts in the form of major empty patches disrupting neighborhood computations, while slide 222 did not contain any REF cells. Cell-type labels $y$ are taken directly from the source study's annotation, and the spatial domain label $d$ is that study's independent histopathological region annotation. For evaluations, we merged the study's fine-grained subtypes (e.g., epithelial subpopulations annotated as Epi1-Epi4) into broad categories, so that $y$ denotes intrinsic cell lineage, independent of spatial domain. This was important since some of the subtypes, specifically epithelial subtypes, are domain-specific, so retaining them would conflate cell type identity with domain (e.g., cancer vs.\ healthy Epithelial cells) and make the REF~$\rightarrow$~CRC prediction task circular.

For the disentanglement benchmark, all sections were merged into a single dataset prior to model fitting. For all other evaluations and downstream applications presented in this work, models were instead fit on one section at a time.

\subsection{Whole-Brain Mouse Cohort}
We downloaded the MERFISH whole-brain mouse cohort \citep{zhang2023molecularly} from the CZI CELLxGENE portal (https://datasets.cellxgene.cziscience.com/93c3bb97-ea05-4ee0-a760-a1508cd04612.h5ad), from which we selected three adjacent coronal slides from the mid-brain (C57BL6J-2.036, C57BL6J-2.039, C57BL6J-2.041). We retained all features without further selection, and the published cell-type and anatomical-region annotations are used unchanged. We further restricted the analysis to three anatomical domains, selected based on (i) their overall abundance, (ii) their consistent representation across the three selected sections, and (iii) the diversity of cell types they contained. Of these, the Thalamus was used as the source domain, while Fiber-tracts and Isocortex served as target domains.

\subsection{Perturb-FISH Tumor Xenograft Dataset}

We downloaded the Perturb-FISH tumor xenograft dataset of \citet{binanSimultaneousCRISPRScreening2025}, a human melanoma xenograft profiled with a 500-gene panel, from \url{https://api.brainimagelibrary.org/web/view?bildid=ace-gem-get}, and use it for the \textit{in vivo} intervention simulation and evaluation. Cells are annotated into three categories (cancer, T cell, and other), and each cancer cell additionally carries a perturbation label indicating which NF$\kappa$B-pathway gene was knocked out in it, or \textit{Control} if unperturbed. We filter cells by a minimum count per cell and remove cells with more than one simultaneous knockout to isolate single-gene effects, yielding 154{,}111 annotated cells. We then subset the 500-gene panel to the 154 genes whose average expression is at least one-fifth of a count per cell. To construct the KO set used during evaluation, we note that the effects of a knockout can only propagate to a neighbor if the KO affects the cancer cell itself. We therefore keep KOs with a noticeable effect in the perturbed cells themselves, i.e.\ those that shift at least 3 genes in the perturbed cancer cells (Mann--Whitney U test with Benjamini--Hochberg FDR correction, $q<0.05$, for significance and Cohen's $d>0.2$ for effect size), and that are surrounded by at least 60 T cells. This leaves 9 KOs: MAP2K2, IRAK1, IRF7, PELI1, IRF3, IRF5, NFKBIA, LBP, and MAP2K3. For each knockout, we score predictions against unperturbed T cells with at least one neighboring cancer cell carrying that knockout and none carrying a different one. During evaluation, we sample 6{,}000 control T cells and simulate a perturbation in their neighborhood by setting neighbors to the mean of the KO-specific cancer population, a choice motivated by the quality of the data. The processed data is available at \url{https://doi.org/10.6084/m9.figshare.34790289}. 

\subsection{Graph Pre-Processing (Shared)}
We process each sample independently to compute a spatial neighbor graph using a Gaussian proximity kernel with bandwidth $\sigma = 100$\,\textmu m,
consistent with the physical length scales at which neighboring cells exchange molecular
signals~\citep{armingol2021deciphering}:
\begin{equation}
    W_{uv} = \exp\!\left(-\frac{d(u,v)^2}{2\sigma^2}\right)
    \label{eq:kernel}
\end{equation}
where $d(u,v)$ is the Euclidean distance between cell centroids. $W_{uv}$ is set to zero for pairs outside the ($k$) nearest neighbors of either cell or when $W_{uv} < \tau = 0.1$; self-loops are excluded ($W_{vv} = 0$). Nearest neighbors parameter $k$ varies per experiment and model: for \textit{Cellina} we use $k=200$ for CRC and MERFISH experiments while for \textit{Cellina-GAT}, we use $k=50$ for both datasets due to the high training cost of \textit{Cellina-GAT}. Both use $k=30$ for the \textit{in vivo} experiment. Furthermore, for \textit{Cellina-GAT} we use binary edge weights ($W_{uv} \in \{0, 1\}$), letting attention learn edge importance during message passing.

\subsection{Edge-Perturbation Donor Sampling}
\label{app:donor_sampling}
In this work, we simulate \textbf{edge perturbation} as an in-silico domain edge rewiring. Let $\mathcal{I}_y \subset \mathcal{V}$ denote the focal cells of type $y$ in the source domain $d$, whose counterfactual expression we wish to predict, and let the \emph{anchors} $\mathcal{A}_y$ be the cells of the same type $y$ in the target domain $d'$. For each focal cell $v \in \mathcal{I}_y$, we drop its original edges and assign it new neighbors $\mathcal{N}'(v)$ taken from the neighborhoods that the anchors have in $d'$, within the same slide. In all leave-one-out benchmarks, cells of type $y$ are removed from these neighborhoods, and the anchors are excluded from the neighborhood graph $W$ during both training and inference, i.e., their expression never enters the model and serves solely as evaluation ground truth; the anchors' neighborhoods $\mathcal{N}(c)$ are read from the unmasked spatial graph.

We implement two draws for $\mathcal{N}'(v)$ (Algorithms~\ref{alg:anchored} and~\ref{alg:rewired}). In the \emph{anchored} draw, each focal cell inherits the observed neighborhood of one anchor $c \sim \mathrm{Uniform}(\mathcal{A}_y)$, sampled with replacement across focal cells, together with its Gaussian-kernel weights (Equation~\ref{eq:kernel}; binary edges for \textit{Cellina-GAT}); i.e., the focal cell is placed into a real neighborhood from $d'$. In the \emph{rewired} draw, we instead pool the neighbors of all anchors into $\mathcal{P}_{\setminus y}$ and sample $n_d = 50$ donors per focal cell from this pool without replacement, attached with unit weights; the resulting neighborhood consists of cells observed around type-$y$ cells in $d'$, but does not correspond to any single observed neighborhood. In both draws, $\varphi(v)$ (\textit{Cellina}) or the local subgraph $\mathcal{G}_v$ (\textit{Cellina-GAT}) is then rebuilt from the new neighbors.

All leave-one-out edge perturbation benchmarks use the anchored draw for both \textit{Cellina} and \textit{Cellina-GAT}. The rewired draw is used where the target composition does not occur in the tissue; for example, for the homotypic neighborhoods of Appendix~\ref{app:neighborhood}, where the donor pool is instead all cells of a single type in $d'$, including the held-out type $y$ itself. The draws are made once per run with a fixed seed and shared between expression and latent predictions.

\begin{algorithm}[h]
\caption{Edge perturbation, \emph{anchored} draw. Focal cells $\mathcal{I}_y$: type $y$ in source domain $d$. Anchors $\mathcal{A}_y$: type-$y$ cells in target domain $d'$, held out from $W$ at training and inference; $\mathcal{N}(c)$ is read from the unmasked spatial graph, so anchors contribute only their neighborhoods, never their expression.}
\label{alg:anchored}
\begin{algorithmic}[1]
\Statex \textbf{Cellina}
\State drop every edge incident on $\mathcal{I}_y$
\For{each focal cell $v \in \mathcal{I}_y$}
  \State $c \sim \mathrm{Uniform}(\mathcal{A}_y)$ \Comment{with replacement across focal cells}
  \State $\mathcal{N}'(v) \gets \mathcal{N}(c) \setminus \{u : y_u = y\}$ with the kernel weights $W_{uc}$ of Equation~\ref{eq:kernel}
  \State $\varphi(v) \gets \sum_{u \in \mathcal{N}'(v)} W_{uc}\, \tilde{x}_u \,/ \sum_{u \in \mathcal{N}'(v)} W_{uc}$ \Comment{recomputed weighted mean}
\EndFor
\Statex \textbf{Cellina-GAT}
\State drop every edge incident on $\mathcal{I}_y$
\For{each focal cell $v \in \mathcal{I}_y$}
  \State $c \sim \mathrm{Uniform}(\mathcal{A}_y)$ \Comment{with replacement across focal cells}
  \State $\mathcal{N}'(v) \gets \mathcal{N}(c) \setminus \{u : y_u = y\}$
  \State add edges $u \to v$ for all $u \in \mathcal{N}'(v)$
\EndFor
\State rebuild the local subgraph $\mathcal{G}_v$ for all $v \in \mathcal{I}_y$ and encode $s \sim q(s \mid \mathcal{G}_v)$ by message passing
\end{algorithmic}
\end{algorithm}

\begin{algorithm}[h]
\caption{Edge perturbation, \emph{rewired} draw. Focal cells $\mathcal{I}_y$ and anchors $\mathcal{A}_y$ as in Algorithm~\ref{alg:anchored}. Donor pool $\mathcal{P}_{\setminus y} = \bigcup_{c \in \mathcal{A}_y} \mathcal{N}(c) \setminus \{u : y_u = y\}$.}
\label{alg:rewired}
\begin{algorithmic}[1]
\Statex \textbf{Cellina}
\State drop every edge incident on $\mathcal{I}_y$
\For{each focal cell $v \in \mathcal{I}_y$}
  \State $\mathcal{N}'(v) \sim \mathrm{Sample}(\mathcal{P}_{\setminus y},\, n_d,\, \text{without replacement})$
  \State add unit-weight edges $v \leftrightarrow \mathcal{N}'(v)$
  \State $\varphi(v) \gets \frac{1}{n_d} \sum_{u \in \mathcal{N}'(v)} \tilde{x}_u$ \Comment{recomputed mean aggregate}
\EndFor
\Statex \textbf{Cellina-GAT}
\State drop every edge incident on $\mathcal{I}_y$
\For{each focal cell $v \in \mathcal{I}_y$}
  \State $\mathcal{N}'(v) \sim \mathrm{Sample}(\mathcal{P}_{\setminus y},\, n_d,\, \text{without replacement})$
  \State add edges $u \to v$ for all $u \in \mathcal{N}'(v)$
\EndFor
\State rebuild the local subgraph $\mathcal{G}_v$ for all $v \in \mathcal{I}_y$ and encode $s \sim q(s \mid \mathcal{G}_v)$ by message passing
\end{algorithmic}
\end{algorithm}

\subsection{Node-Perturbation Shift and Gene-Set Sweep}
\label{app:convergence}

Recall that node perturbation modifies a subset $\mathcal{S}$ of $m = |\mathcal{S}|$ genes per neighbor cell. In our experiments, we alter the genes of neighbors in source domains by the log fold-change (logFC) shift $\delta$, representing the average difference between pseudobulked cell populations. Concretely, $\delta$ is added to the log-normalized count $\tilde{x}_{u,g}$ of each neighbor $u$, i.e.\ $\tilde{x}_{u,g}^{\mathrm{cf}} = \tilde{x}_{u,g} + \delta_{g}$ for $g \in \mathcal{S}$ (with $\delta_{g}$ replaced by $\delta_{y(u),g}$ in the cell-type-specific variant below). Specifically, we calculate a \textbf{global} and a \textbf{cell-type-specific} $\delta$. Let $\ell_0 = 10^{4}$ denote the target library size, $\mathcal{C}^{d}$ the set of cells in domain $d \in \{\text{REF}, \text{CRC}\}$, $\mathcal{C}^{d}_{y} \subseteq \mathcal{C}^{d}$ its restriction to cell type $y$, and $x_{i,g}$ the raw count of gene $g$ in cell $i$.

\begin{itemize}
    \item \textbf{Global.} Pseudobulk across all cells per domain, $b^{d}_{g} = \sum_{i \in \mathcal{C}^{d}} x_{i,g}$, library-normalize, and log-transform: $\tilde{b}^{d}_{g} = \log\!\left(1 + \ell_0\, b^{d}_{g} / \sum_{g'} b^{d}_{g'}\right)$. The global shift is $\delta_{g} = \tilde{b}^{\,\text{CRC}}_{g} - \tilde{b}^{\,\text{REF}}_{g}$.
    \item \textbf{Cell-type-specific.} Pseudobulk within each cell type $y$ and domain, $b^{d}_{y,g} = \sum_{i \in \mathcal{C}^{d}_{y}} x_{i,g}$, library-normalize, and log-transform: $\tilde{b}^{d}_{y,g} = \log\!\left(1 + \ell_0\, b^{d}_{y,g} / \sum_{g'} b^{d}_{y,g'}\right)$. The cell-type-specific shift is $\delta_{y,g} = \tilde{b}^{\,\text{CRC}}_{y,g} - \tilde{b}^{\,\text{REF}}_{y,g}$.
\end{itemize}

Note that \textit{Cellina}'s neighborhoods aggregate log1p-normalized expression, while \textit{Cellina-GAT} operates on raw count data. Accordingly, for \textit{Cellina}, the perturbation is an additive shift in log-normalized space ($\tilde{x}^{\mathrm{cf}} = \tilde{x} + \delta$), whereas for \textit{Cellina-GAT} the corresponding operation is a multiplicative scaling on counts ($x^{\mathrm{cf}} = x \cdot e^{\delta}$), obtained by exponentiating the logFC shift. As such, \textit{Cellina-GAT} applies $T_g$ directly in count space, $T_g : \mathbb{R}_{\geq 0} \to \mathbb{R}_{\geq 0}$. Also, to preserve a strict counterfactual setting, for the held-out cell type $y$ we assign the global shift $\delta_{\setminus y}$, pseudobulked with type $y$ removed, so the perturbed input carries the domain-level shift only and never the held-out population's own signature.

\begin{figure}[htbp]
    \centering
    \includegraphics[width=\textwidth]{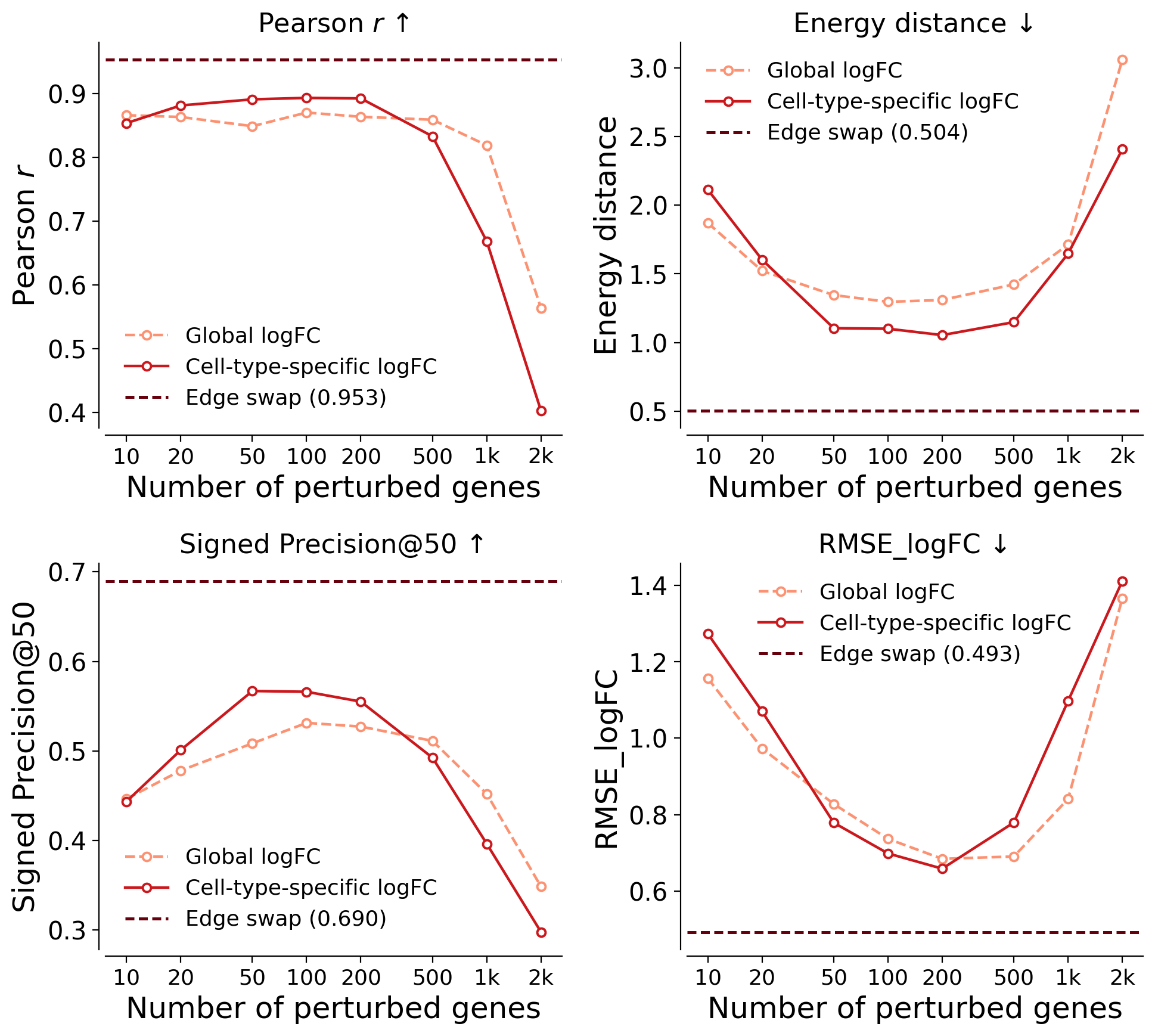}
    \caption{\textbf{Node perturbation as a function of the perturbed gene set.} We sweep over a range of gene sets and evaluate in-distribution node perturbations against the target domain under two choices of the logFC shift $\delta_{g}$: a \emph{global} logFC pooled across all cells (dashed, light), and a \emph{cell-type-specific} $\delta_{y,g}$ logFC computed within cell-type groups (solid, dark). The maroon dashed line marks the edge perturbation reference, i.e.\ a complete neighborhood swap, plus sampling from the domain target neighbors.}
    \label{fig:convergence}
\end{figure}

In this sweep (Figure~\ref{fig:convergence}), we tested the performance of \textit{Cellina} as $m$ approaches the total number of genes $G$, i.e., every gene of every neighbor is altered by $\delta_g$ or $\delta_{y,g}$, producing a transcriptome-wide shift for all neighborhoods toward the target domain profile. Here, we see that all four metrics improve with $m$ for both logFC variants and saturate by $m\!\approx\!100$--$200$, collapsing shortly after (Figure~\ref{fig:convergence}). At $m=200$ (in-distribution), the cell-type-specific variant consistently outperforms the global one and reaches Pearson $r \approx 0.89$, Signed Precision $\approx 0.56$, E-distance $\approx 1.05$, and $\text{RMSE}_\text{LFC}$ of $0.66$, compared with the edge perturbation ceiling of $0.95$, $0.69$, $0.50$, and $0.49$ respectively. Node perturbation therefore saturates \emph{below} the edge-perturbation reference for every gene set, because $\delta$ is a population-level logFC: neighbors are shifted toward an average target profile rather than onto the profiles of real observed cells, so no choice of $\mathcal{S}$ recovers a real neighborhood's composition (Appendix~\ref{app:operators}). We assume that this is a biologically meaningful result: biological perturbations are often assumed to elicit relatively sparse shifts on a few genes, rather than in full expression space \citep{lopez2023learning}.

\section{Hyperparameters, Architecture, and Running Time Details}
\label{app:hyperparams}
We implement \textit{Cellina} using the \texttt{scvi-tools} API \citep{gayoso2022python} -- a standard framework in single-cell genomics. Hence, we adopt most default hyperparameters and best practices recommended by \texttt{scvi-tools}. To ensure a consistent comparison, all methods are evaluated using their default settings across experiments and whenever possible using count likelihoods.

\subsection{Compute Resources}
\label{app:compute}
All experiments were conducted on GPU machines with four NVIDIA GeForce RTX 4090, AMD Ryzen Threadripper PRO 7975WX 32-Cores, and 500GB of RAM. 

We provide training times for competitors and \textit{Cellina} variants in Figure \ref{fig:scalability}. A single training and inference run of \textit{Cellina} takes approximately one hour on the CRC slides from \citet{crowell2025tracing} and approximately 20 minutes on the mouse brain data from \citet{zhang2023molecularly}. The in-silico perturbation evaluations reported in Tables~\ref{tab:loo_summary_crc} and~\ref{tab:loo_summary_merfish} are the most expensive experiments, requiring roughly ten and two days of compute, respectively. The ablation sweeps in the Appendix require three to five days, while the remaining appendix experiments each complete within one day. In total, all reported experiments required approximately two to three weeks of GPU time on the hardware described above.

\begin{table}[htbp]
\centering
\small
\caption{Architecture and training hyperparameters. \textit{Cellina} uses a VAE architecture, with standard scvi-tools defaults for single-cell data, trained with unit weights on all active regularization terms, each normalized to the VAE loss scale (the graph contrastive term applies to \textit{Cellina-GAT} only).}
\label{tab:hyperparams}
\begin{tabular}{llcc}
\toprule
Component & Parameter & Cellina & Cellina-GAT \\
\midrule
$z$-encoder   & Hidden dim            & 128 & 128 \\
$z$-encoder   & Layers                & 2   & 3   \\
$z$-encoder   & Latent dim $m$        & 64  & 64  \\
$s$-encoder   & Hidden dim            & 128 & 128 \\
$s$-encoder   & Layers                & 2   & 3   \\
$s$-encoder   & Latent dim $m$        & 64  & 64  \\
Decoder       & Hidden dim            & 128 & 128 \\
Decoder       & Layers                & 2   & 3   \\
Discriminator & Hidden dim            & 32  & 32  \\
Discriminator & Layers                & 2   & 2   \\
GNN           & Convolution           & —   & GATv2 \\
Training      & Batch size            & 2048 & 256 \\
Training      & Max epochs            & 100 & 100 \\
Training      & Optimizer             & Adam & Adam \\
Training      & Learning rate         & $10^{-3}$ & $10^{-3}$ \\
Training      & Weight decay          & $10^{-6}$   & $10^{-4}$ \\
Training      & KL warmup             & linear, $0\to1$ & linear, $0\to1$ \\
Training      & $\lambda_\mathrm{clf}$     & 1 & 1 \\
Training      & $\lambda_\mathrm{disc} = \lambda_\mathrm{adv}$ (tied) & 1 & 1 \\
Training      & $\lambda_\mathrm{spatial}$    & 0 & 1 \\
Spatial graph & Bandwidth             & 100 $\mu$m & 100 $\mu$m \\
Spatial graph & Kernel                & Gaussian & — \\
Spatial graph & Cutoff                & 0.1 & — \\
Count Decoding & Distribution          & Negative Binomial & Negative Binomial \\
\bottomrule
\end{tabular}
\end{table}

\section{Model Details and Training Objective}
\label{app:model_details}

\subsection{Generative Model}
\label{app:generative}

\textit{Cellina} is a (graph) variational autoencoder \citep{kingma2013auto, kipf2016variational} with two latent variables:

\begin{itemize}[noitemsep]
\item $z \in \mathbb{R}^m$: intrinsic cell identity, capturing variation independent of spatial context
\item $s \in \mathbb{R}^m$: spatial niche representation, capturing microenvironmental variation
\end{itemize}

The joint generative model is:

\[p(x, z, s) = p(x \mid z, s)\, p(z)\, p(s)\]

with standard normal priors $p(z) = p(s) = \text{Normal}(0, I_m)$.

The likelihood used in the evaluations throughout this study is a Negative Binomial (NB) distribution over counts, a standard choice in modeling single-cell data \citep{lopez2018deep, gayoso2022python}, with parameters produced by a decoder operating on $[z;\, s] \in \mathbb{R}^{2m}$:

\[
x \sim \mathrm{NB}(\mu, r), \quad 
\text{with } \mu = \exp(\ell)\,\mu_\theta([z;\,s], b), \; r = r_\theta([z;\,s], b)
\]

where $b$ is a one-hot-encoded sequencing batch covariate injected into the decoder,
$\mu_{\theta} \in \mathbb{R}^G_{>0}$, $r_{\theta} \in \mathbb{R}^G_{>0}$ are learnable per-gene mean and inverse dispersion parameters,
and $\ell = \log \sum_g x_g$ is the observed log-library size used to scale the
NB rate. The decoder input dimensionality is $2m$, reflecting the concatenation of
both latent variables. The model supports conditioning on $b$ in both the encoders
and the decoder, but all experiments reported in this work use a single batch, so
$b$ is fixed and omitted from the main-text notation.

\subsection{Inference Model}
\label{app:inference}

Both \textit{Cellina} variants share the $z$ encoder (an MLP with counts as input); they differ only in how the niche input is constructed and how $s$ is encoded, as detailed below.

The approximate posterior factorizes differently per variant:
\begin{align*}
\text{Cellina:} \quad &q(z, s \mid x, \varphi_v) = q(z \mid x)\, q(s \mid \varphi_v) \\
\text{Cellina-GAT:} \quad &q(z, s \mid x, \mathcal{G}_v) = q(z \mid x)\, q(s \mid \mathcal{G}_v)
\end{align*}
where $\mathcal{G}_v = (x_v,\, \{x_u\}_{u \in \mathcal{N}(v)},\, \mathcal{E}_v)$ denotes $v$'s local subgraph. Both variants share how $z$ is encoded: an MLP parameterizing a diagonal Gaussian over counts,

\[q(z \mid x) = \text{Normal}(\mu_z(x,\, b),\; \sigma^2_z(x,\, b))\]

\textbf{Cellina (base).} The spatial input is the degree-normalized aggregation of
log-normalized neighbor expression. Let $\mathcal{N}(v) = \{u : W_{uv} > 0\}$ and
$\tilde{X} \in \mathbb{R}^{N \times G}$ the matrix of log-normalized counts, where
\[
  \tilde{x}_{u,g} = \log\!\left(1 + \frac{x_{u,g}}{\ell_u} \cdot \ell_0\right), \qquad \ell_u = \sum_{g} x_{u,g},\quad \ell_0 = 10^4,
\]
with $\ell_u$ denoting the library size (total counts per cell). Spatial weights are
given by a Gaussian proximity kernel as described in equation \eqref{eq:kernel}. The niche feature vector is then:

\[\varphi(v) = \frac{\sum_{u \in \mathcal{N}(v)} W_{uv}\, \tilde{x}_u}{\sum_{u \in \mathcal{N}(v)} W_{uv}}\]

a simple degree-normalized aggregation of neighbor expression. The $s$ encoder is an MLP:

\[q(s \mid \varphi_v) = \text{Normal}(\mu_s(\varphi_v,\, b),\; \sigma^2_s(\varphi_v,\, b))\]

\textbf{Cellina-GAT.} The $s$ encoder is replaced by a graph neural network $f_s$
that processes $v$'s local subgraph $(x_v,\, \{x_u\}_{u \in \mathcal{N}(v)},\, \mathcal{E}_v)$
directly (log1p-transformed inside $f_s$), where $\mathcal{E}_v$ is the local edge set derived from the same proximity
graph $W$ (i.e.\ $\mathcal{E}_v = \{\{u,v\} : W_{uv} > 0\}$); edges are binarized
($W_{uv} \in \{0,1\}$), so GATv2 receives only the graph topology --- attention
weights are learned entirely from gene expression. $f_s$ is a multi-layer GATv2
\citep{brody2021attentive}, implemented via \texttt{pytorch-geometric} \citep{fey2019fast}. Self-loops are
excluded, so $f_s$ does not directly aggregate $v$'s own expression, which is
captured by $z$. The posterior conditions on the local subgraph:

\[q(s \mid \mathcal{G}_v)
= \text{Normal}\!\bigl(\mu_s^{\mathcal{G}}(\mathcal{G}_v),\; \sigma^2_s{}^{\mathcal{G}}(\mathcal{G}_v)\bigr)\]

where $\mu_s^{\mathcal{G}}$ and $\sigma^2_s{}^{\mathcal{G}}$ are linear projections
applied to the seed-node representation after $L$ rounds of message passing. The learned attention weights can be extracted per edge via \textit{Cellina-GAT}'s API, providing a neighbor-level interpretability readout unavailable to the pre-aggregated encoder.

Samples are drawn via the reparameterization trick \citep{kingma2013auto}.

\subsection{Training Objective}
\label{app:training_objective}

\textbf{ELBO.} The variational lower bound is:

\[\mathrm{ELBO} = \mathbb{E}_{q}\bigl[\log p(x \mid z, s)\bigr]
- \beta_t\Bigl[\mathrm{KL}\bigl(q(z \mid x) \| p(z)\bigr)
+ \mathrm{KL}\bigl(q(s \mid \varphi_v) \| p(s)\bigr)\Bigr]\]

where $\beta_t$ is a KL warmup schedule increasing linearly from 0 to 1 \citep{gayoso2022python}, and for \textit{Cellina-GAT}, we replace $\varphi_v$ with $\mathcal{G}_v$.
The size of the library is treated as observed ($\ell = \log \sum_g x_g$), so no KL term appears in the library. We minimize the negative ELBO:
\[\mathcal{L}_\mathrm{VAE} = -\mathrm{ELBO}\]

\textbf{Dual supervised disentanglement on $z$.} Optimizing $\mathcal{L}_\mathrm{VAE}$ alone does not prevent
$z$ from absorbing spatial variation. To enforce a meaningful partition,
we apply two additional objectives exclusively to $z$.

We write $\Delta^K = \{p \in \mathbb{R}^K_{\geq 0} : \sum_i p_i = 1\}$ for the $(K\!-\!1)$-dimensional probability simplex and $\mathrm{sg}(\cdot)$ for the stop-gradient operator $\mathrm{sg}$.

A cell-type classifier $f_\mathrm{clf}: \mathbb{R}^m \to \Delta^C$ is trained jointly
to predict the cell-type label $y$ from $z$:

\[\mathcal{L}_\mathrm{clf} = \mathbb{E}\bigl[-\log f_\mathrm{clf}(y \mid z)\bigr]\]

An adversarial domain discriminator $f_\mathrm{disc}: \mathbb{R}^m \to \Delta^D$ is
trained in a two-step alternating procedure. In step 1 (VAE frozen), the discriminator
is trained to predict the spatial domain label $d$ from a detached $z$:

\[\mathcal{L}_\mathrm{disc} = \mathbb{E}\bigl[-\log f_\mathrm{disc}(d \mid \mathrm{sg}(z))\bigr]\]

In step 2 (discriminator frozen), the VAE is trained to fool the discriminator by
maximizing its cross-entropy, that is, minimizing the negated cross-entropy with weight
$-1$:

\[\mathcal{L}_\mathrm{adv} = \mathbb{E}\bigl[-\log f_\mathrm{disc}(d \mid z)\bigr]\]

In the base \textit{Cellina} variant, $s$ is unsupervised; \textit{Cellina-GAT} additionally applies a graph contrastive loss on $s$.

\textbf{Spatial contrastive loss for \textit{Cellina-GAT}.}
For the \textit{Cellina-GAT} variant we additionally consider a modified graph-supervised
contrastive loss on $s$. The loss operates on the $\ell_2$-normalized spatial-latent
posterior mean
\[
\hat{s}_v \;=\; \mu_s(v) \,/\, \lVert\mu_s(v)\rVert_2,
\]
and uses scaled cosine similarity $\mathrm{sim}_\tau(v,u) = \hat{s}_v^{\!\top}\hat{s}_u / \tau$
with temperature $\tau = 0.25$.
For a mini-batch of anchor cells $\mathcal{B} \subseteq \mathcal{V}$ and an anchor $v \in \mathcal{B}$, define
\begin{align*}
\mathcal{P}(v) &\;=\; \mathcal{N}(v) && \text{(positives: spatial neighbors of $v$)} \\
\mathcal{Q}(v) &\;=\; \{\,u \in \mathcal{V} : d_u \neq d_v,\; u \notin \mathcal{N}(v) \cup \{v\}\,\} && \text{(negatives: different-domain non-neighbors)} \\
\mathcal{A}(v) &\;=\; \mathcal{P}(v)\,\cup\,\mathcal{Q}(v).
\end{align*}

The per-anchor loss is
\[
\ell_{\mathrm{SupCon}}(v)
\;=\;
-\,\frac{1}{\lvert\mathcal{P}(v)\rvert}
\sum_{p\,\in\,\mathcal{P}(v)}
\log
\frac{\exp\!\bigl(\mathrm{sim}_\tau(v,p)\bigr)}
     {\displaystyle\sum_{a\,\in\,\mathcal{A}(v)} \exp\!\bigl(\mathrm{sim}_\tau(v,a)\bigr)},
\]
and the batch-level loss averages over anchors with at least one valid positive and negative, both subsampled from $\mathcal{B}$:
\[
\mathcal{L}_{\mathrm{spatial}}
\;=\;
\frac{1}{\lvert\mathcal{B}^{\star}\rvert}
\sum_{v \in \mathcal{B}^{\star}}
\ell_{\mathrm{SupCon}}(v),
\qquad
\mathcal{B}^{\star} \;=\; \{\,v\in\mathcal{B} : \mathcal{P}(v)\neq\emptyset,\; \mathcal{Q}(v)\neq\emptyset\,\}.
\]

This is a variant of SupCon~\citep{khosla2020supervised}, in which positives are induced by
the spatial graph $\mathcal{E}$ rather than by class label, while supervision enters
through the negative set via domain labels $d_v$. We motivate grounding positives in the spatial graph rather than in labels to avoid the collapse behavior of label-supervised contrastive objectives: with domain labels as positives, $s$ could degenerate toward one embedding per domain, erasing exactly the within-domain neighborhood variation it is meant to capture. Treating local neighborhood
membership as the positive criterion encodes a biologically motivated inductive bias --- a cell's relevant microenvironment is its immediate spatial context, not the coarse domain partition. Same-domain non-neighbors are excluded from both sets, avoiding ambiguous supervision from cells whose spatial relationship to the anchor is uninformative. For example, cells at the extremes of the same domain (or tissue region) may be completely unrelated.

\subsection{Loss Normalization}
\label{app:loss_norm}

The $\lambda_\mathrm{clf}$, $\lambda_\mathrm{disc}$, $\lambda_\mathrm{adv}$, and $\lambda_\mathrm{spatial}$ weights (Table~\ref{tab:hyperparams}) control the relative importance of each auxiliary objective but do not account for the inherent scale differences between the reconstruction loss and the auxiliary terms. To prevent any auxiliary objective from dominating training and to simplify the choice of weights, we additionally compute fixed normalization scales $\alpha$ from the raw loss values observed during the first training epoch:
\[\alpha_\mathrm{clf} = \frac{\overline{|\mathcal{L}_\mathrm{VAE}|}}{\overline{|\mathcal{L}_\mathrm{clf}|} + \epsilon}, \quad
\alpha_\mathrm{adv} = \frac{\overline{|\mathcal{L}_\mathrm{VAE}|}}
{\overline{|\mathcal{L}_\mathrm{adv}|} + \epsilon}, \quad
\alpha_\mathrm{disc} = \frac{\overline{|\mathcal{L}_\mathrm{VAE}|}}
{\overline{|\mathcal{L}_\mathrm{disc}|} + \epsilon}, \quad
\alpha_\mathrm{spatial} = \frac{\overline{|\mathcal{L}_\mathrm{VAE}|}}{\overline{|\mathcal{L}_\mathrm{spatial}|} + \epsilon}\]
where overbars denote epoch-0 means and $\epsilon = 10^{-8}$. These scales are fixed after the first epoch.
The full training objective in step 2 is then:
\[\mathcal{L} = \mathcal{L}_\mathrm{VAE}
+ \lambda_\mathrm{clf}\, \alpha_\mathrm{clf}\, \mathcal{L}_\mathrm{clf}
+ \lambda_\mathrm{spatial}\, \alpha_\mathrm{spatial}\, \mathcal{L}_\mathrm{spatial}
- \lambda_\mathrm{adv}\, \alpha_\mathrm{adv}\, \mathcal{L}_\mathrm{adv}\]
minimized over encoder and decoder parameters, with the discriminator frozen.
The ablation studies and sweeps in Appendix~\ref{app:ablation_disc} show that, with this
normalization in place, unit weights, used throughout our work, are a robust
default for $\lambda_\mathrm{clf}$, $\lambda_\mathrm{disc}$, and $\lambda_\mathrm{spatial}$.

\subsection{Adversarial Training Procedure}
\label{app:training}

The two-step alternating training is implemented via PyTorch Lightning's manual
optimization (\url{https://github.com/Lightning-AI/pytorch-lightning}). Both optimizers are Adam. The VAE optimizer covers all parameters except the discriminator head;
the discriminator optimizer covers the discriminator head only. In each training step:

\textbf{Step 1.} Freeze VAE. Sample $z$ without gradients. Minimize discriminator
cross-entropy:

\[\theta_\text{disc} \leftarrow \theta_\text{disc} - \eta \nabla_{\theta_\text{disc}} \left[\lambda_\text{disc} \cdot \mathbb{E}[-\log f_\text{disc}(d \mid \mathrm{sg}(z))]\right]\]

\textbf{Step 2.} Freeze the discriminator. Minimize VAE + classifier + adversarial loss:

\[\theta_\text{VAE} \leftarrow \theta_\text{VAE} - \eta \nabla_{\theta_\text{VAE}} \left[\mathcal{L}_\text{VAE} + \lambda_\text{clf}\mathcal{L}_\text{clf} + \lambda_\text{spatial}\mathcal{L}_\text{spatial} - \lambda_\text{adv}\mathcal{L}_\text{adv}\right]\]

In step 2, the adversary $\lambda$ is negated: minimizing
$-\lambda_\text{adv} \cdot \mathbb{E}[-\log f_\text{disc}(d \mid z)]$ is equivalent
to maximizing the adversary's cross-entropy --- i.e., encoding $z$ such that the
adversary cannot recover $d$.

\section{Component Ablations and Loss-Weight Sensitivity}
\label{app:ablation_disc}

\begin{figure}[htbp]
    \centering
    \includegraphics[width=0.9\textwidth]{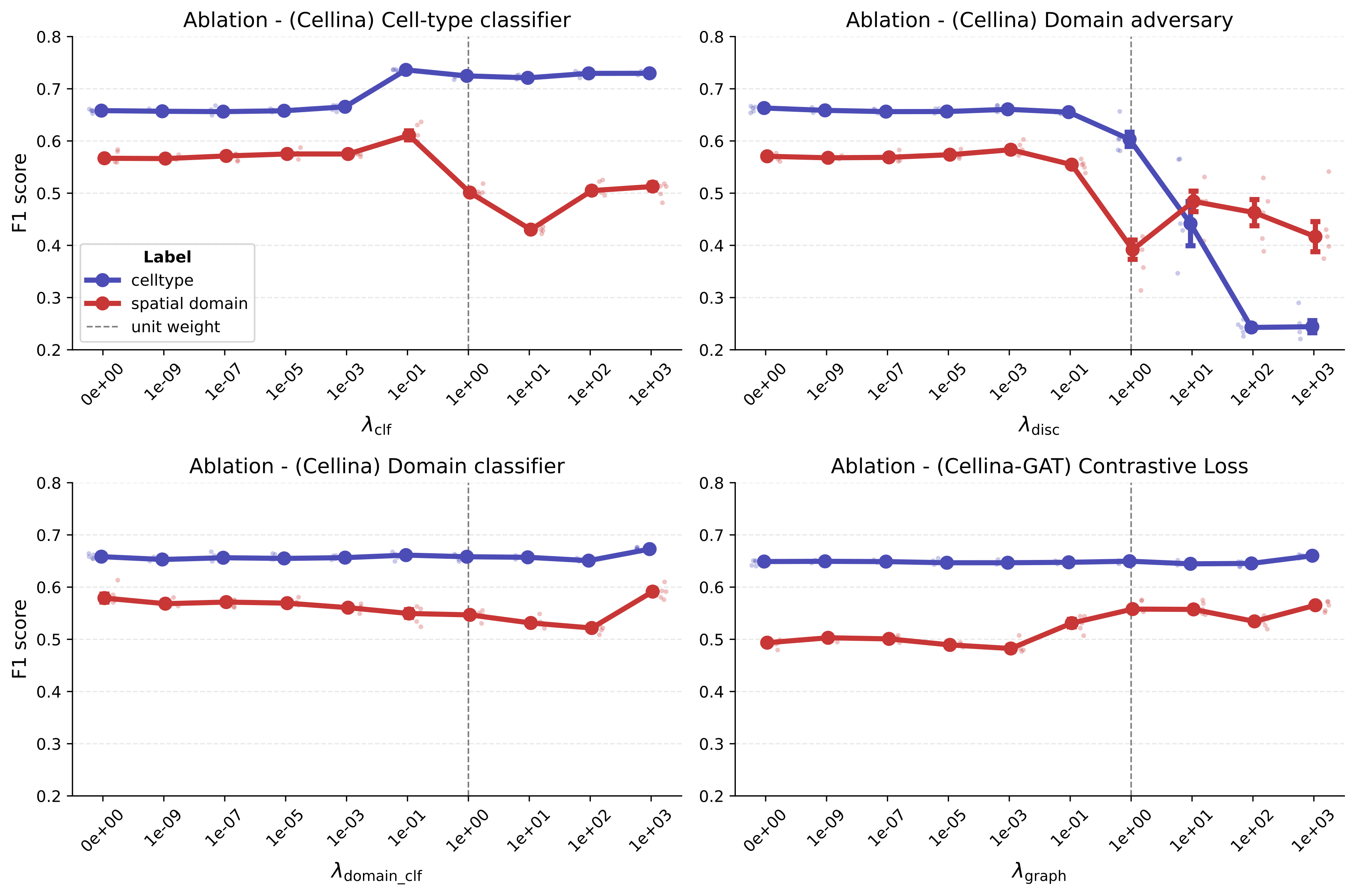}
    \vspace{0.5em}
    \includegraphics[width=0.9\textwidth]{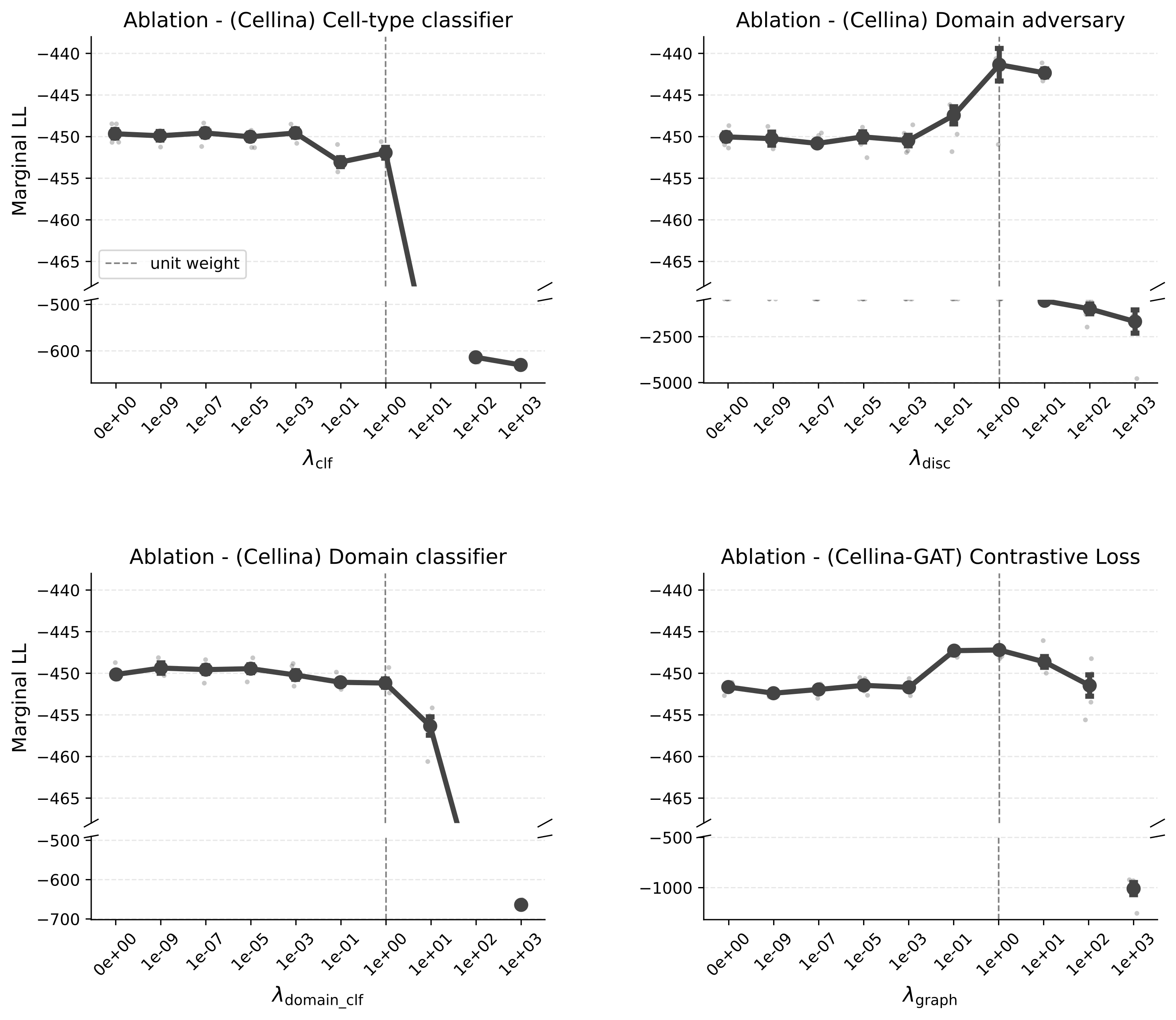}
    \caption{Lambda sweep ablations. Each panel varies one $\lambda$ (others set to $10^{-7}$). \emph{Top:} macro-F1 from logistic-regression probes on the frozen latents, evaluated against cell-type $y$ and spatial-domain $d$ labels; the top row (cell-type classifier, domain adversary) probes $z$, and the bottom row (domain classifier on \textit{Cellina}'s $s$, \textit{Cellina-GAT} contrastive loss) probes $s$. \emph{Bottom:} marginal log-likelihood. Lines connect median values; error bars show standard error over five seeds; dots show individual seeds. Overall, we see that unit weights provide a good balance between disentanglement and fit quality. The axis label $\lambda_\mathrm{graph}$ denotes $\lambda_\mathrm{spatial}$. }
    \label{fig:ablations}
\end{figure}

We ablate each model component independently by sweeping its loss weight, one $\lambda$ at a time, over
$\{0,\, 10^{-9},\, 10^{-7},\, 10^{-5},\, 10^{-3},\, 0.1,\, 1,\, 10,\, 100,\, 10^3\}$ on a
single CRC slide, while holding all other $\lambda$ values at $10^{-7}$; the $\lambda = 0$ endpoint removes the component entirely.
We train five random seeds per setting and evaluate on a 10\% holdout via
macro-F1 from a logistic-regression probe trained on the frozen latents (measuring the
cell-type and spatial-domain information each latent carries) and marginal log-likelihood (MLL, $N\!=\!500$ samples). Results are shown in Figure~\ref{fig:ablations}.

\textbf{Cell-type classifier ($\lambda_\mathrm{clf}$).}
As we increase $\lambda_\mathrm{clf}$, cell-type F1 on $z$ improves, indicating
that the classifier successfully anchors $z$ to cell identity. We also observe a
drop in spatial-domain information carried by $z$, which we take as evidence that
the disentanglement is working as intended. Both effects level off around
$\lambda_\mathrm{clf} = 1$, and there is only a modest MLL cost for them.

\textbf{Domain adversary ($\lambda_\mathrm{disc} = \lambda_\mathrm{adv}$).}
Larger $\lambda_\mathrm{disc}$ values push spatial-domain accuracy on $z$ down, showing
that the adversary is fulfilling its purpose and stripping domain-level signal from $z$:
domain macro-F1 decodable from frozen $z$ falls from ${\approx}0.57$ with the adversary off
to ${\approx}0.40$ at the deployed unit weight. Cell-type F1 is largely retained
(${\approx}0.66 \to 0.60$); we attribute this small dip to the entanglement between
cell-type composition and spatial domain (i.e., domains are to a certain extent defined by their cell-type composition).
What the adversary removes at unit weight is thus largely domain-predictive, not lineage-predictive,
information. Beyond unit weight this trade-off degrades, with cell-type F1 dropping to
${\approx}0.24$ at a high cost in MLL.

\textbf{Domain classifier on $s$ ($\lambda_\mathrm{domain\_clf}$).}
We also experimented with a supervised domain-classification loss on $s$, but it did
not improve spatial-domain F1 from $s$ and introduced a small MLL penalty at
$\lambda = 1$. We therefore drop this term from \textit{Cellina} and leave $s$ entirely
unsupervised in the base variant.

\textbf{Graph contrastive loss ($\lambda_\mathrm{spatial}$, \textit{Cellina-GAT} only).}
The modified contrastive loss described above raises spatial-domain F1 from $s$
(${\approx}0.49 \to 0.56$ at unit weight) and gives a slight MLL improvement, which we view as evidence that it provides
a useful inductive bias for \textit{Cellina}'s GAT variant.

Taken together, MLL stays largely flat around $\lambda = 1$ and only degrades for
substantially larger values, which suggests that our normalization scheme
(Appendix~\ref{app:loss_norm}) gives a robust default at unit weights and removes
the need for per-dataset tuning. As such, for all experiments reported in this paper, we set $\lambda = 1$, and thus omit it from the loss definition in the main text.

\textbf{Architecture ablations.}
We also check whether the encoders need to be variational, and whether attention contributes to performance over plain graph convolution. We test both on the CRC edge perturbation task, with the same protocol as Table~\ref{tab:loo_summary_crc} (Table~\ref{tab:ae_gcn}).
\textit{Cellina-AE} replaces the variational encoders with a deterministic autoencoder, i.e.\ it outputs point embeddings and drops the KL terms from the objective (Appendix~\ref{app:training_objective}).
\textit{Cellina-GCN} swaps the GATv2 layers of \textit{Cellina-GAT} for graph-convolution layers of the same depth and width.
The deterministic encoder \textit{Cellina-AE} does slightly better on three of the four metrics and ties on Signed Precision. If anything, this variant would widen the margins over the baselines in Table~\ref{tab:loo_summary_crc}; we keep the variational formulation, which all reported experiments used.
GAT and GCN are indistinguishable, which is in line with the small gap between \textit{Cellina-GAT} and \textit{Cellina} in Table~\ref{tab:loo_summary_crc}.
On these data, the gains largely come from conditioning on the neighborhood and from the supervised disentanglement (\textit{Cellina (ablated)} in Table~\ref{tab:loo_summary_crc}), not from how the neighbors are aggregated.
We, nevertheless, keep the attention variant as default, as its per-edge weights give a neighbor-level readout of which cells drive a prediction (Appendix~\ref{app:inference}), which GCN does not offer.

\begin{table}[htbp]
\centering
\caption{\textit{Cellina} architecture comparison. CRC edge perturbation results for variational vs.\ standard (deterministic) autoencoder (AE) variants of \textit{Cellina}, and Graph Attention (GAT) vs.\ Graph Convolution (GCN) encoders for \textit{Cellina-GAT}. Same protocol and reporting as Table~\ref{tab:loo_summary_crc}.}
\label{tab:ae_gcn}
\begin{tabular}{lcccc}
\toprule
Model & Pearson $\uparrow$ & Precision$_{\text{signed}}$ $\uparrow$ & E-distance $\downarrow$ & RMSE\textsubscript{LFC} $\downarrow$ \\
\midrule
Cellina & 0.86 $\pm$ 0.17 & 0.45 $\pm$ 0.19 & 7.00 $\pm$ 1.05 & 1.14 $\pm$ 0.56 \\
Cellina-AE & 0.89 $\pm$ 0.10 & 0.45 $\pm$ 0.17 & 6.29 $\pm$ 1.13 & 1.07 $\pm$ 0.44 \\
\midrule
Cellina-GAT & 0.89 $\pm$ 0.09 & 0.42 $\pm$ 0.20 & 8.59 $\pm$ 1.90 & 1.09 $\pm$ 0.61 \\
Cellina-GCN & 0.89 $\pm$ 0.10 & 0.43 $\pm$ 0.20 & 8.64 $\pm$ 1.34 & 1.07 $\pm$ 0.60 \\
\bottomrule
\end{tabular}
\end{table}

\section{Related Methods} 
\label{app:baselines}

\textbf{scVI} \citep{lopez2018deep} is a conditional VAE 
for single-cell RNA-seq data. It models raw counts with a negative 
binomial (or zero-inflated negative binomial) likelihood, decomposing each 
cell's expression into a low-dimensional latent representation and a 
separately inferred library-size factor. Batch effects (and other covariates) are mitigated by conditioning the encoder and 
decoder on the corresponding labels, typically encoded as one-hot vectors. scVI 
is a popular choice for dimensionality reduction in the single-cell field, and its code base has since been expanded into 
\textbf{scvi-tools} \citep{gayoso2022python}, a probabilistic modeling 
framework for single-cell omics; several of the related methods in this section, as well as \textit{Cellina}, are built on scvi-tools.

\textbf{scANVI} \citep{xu2021probabilistic} is a semi-supervised extension 
of scVI in which the latent space is structured by cell-type identity. 
On top of scVI's per-cell latent representation, scANVI introduces a 
second, label-conditional latent variable and a classifier that predicts 
the cell-type label from scVI's latent. During training, labeled cells 
contribute an additional cross-entropy term, resulting in an explicitly 
supervised latent. scANVI is widely used for cell-type label transfer, particularly when 
annotating a query dataset against a labeled reference.

\textbf{scGen} \citep{lotfollahi2019scGen} is a VAE that predicts perturbation responses via latent-space arithmetic. The model is 
trained to reconstruct normalized gene expression profiles through a low-dimensional 
latent space; the effect of a perturbation is then summarized by a difference 
vector $\delta$, calculated as the difference between the mean latent 
representations of perturbed and unperturbed training cells. To predict the 
response of an unseen (test) cell-type population, $\delta$ is added to the 
latent representation of each unperturbed test cell, and the resulting vector 
is decoded back to gene expression space. For the two largest slides (120 
and 210), scGen's \texttt{predict} functions failed with internal errors, which were resolved by training on subsets of the slides; we therefore sub-sampled the data for scGen to 30\% on these slides.

\textbf{CPA} \citep{lotfollahi2023predicting} models single-cell gene 
expression as additive compositions of disentangled latent factors: 
a basal cell state plus separate embeddings for each perturbation and 
covariate (e.g.,\ cell type, drug). Only the basal state is produced by 
the encoder; perturbation and covariate embeddings are learned per 
label (or dosage) and added before decoding. Disentanglement is enforced 
adversarially: auxiliary classifiers are trained to recover the labels 
from the basal embedding, and the encoder is penalized whenever they 
succeed.

\textbf{scVIVA} \citep{levy2025scviva} is a VAE-based model designed for spatial transcriptomics data. 
It learns a shared embedding that captures both intrinsic cell state and microenvironment context. 
To effectively add spatial information in the latent embedding, scVIVA predicts niche 
gene expression and neighbor composition (proportion of each cell type in a given neighborhood) in addition
to denoised gene expression of query cells. It has no mechanism for disentanglement and cannot answer counterfactual questions; it therefore cannot generate predictions under an altered neighborhood and is excluded from the in-silico perturbation benchmarks.

\textbf{SIMVI} \citep{dong2025simvi} is a spatially-informed VAE that disentangles gene expression variability into two latent factors: an intrinsic variable $z$ and a spatial variable $s$. The spatial latent $s$ is inferred by aggregating intrinsic representations of neighboring cells via a Graph Attention Network. To promote independence between $z$ and $s$, SIMVI uses an additional unsupervised regularization: an MMD-based term that promotes independence between $(z, s)$ or, alternatively, a mutual-information penalty. We excluded SIMVI from our counterfactual benchmarks because it exceeds available memory at $10^5$ cells on an NVIDIA GeForce RTX 4090 (24~GB VRAM) in our scalability tests (Figure~\ref{fig:scalability}), well below the size of our slides, and does not natively support counterfactual donor-swap or neighbor-feature interventions.

\textbf{MintFlow} \citep{akbarnejad2025mapping} is a flow-matching generative model with an underlying graphVAE-style encoding and count-specific decoding that disentangles single-cell gene expression into intrinsic and microenvironment-induced components. It learns three latent variables per cell (an intrinsic state and incoming/outgoing spatial signals). MintFlow supports in-domain counterfactual queries via in-silico perturbation of the tissue (e.g., relabeling or deleting cells and re-sampling from the generative model), but conditions each cell on its cell-type label and therefore cannot extrapolate to cell types or contexts unseen during training, whereas \textit{Cellina} takes no categorical code at inference. Consequently, we provide MintFlow with all cells, including those held out for other methods during training, giving it a strictly in-domain evaluation setting that constitutes an advantage over the other benchmarked approaches. Because MintFlow does not natively support edge swapping of neighbors, we adapted its inference procedure to align it as closely as possible with \textit{Cellina}: for each target cell, we replace its gene expression vector with that of a randomly selected control cell of the same cell type and then call \texttt{\detokenize{generate_insilico_ST_data()}} to produce counterfactual counts (matching \textit{Cellina}'s sampling procedure). We note that certain trained models raised an internal numerical error (\texttt{\detokenize{Expected parameter rate to satisfy the constraint GreaterThan(lower_bound=0.0)}}) at arbitrary checkpoints; in these cases, we evaluated on the last stable checkpoint preceding the error.

\textbf{SpatialProp} \citep{sun2025spatialprop} is a graph neural network model that predicts how perturbations to neighboring cells propagate to a center cell by inferring its gene expression from masked k-hop (2-hop by default) neighborhood graphs. SpatialProp is tailored to in-domain predictions, and its SparseRenorm post-processing calibrates predictions using an empirical error distribution derived from unperturbed (in-distribution) base predictions, which we omit for out-of-distribution perturbation evaluated in this work. Like \textit{Cellina}, SpatialProp provides functionality to predict downstream effects of perturbations on the spatial microenvironment, enabling direct comparison in the neighbor node perturbation setting, via adaptation of this vignette: \url{https://github.com/abuendia/spatial-prop/blob/main/notebooks/api_demo.ipynb}. We also implemented a custom edge perturbation API for SpatialProp modifying its neighborhood steering function (\texttt{\detokenize{batch_steering_cell()}}) and evaluated only on the MERFISH cohort due to the high computational cost of the CRC dataset, seeing similar results to the node perturbation experiment on the same cohort (Pearson $0.68$ for edge vs.\ $0.66$ for node perturbation).

\textbf{Concert} \citep{lin2025concert} predicts spatially-resolved perturbation responses by disentangling spot-level expression into learnable embeddings for basal cell state and perturbation or covariate identities, following the LORD framework \citep{gabbay2019demystifying, piran2024disentanglement}. Perturbation effects are propagated across the tissue via Gaussian process priors with perturbation-specific Cauchy kernels. Concert simulates perturbations by swapping embeddings for categorical attributes (e.g., perturbation identity, disease state) or interpolating learned projections of continuous attributes (e.g., time, dose). Such swaps can be applied to the cells surrounding a target and propagated inward through the learned kernel. Unlike \textit{Cellina}, however, Concert defines no operation on the tissue graph itself: propagation is a continuous function of cell coordinates rather than of neighborhood membership, and only a subset of its latent dimensions is spatially coupled. Neither a cell's neighbors (the edges) nor their expression (the nodes) can be altered, so neither of our perturbation operators is natively available to it. Concert therefore admits only a domain identity swap (e.g.\ REF $\rightarrow$ CRC), which we adopt as its closest available analogue of an edge perturbation; this restriction follows from Concert's design rather than from any simplification on our part, and Concert is otherwise run at its default settings.

\textbf{Celcomen} \citep{megas2025estimation} is a generative graph neural network model that disentangles intra- and inter-cellular gene regulation in spatial transcriptomics via a maximum-entropy formulation. The model's parameters are guaranteed to yield identifiability for the gene-gene interaction matrices. It consists of an inference module that learns intra- and inter-cellular gene-gene interaction matrices (under an acyclic regulatory assumption), and a generative simulation module that produces counterfactual spatial samples by intervening on selected nodes (e.g., gene knockouts in specific tissue locations). As such, unlike \textit{Cellina}, Celcomen targets in-silico gene-level perturbations rather than perturbations directly on tissue graphs: its interventions act through a single tissue-wide gene--gene interaction matrix rather than on the expression of a focal cell's own neighbors, so it can be modified to approximate node perturbation only at the level of global gene regulation and offers no edge perturbation at all; its generative module (Simcomen) likewise samples tissue-wide counterfactuals rather than responses of a focal cell to its own neighbors. Our best-effort adaptation also did not scale, attempting to allocate ${\sim}42.6$~GiB on the smallest slide (232) from \citet{crowell2025tracing}.

Table~\ref{tab:method_comparison} contrasts these methods component by component.

\begin{table}[htbp]
  \centering
  \caption{Comparison of related methods against the requirements of tissue graph
  counterfactual inference. $\checkmark$ indicates the property is satisfied;
  --- indicates it is not. $\checkmark^{*}$ indicates partial or adapted support
  (see text for details). \textit{Italicized} methods are discussed in this section
  but are not included in any benchmark (counterfactual, disentanglement, or scalability).}
  \label{tab:method_comparison}
  \begin{tabular}{lcccccc}
    \toprule
    Method
      & \makecell{Spatial\\graph}
      & \makecell{Disen-\\tanglement}
      & \makecell{Counterfactual\\inference}
      & \makecell{Edge\\pert.}
      & \makecell{Node\\pert.}
      & \makecell{Unseen cell\\type $\times$ domain} \\
    \midrule
    scVI   & --- & ---            & ---            & --- & --- & --- \\
    scANVI & --- & ---            & ---            & --- & --- & --- \\
    scGen  & --- & ---            & $\checkmark$   & --- & --- & --- \\
    CPA    & --- & $\checkmark$   & $\checkmark$   & --- & --- & --- \\
    \midrule
    \textit{MISTy}  & $\checkmark$ & --- & --- & --- & --- & --- \\
    \textit{NCEM}   & $\checkmark$ & --- & --- & --- & --- & --- \\
    scVIVA          & $\checkmark$ & --- & --- & --- & --- & --- \\
    SIMVI           & $\checkmark$ & $\checkmark$ & --- & --- & --- & --- \\
    \midrule
    MintFlow              & $\checkmark$ & $\checkmark$ & $\checkmark$ & $\checkmark^{*}$ & --- & --- \\
    SpatialProp           & $\checkmark$ & --- & $\checkmark$ & ---             & $\checkmark$ & $\checkmark^{*}$ \\
    Concert               & $\checkmark$ & $\checkmark$ & $\checkmark$ & $\checkmark^{*}$ & --- & $\checkmark$ \\
    \textit{Celcomen}     & $\checkmark$ & --- & $\checkmark$ & ---              & $\checkmark^{*}$ & --- \\
    \midrule
    \rowcolor{gray!10}
    \textbf{Cellina}     & $\checkmark$ & $\checkmark$ & $\checkmark$ & $\checkmark$ & $\checkmark$ & $\checkmark$ \\
    \rowcolor{gray!10}
    \textbf{Cellina-GAT} & $\checkmark$ & $\checkmark$ & $\checkmark$ & $\checkmark$ & $\checkmark$ & $\checkmark$ \\
    \bottomrule
  \end{tabular}
\end{table}

\textbf{Additional notes.} All competing methods are evaluated using their default parameters under the 
same protocol as \textit{Cellina} (Section~\ref{sec:benchmark}) with identical 
leave-one-cell-type-out splits, except for MintFlow as described in Appendix~\ref{app:baselines} where no cells are held out. Each 
method receives the input its formulation supports: scGen 
and CPA are given only cell expression and domain labels, as in their 
original formulations, and are not provided neighbor composition features. 
SpatialProp, which takes spatial neighborhoods as input, is evaluated 
in \textit{the node perturbation} task: a partial neighbor-node perturbation restricted to 
the top 200 genes, matching the setting used for \textit{Cellina}$_{\text{node-pert}}$ is applied in inference.

\section{Extended Results}
\label{app:extended-results}

\begin{figure}[htbp]
    \centering
    \includegraphics[width=\textwidth]{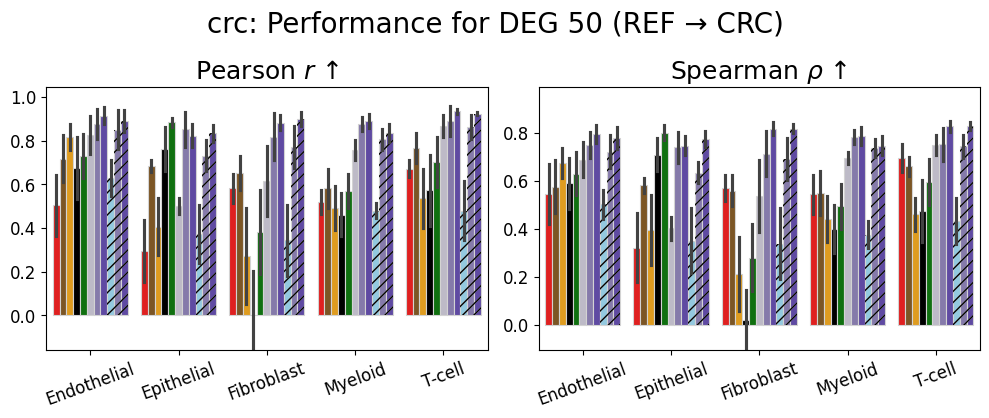}
    \vspace{0.5em}
    \includegraphics[width=\textwidth]{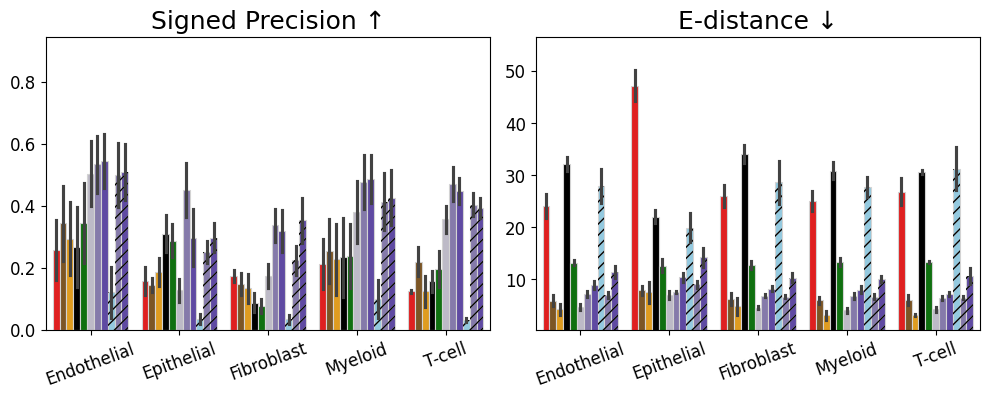}
    \vspace{0.5em}
    \includegraphics[width=\textwidth]{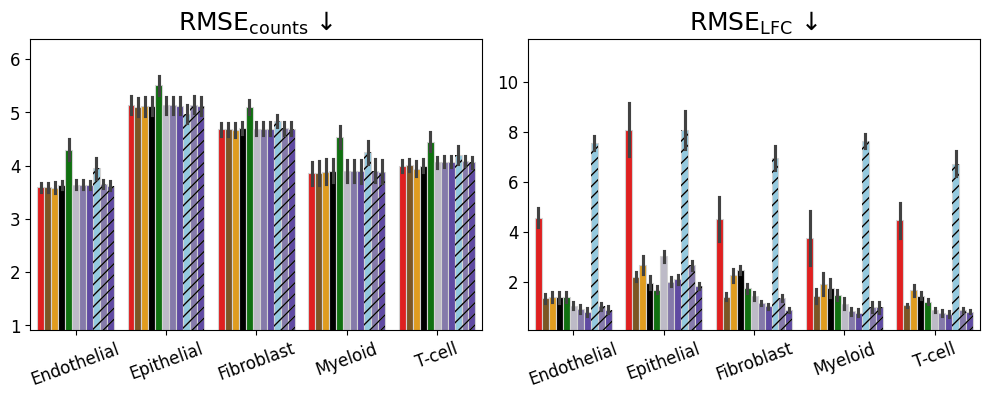}
    \includegraphics[width=\textwidth]{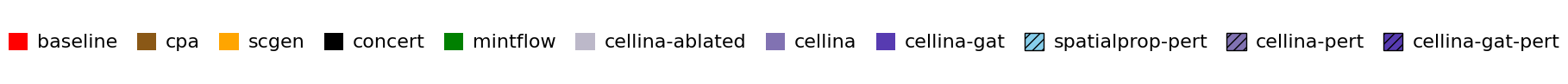}
    \caption{Colorectal cancer per-cell-type results with extended metrics for CRC held-out region. Striped bars with \textbf{-pert} in method names denote evaluation under \textit{node perturbation} on a subset of features. The mean shift baseline is denoted as ``baseline''. Bar heights indicate mean over slides, error bars indicate standard error of the mean.}
    \label{fig:crc_barplots}
\end{figure}

\begin{figure}[htbp]
    \centering
    \includegraphics[width=\textwidth]{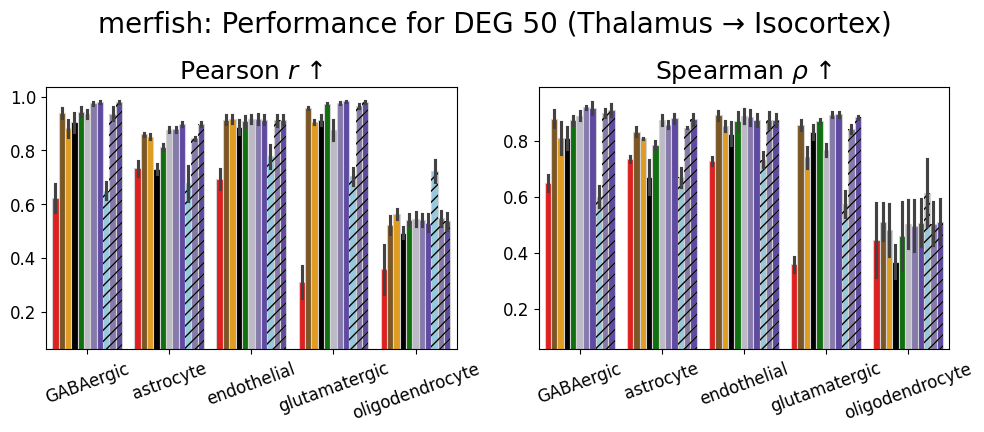}
    \vspace{0.5em}
    \includegraphics[width=\textwidth]{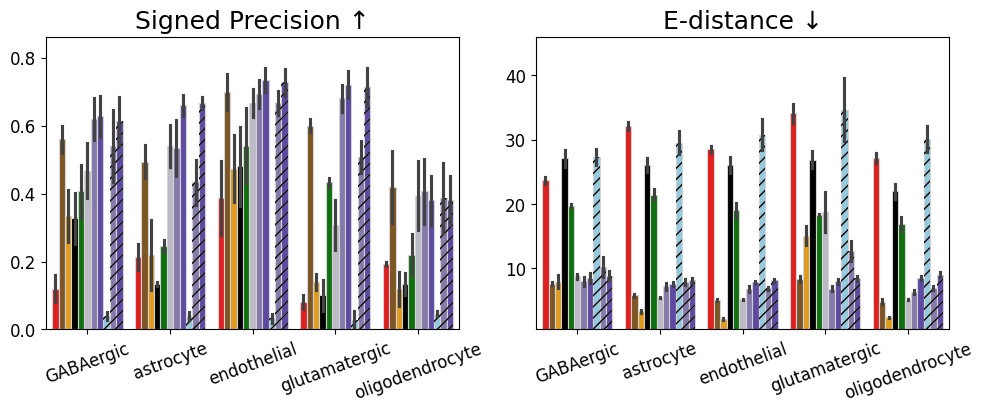}
    \vspace{0.5em}
    \includegraphics[width=\textwidth]{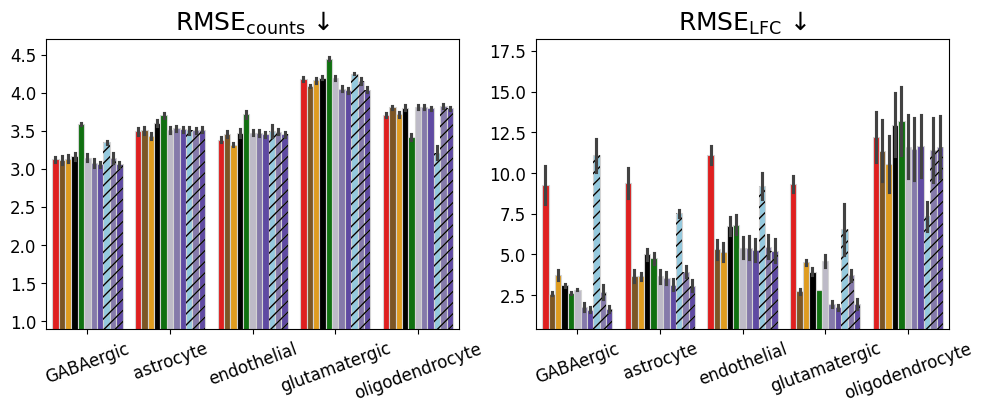}
    \includegraphics[width=\textwidth]{figures/merfish_50_bars_legend.png}
    \caption{MERFISH per-cell-type results with extended metrics for Isocortex held-out region. Striped bars with \textbf{-pert} in method names denote evaluation under \textit{node perturbation} on a subset of features. The mean shift baseline is denoted as ``baseline''. Bar heights indicate mean over slides, error bars indicate standard error of the mean.}
    \label{fig:merfish_barplots_isocortex}
\end{figure}

\begin{figure}[htbp]
    \centering
    \includegraphics[width=\textwidth]{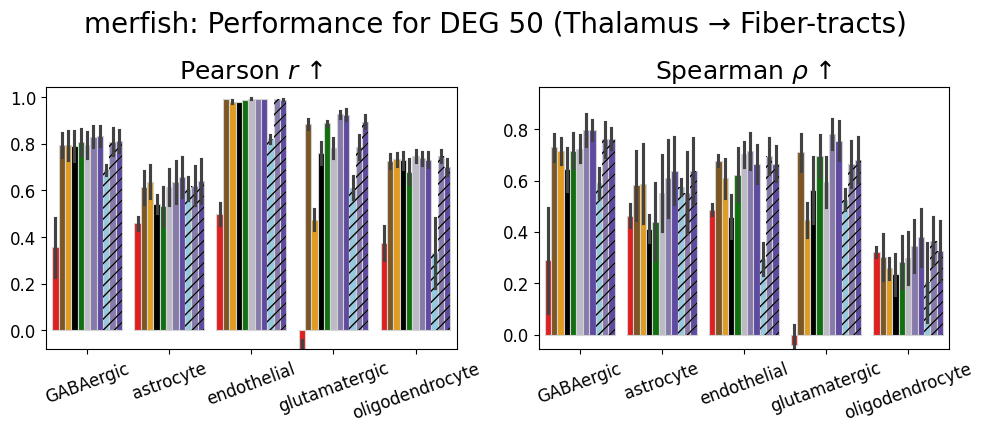}
    \vspace{0.5em}
    \includegraphics[width=\textwidth]{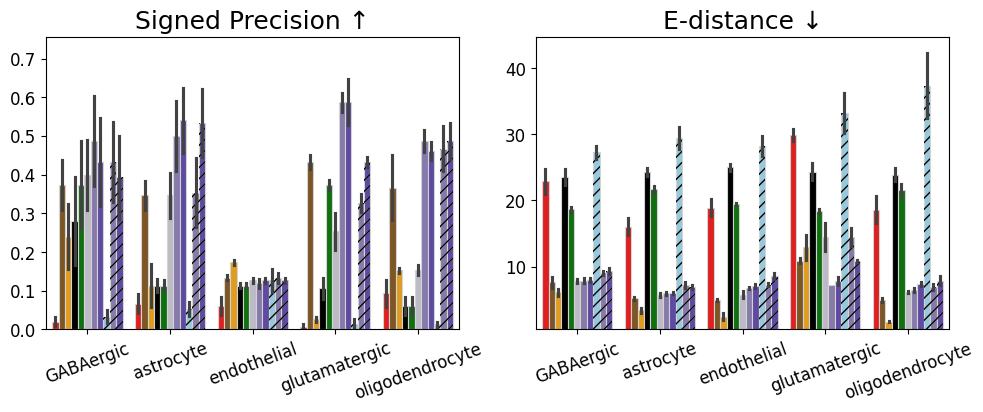}
    \vspace{0.5em}
    \includegraphics[width=\textwidth]{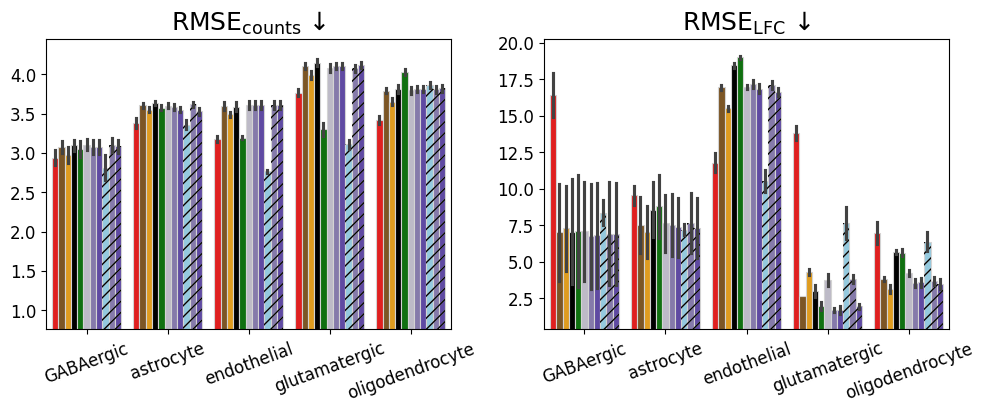}
    \includegraphics[width=\textwidth]{figures/merfish_50_bars_legend.png}
    \caption{MERFISH per-cell-type results with extended metrics for Fiber-tracts held-out region. \textbf{-pert} denotes node perturbation models. The mean shift baseline is denoted as ``baseline''. Bar heights indicate mean over slides, error bars indicate standard error of the mean.}
    \label{fig:merfish_barplots_fiber-tracts}
\end{figure}

\textbf{Extended Metrics for Leave-One-Cell-Type-Out Evaluation.}
Here, we discuss results at the cell-type level for CRC and MERFISH datasets across an extended list of metrics. As a reminder, for the colorectal cancer data, we design our experiments around the in-silico perturbation query to predict the effect of the cancer region on healthy cells (REF $\rightarrow$ CRC). We employ a leave-one-cell-type-out strategy for a comprehensive evaluation of models across samples from six patients. In CRC, \textit{Cellina} consistently achieves the best performance across all held-out cell types on \textit{Pearson} and \textit{Spearman} $\rho$, with the exception of Epithelial, where MintFlow is ahead, likely because this is the most abundant cell type, and MintFlow observes it during training. On \textit{Signed Precision}, \textit{Cellina}'s variants lead all competing methods for every cell type, and on $\text{RMSE}_{\text{LFC}}$ for every cell type except Epithelial, where MintFlow and Concert are ahead. $\text{RMSE}_{\text{counts}}$ is largely comparable across methods, with MintFlow worse; E-distance instead separates the mean shift, Concert and SpatialProp, which score several-fold higher, from the remaining methods (Figure \ref{fig:crc_barplots}). Notably, the contribution of supervision (i.e., the gain of the full \textit{Cellina} over \textit{Cellina (ablated)}) is largest for Epithelial ($+0.35$ Pearson) and Fibroblast ($+0.20$) cells, the two types whose predictions depend most on their neighbors (Tables~\ref{tab:homotypic} and~\ref{tab:k_sweep}), whereas the gain is marginal for T cells, whose state is largely intrinsic. To further assess whether \textit{Cellina}'s advantage over competitors generalizes to other datasets, we repeat the same evaluation strategy on a whole mouse brain dataset (3 slides chosen, most abundant cell types). We use Thalamus as the control region, while Fiber-tracts and Isocortex are held out. Both settings are discussed in the main text. Similar results are seen on the MERFISH mouse brain data (Figures \ref{fig:merfish_barplots_isocortex} and \ref{fig:merfish_barplots_fiber-tracts}), suggesting that \textit{Cellina} maintains the highest scores in log-fold change correlation metrics (as measured by Spearman and Pearson) as well as Signed Precision and $\text{RMSE}_{\text{LFC}}$, while remaining competitive in distributional metrics ($\text{RMSE}_{\text{counts}}$ and E-distance).

Both distributional metrics above are computed on log-normalized expression, which is the space scGen is trained to reconstruct, whereas the count-based methods are fit to a
negative-binomial likelihood on raw counts. To represent both expression distributions, we additionally score the same predictions with NB deviance
(Appendix~\ref{app:metrics}), which compares predicted means against observed counts under a common per-gene
dispersion. Under this measure scGen's deviance is roughly twice \textit{CPA}'s (Table~\ref{tab:nb_deviance}) and roughly $3\times$ \textit{Cellina}'s on CRC, with the gap widening even further on MERFISH, where scGen's deviance is roughly tenfold \textit{Cellina}'s ($0.33$ vs.\ $0.03$, averaged over both held-out domains). We therefore read the two metrics as complementary, since each rewards its matched likelihood family.

Table~\ref{tab:loo_summary_merfish_full} extends the MERFISH evaluation of Table~\ref{tab:loo_summary_merfish} (Section~\ref{sec:results}) with the results disaggregated by held-out domain, and Table~\ref{tab:invivo_full_results} extends the \textit{in vivo} Perturb-FISH evaluation of Table~\ref{tab:cellina-invivo} with the KO-specific decomposition and the randomized-control comparisons discussed in Section~\ref{sec:results}.

\begin{table}[htbp]
\centering
\caption{Leave-one-cell-type-out performance (top 50 DEGs) for the in-silico perturbations Thalamus $\rightarrow$ Isocortex, Fiber-tracts (MERFISH). Mean $\pm$ std across cell types and three slides, averaged over both held-out domains (\textit{Average}) and per held-out domain. Best per metric within each block (edge- vs.\ node-perturbation) in \textbf{bold}; \textbf{np} refers to the node perturbation task; \textit{Cellina} variants in gray, spatially-informed competitors in orange. On average, \textit{Cellina} and \textit{Cellina-GAT} perform largely on par, and Concert scores similarly to the spatially-uninformed methods ($0.18$ Signed Precision against scGen's $0.20$; E-distance $24.91$ against the mean shift's $25.18$).}
\label{tab:loo_summary_merfish_full}
\begin{tabular}{llcccc}
\toprule
Holdout & Method & Pearson $\uparrow$ & Precision\textsubscript{signed} $\uparrow$ & E-distance $\downarrow$ & RMSE\textsubscript{LFC} $\downarrow$ \\
\midrule
& Mean shift & 0.43 $\pm$ 0.23 & 0.12 $\pm$ 0.09 & 25.18 $\pm$ 4.75 & 10.98 $\pm$ 2.78 \\
 & CPA & 0.82 $\pm$ 0.16 & 0.44 $\pm$ 0.13 & 6.53 $\pm$ 1.99 & 6.36 $\pm$ 4.71 \\
 & scGen & 0.77 $\pm$ 0.16 & 0.20 $\pm$ 0.14 & \textbf{5.73 $\pm$ 4.84} & 6.48 $\pm$ 3.97 \\
\rowcolor{orange!10}
 & Concert & 0.77 $\pm$ 0.16 & 0.18 $\pm$ 0.15 & 24.91 $\pm$ 2.08 & 7.45 $\pm$ 4.99 \\
\rowcolor{orange!10}
 & MintFlow & 0.81 $\pm$ 0.17 & 0.29 $\pm$ 0.16 & 19.52 $\pm$ 1.81 & 7.27 $\pm$ 5.43 \\
\rowcolor{gray!10}
 & Cellina (ablated) & 0.81 $\pm$ 0.15 & 0.37 $\pm$ 0.15 & 8.33 $\pm$ 4.76 & 6.78 $\pm$ 4.54 \\
\rowcolor{gray!10}
 & Cellina & \textbf{0.84 $\pm$ 0.16} & 0.51 $\pm$ 0.17 & 6.93 $\pm$ 0.83 & 6.08 $\pm$ 5.06 \\
\rowcolor{gray!10}
 & Cellina-GAT & \textbf{0.84 $\pm$ 0.16} & \textbf{0.53 $\pm$ 0.17} & 7.67 $\pm$ 0.83 & \textbf{5.97 $\pm$ 5.07} \\
 \midrule
\rowcolor{gray!10}
 & Cellina\textsubscript{np} & 0.82 $\pm$ 0.15 & 0.42 $\pm$ 0.15 & 8.96 $\pm$ 2.92 & 6.65 $\pm$ 4.63 \\
\rowcolor{gray!10}
 & Cellina-GAT\textsubscript{np} & \textbf{0.83 $\pm$ 0.16} & \textbf{0.51 $\pm$ 0.16} & \textbf{8.61 $\pm$ 1.15} & \textbf{5.98 $\pm$ 4.98} \\
\rowcolor{orange!10}
\multirow[c]{-11}{*}{Average} & SpatialProp\textsubscript{np} & 0.66 $\pm$ 0.14 & 0.04 $\pm$ 0.04 & 30.79 $\pm$ 5.18 & 8.18 $\pm$ 2.05 \\
\midrule

& Mean shift & 0.31 $\pm$ 0.26 & 0.05 $\pm$ 0.05 & 21.29 $\pm$ 5.51 & 11.71 $\pm$ 3.64 \\
 & CPA & 0.80 $\pm$ 0.15 & 0.33 $\pm$ 0.13 & 6.70 $\pm$ 2.48 & 7.58 $\pm$ 5.80 \\
 & scGen & 0.72 $\pm$ 0.19 & 0.14 $\pm$ 0.10 & \textbf{5.28 $\pm$ 4.44} & 7.44 $\pm$ 5.02 \\
\rowcolor{orange!10}
 & Concert & 0.76 $\pm$ 0.16 & 0.13 $\pm$ 0.11 & 24.23 $\pm$ 1.53 & 8.55 $\pm$ 6.10 \\
\rowcolor{orange!10}
 & MintFlow & 0.78 $\pm$ 0.18 & 0.21 $\pm$ 0.16 & 19.95 $\pm$ 1.64 & 8.50 $\pm$ 6.59 \\
\rowcolor{gray!10}
 & Cellina (ablated) & 0.79 $\pm$ 0.14 & 0.26 $\pm$ 0.13 & 7.98 $\pm$ 3.75 & 7.93 $\pm$ 5.58 \\
\rowcolor{gray!10}
 & Cellina & \textbf{0.83 $\pm$ 0.15} & \textbf{0.44 $\pm$ 0.20} & 6.82 $\pm$ 0.75 & 7.34 $\pm$ 6.19 \\
\rowcolor{gray!10}
 & Cellina-GAT & \textbf{0.83 $\pm$ 0.14} & 0.43 $\pm$ 0.19 & 7.23 $\pm$ 0.91 & \textbf{7.26 $\pm$ 6.07} \\
 \midrule
\rowcolor{gray!10}
 & Cellina\textsubscript{np} & 0.79 $\pm$ 0.15 & 0.34 $\pm$ 0.16 & 8.99 $\pm$ 3.11 & 7.85 $\pm$ 5.74 \\
\rowcolor{gray!10}
 & Cellina-GAT\textsubscript{np} & \textbf{0.81 $\pm$ 0.15} & \textbf{0.39 $\pm$ 0.18} & \textbf{8.68 $\pm$ 1.54} & \textbf{7.25 $\pm$ 5.93} \\
\rowcolor{orange!10}
\multirow[c]{-11}{*}{\shortstack{Fiber-\\tracts}} & SpatialProp\textsubscript{np} & 0.61 $\pm$ 0.20 & 0.05 $\pm$ 0.05 & 31.12 $\pm$ 5.62 & 8.02 $\pm$ 1.87 \\
\midrule

& Mean shift & 0.54 $\pm$ 0.20 & 0.20 $\pm$ 0.14 & 29.07 $\pm$ 3.99 & 10.25 $\pm$ 1.92 \\
 & CPA & 0.84 $\pm$ 0.17 & 0.55 $\pm$ 0.13 & 6.35 $\pm$ 1.51 & 5.14 $\pm$ 3.63 \\
 & scGen & 0.82 $\pm$ 0.14 & 0.26 $\pm$ 0.18 & \textbf{6.17 $\pm$ 5.23} & 5.52 $\pm$ 2.92 \\
\rowcolor{orange!10}
 & Concert & 0.79 $\pm$ 0.17 & 0.23 $\pm$ 0.18 & 25.60 $\pm$ 2.64 & 6.36 $\pm$ 3.89 \\
\rowcolor{orange!10}
 & MintFlow & 0.84 $\pm$ 0.16 & 0.37 $\pm$ 0.16 & 19.10 $\pm$ 1.97 & 6.04 $\pm$ 4.26 \\
\rowcolor{gray!10}
 & Cellina (ablated) & 0.83 $\pm$ 0.15 & 0.47 $\pm$ 0.17 & 8.67 $\pm$ 5.77 & 5.62 $\pm$ 3.51 \\
\rowcolor{gray!10}
 & Cellina & \textbf{0.86 $\pm$ 0.17} & 0.59 $\pm$ 0.15 & 7.05 $\pm$ 0.91 & 4.83 $\pm$ 3.94 \\
\rowcolor{gray!10}
 & Cellina-GAT & \textbf{0.86 $\pm$ 0.18} & \textbf{0.62 $\pm$ 0.15} & 8.10 $\pm$ 0.74 & \textbf{4.68 $\pm$ 4.07} \\
 \midrule
\rowcolor{gray!10}
 & Cellina\textsubscript{np} & 0.84 $\pm$ 0.16 & 0.51 $\pm$ 0.15 & 8.94 $\pm$ 2.73 & 5.45 $\pm$ 3.51 \\
\rowcolor{gray!10}
 & Cellina-GAT\textsubscript{np} & \textbf{0.86 $\pm$ 0.17} & \textbf{0.62 $\pm$ 0.15} & \textbf{8.54 $\pm$ 0.75} & \textbf{4.71 $\pm$ 4.03} \\
\rowcolor{orange!10}
\multirow[c]{-11}{*}{Isocortex} & SpatialProp\textsubscript{np} & 0.71 $\pm$ 0.08 & 0.04 $\pm$ 0.02 & 30.45 $\pm$ 4.74 & 8.34 $\pm$ 2.23 \\
\bottomrule
\end{tabular}
\end{table}

\begin{table}[htbp]
\centering
\caption{Negative-binomial deviance of the same in-silico perturbation predictions scored in count space
(Appendix~\ref{app:metrics}), for the leave-one-cell-type-out evaluation on both cohorts. Lower is better; best per metric within each block (edge- vs.\ node-perturbation) in \textbf{bold}; \textit{Cellina} variants in gray, spatially-informed competitors in orange. All methods are scored under a common per-gene dispersion estimated from the
observed target population, so no method contributes its own. Mean $\pm$ std across cell type-slide experiments.}
\label{tab:nb_deviance}
\begin{tabular}{lcccc}
\toprule
& \multicolumn{2}{c}{CRC} & \multicolumn{2}{c}{MERFISH} \\
\cmidrule(lr){2-3} \cmidrule(lr){4-5}
Method & NB deviance $\downarrow$ & E-distance $\downarrow$ & NB deviance $\downarrow$ & E-distance $\downarrow$ \\
\midrule
Mean shift & 0.44 $\pm$ 0.13 & 29.89 $\pm$ 10.49 & 2.78 $\pm$ 0.97 & 25.18 $\pm$ 4.75 \\
CPA & 0.04 $\pm$ 0.05 & 6.44 $\pm$ 2.27 & 0.05 $\pm$ 0.04 & 6.53 $\pm$ 1.99 \\
scGen & 0.09 $\pm$ 0.03 & \textbf{4.65 $\pm$ 3.49} & 0.33 $\pm$ 0.05 & \textbf{5.73 $\pm$ 4.84} \\
\rowcolor{orange!10}
MintFlow & 0.50 $\pm$ 0.14 & 13.01 $\pm$ 1.76 & 2.44 $\pm$ 0.83 & 19.52 $\pm$ 1.81 \\
\rowcolor{orange!10}
Concert & 0.05 $\pm$ 0.04 & 29.98 $\pm$ 5.22 & 0.19 $\pm$ 0.07 & 24.91 $\pm$ 2.08 \\
\rowcolor{gray!10}
Cellina (ablated) & 0.04 $\pm$ 0.04 & 4.93 $\pm$ 1.66 & 0.08 $\pm$ 0.09 & 8.33 $\pm$ 4.76 \\
\rowcolor{gray!10}
Cellina & \textbf{0.03 $\pm$ 0.03} & 7.00 $\pm$ 1.05 & 0.03 $\pm$ 0.02 & 6.93 $\pm$ 0.83 \\
\rowcolor{gray!10}
Cellina-GAT & \textbf{0.03 $\pm$ 0.04} & 8.59 $\pm$ 1.90 & \textbf{0.02 $\pm$ 0.02} & 7.67 $\pm$ 0.83 \\
\midrule
\rowcolor{gray!10}
Cellina\textsubscript{node-pert} & \textbf{0.03 $\pm$ 0.04} & \textbf{7.20 $\pm$ 1.45} & 0.06 $\pm$ 0.07 & 8.96 $\pm$ 2.92 \\
\rowcolor{gray!10}
Cellina-GAT\textsubscript{node-pert} & \textbf{0.03 $\pm$ 0.05} & 11.35 $\pm$ 3.26 & \textbf{0.03 $\pm$ 0.02} & \textbf{8.61 $\pm$ 1.15} \\
\rowcolor{orange!10}
SpatialProp\textsubscript{node-pert} & 6.60 $\pm$ 4.22 & 27.12 $\pm$ 8.51 & 9.92 $\pm$ 5.36 & 30.79 $\pm$ 5.18 \\
\bottomrule
\end{tabular}
\end{table}

\begin{table}[htbp]
\centering
\caption{Extended table for the \textit{in vivo} Perturb-FISH tumor
xenograft cohort. \textit{Full} scores the complete
predicted T cell response; \textit{KO-specific} isolates the component attributable to knockout identity. \textit{-rand} are specificity controls in which neighbors are set to the mean of randomly drawn cancer cells (300 cells, 5 draws) instead of the KO-specific population, decoupling the predicted shift from the true intervention. Values are means over 9 KOs and $\pm$ is the std across 5 random seeds. Best per metric per setting is
highlighted in \textbf{bold}; \textit{Cellina} variants in gray, spatially-informed competitors in orange. All models drop in Pearson when the correct knockout identity is withheld (random control), confirming that the predicted shift is tied to the specific intervention. }
\label{tab:invivo_full_results}
\begin{tabular}{llcccc}
\toprule
Type & Method & Pearson $\uparrow$ & Precision\textsubscript{signed} $\uparrow$ & E-distance $\downarrow$ & RMSE\textsubscript{LFC} $\downarrow$ \\
\midrule
\rowcolor{gray!10}
& Cellina & 0.33 $\pm$ 0.08 & \textbf{0.68 $\pm$ 0.03} & 4.13 $\pm$ 0.15 & 0.43 $\pm$ 0.01 \\
\rowcolor{gray!10}
& Cellina-GAT & \textbf{0.40 $\pm$ 0.07} & 0.66 $\pm$ 0.03 & 4.89 $\pm$ 0.19 & \textbf{0.41 $\pm$ 0.01} \\
\rowcolor{orange!10}
& SpatialProp & 0.16 $\pm$ 0.06 & 0.55 $\pm$ 0.06 & 7.44 $\pm$ 0.04 & 0.64 $\pm$ 0.04 \\
& Mean shift & 0.06 $\pm$ 0.06 & 0.51 $\pm$ 0.03 & \textbf{0.13 $\pm$ 0.01} & 0.46 $\pm$ 0.02 \\
\rowcolor{gray!10}
& Cellina-rand & -0.00 $\pm$ 0.05 & 0.48 $\pm$ 0.04 & 4.13 $\pm$ 0.18 & 0.47 $\pm$ 0.03 \\
\rowcolor{gray!10}
& Cellina-GAT-rand & 0.07 $\pm$ 0.03 & 0.51 $\pm$ 0.03 & 4.88 $\pm$ 0.19 & 0.46 $\pm$ 0.03 \\
\rowcolor{orange!10}
\multirow{-7}{*}{\shortstack{KO-\\specific}} & SpatialProp-rand & 0.11 $\pm$ 0.05 & 0.52 $\pm$ 0.06 & 7.48 $\pm$ 0.02 & 0.66 $\pm$ 0.04 \\
\midrule
\rowcolor{gray!10}
Full & Cellina & 0.53 $\pm$ 0.05 & \textbf{0.76 $\pm$ 0.03} & 4.14 $\pm$ 0.16 & \textbf{0.50 $\pm$ 0.02} \\
\rowcolor{gray!10}
& Cellina-GAT & \textbf{0.55 $\pm$ 0.10} & \textbf{0.76 $\pm$ 0.05} & 4.87 $\pm$ 0.22 & \textbf{0.50 $\pm$ 0.03} \\
\rowcolor{orange!10}
& SpatialProp & 0.26 $\pm$ 0.16 & 0.61 $\pm$ 0.08 & 7.45 $\pm$ 0.02 & 0.68 $\pm$ 0.09 \\
& Mean shift & 0.51 $\pm$ 0.13 & 0.73 $\pm$ 0.08 & \textbf{0.13 $\pm$ 0.01} & 0.54 $\pm$ 0.03 \\
\rowcolor{gray!10}
& Cellina-rand & 0.11 $\pm$ 0.23 & 0.55 $\pm$ 0.06 & 4.12 $\pm$ 0.18 & 0.59 $\pm$ 0.03 \\
\rowcolor{gray!10}
& Cellina-GAT-rand & 0.14 $\pm$ 0.30 & 0.55 $\pm$ 0.12 & 4.88 $\pm$ 0.18 & 0.59 $\pm$ 0.03 \\
\rowcolor{orange!10}
& SpatialProp-rand & 0.19 $\pm$ 0.17 & 0.58 $\pm$ 0.09 & 7.47 $\pm$ 0.02 & 0.70 $\pm$ 0.09 \\
\bottomrule
\end{tabular}
\end{table}

\section{Colorectal Cancer Application: CRC Subdomains and Pathway-Specific Perturbations}
\label{app:application}

\textbf{Model fitting.} \textit{Cellina} was fit in-domain to the largest CRC section (210), using the shared graph
pre-processing of Appendix~\ref{app:data} ($\sigma = 100$\,\textmu m, $k = 200$). For this analysis, cells were split at random; no cell type or domain was withheld.
The model used $64$ latent dimensions per encoder, two hidden layers, a negative-binomial likelihood with
observed library size, and unit classifier and adversary weights, and was trained for at most $100$ epochs
(batch size $2048$, learning rate $10^{-3}$, early stopping on validation reconstruction loss with patience
$5$, best checkpoint retained). Both latents were then extracted for every cell.

\textbf{Subdomain identification.} Cells whose histopathological annotation is CRC (all cell types) were
clustered with Hotspot~\citep{detomaso2021hotspot} over the spatial latent $s$: an unweighted
$k$-nearest-neighbor graph ($k = 30$) in $s$, an autocorrelation test on raw counts under a depth-adjusted
negative-binomial null (top $1{,}200$ genes at FDR $< 0.05$), and module detection from the pairwise local
correlations among those genes (core genes only, $\geq 100$ genes per module, FDR $< 0.05$). This yields
three modules; each CRC-domain cell is assigned to its highest-scoring module.

\textbf{Pathway annotation.} A module $\times$ gene matrix was assembled from the Hotspot local-correlation
coefficients of module-assigned genes and scored against PROGENy~\citep{schubert2018perturbation} and MSigDB
Hallmark gene sets using the univariate linear model of \texttt{decoupler}~\citep{badia2022decoupler}. The
two modules with the most distinct pathway profiles, denoted CRC1 (TGF$\beta$-dominant) and CRC2
(NF$\kappa$B/MAPK-dominant), are those analyzed in Section~\ref{sec:subtypes}; the third is not examined
further.

\textbf{Subdomain-conditioned edge perturbations.} For each of the five cell types, control cells are the
cells of that type outside the CRC domain, and their neighborhoods are rewired (Definition~1) to donors drawn
either from the whole CRC domain or from a single subdomain, with predictions returned at a library size of
$10^4$. Predictions are scored by the correlation between predicted and observed $\log_2$ fold changes over
the $200$ genes with the largest observed $|\log_2\mathrm{FC}|$. The observed reference is the pseudobulk of
all cells residing in the CRC domain (global arm) or in the subdomain (subdomain arm); both arms therefore
share the same control population and reference construction, and differ only in the neighborhoods on which
the prediction is conditioned.

\textbf{Pathway-guided node perturbations.} Alteration vectors $\delta_g$ were taken directly from PROGENy
weights ($|\text{weight}| > 0.5$, clipped to $\pm 5$), restricted to genes with pseudobulk
log fold change between the subdomain and the control region ($|\log\mathrm{FC}| > 0.01$): TGF$\beta$ for CRC1
and NF$\kappa$B for CRC2. The shift is applied to the log-normalized expression of every neighbor node on fixed
topology (Definition~2) and scored as above over the top $100$ genes; Figure~\ref{fig:application}(d) highlights the $50$ most affected genes.

\section{Seed Stability}
\label{app:seed-stability}

To assess the stability of the evaluations across random seeds, we conduct a multi-seed analysis on the edge perturbation task of Tables \ref{tab:loo_summary_crc} and \ref{tab:loo_summary_merfish}.
The goal of this analysis is to evaluate whether \textit{Cellina}'s performance advantages are random (as the result of a lucky seed value) or
systematic (consistent over many seed values).
We compare \textit{Cellina} to CPA and scGen, training each on three different seeds per slide per cell type.
As seen in Table~\ref{tab:multi-seed}, \textit{Cellina} maintains its performance advantage
over competitors and the variance across seeds is negligible. The \textit{in vivo} Perturb-FISH evaluation (Tables~\ref{tab:cellina-invivo} and~\ref{tab:invivo_full_results}) is likewise reported as the mean over five random seeds for every method, since we observed seed-to-seed variation in that smaller dataset.

\begin{table}[htbp]
\centering
\caption{Evaluation metrics on random seeds. Mean $\pm$ standard deviation is reported across three random seeds for each model; means are therefore averaged over seeds and differ from the corresponding single-run values in Tables~\ref{tab:loo_summary_crc} and~\ref{tab:loo_summary_merfish}. \textit{Cellina} in gray. Low variance indicates seed selection does not contribute to model performance.}
\label{tab:multi-seed}
\begin{tabular}{l l cccc}
\toprule
Dataset & Method & Pearson $\uparrow$
& Spearman $\uparrow$
& Precision\textsubscript{signed} $\uparrow$
& RMSE\textsubscript{LFC} $\downarrow$ \\
\midrule
\rowcolor{gray!10}
\multirow{3}{*}{CRC}
& Cellina & $0.87 \pm 0.01$ & $0.76 \pm 0.01$ & $0.45 \pm 0.01$ & $1.13 \pm 0.01$ \\
& CPA     & $0.69 \pm 0.02$ & $0.59 \pm 0.02$ & $0.23 \pm 0.01$ & $1.49 \pm 0.02$ \\
& scGen   & $0.50 \pm 0.01$ & $0.43 \pm 0.01$ & $0.19 \pm 0.00$ & $2.09 \pm 0.09$ \\
\midrule
\rowcolor{gray!10}
\multirow{3}{*}{MERFISH}
& Cellina & $0.84 \pm 0.00$ & $0.73 \pm 0.00$ & $0.52 \pm 0.01$ & $6.08 \pm 0.04$ \\
& CPA     & $0.82 \pm 0.00$ & $0.70 \pm 0.01$ & $0.44 \pm 0.00$ & $6.36 \pm 0.01$ \\
& scGen   & $0.77 \pm 0.01$ & $0.63 \pm 0.00$ & $0.21 \pm 0.02$ & $6.47 \pm 0.02$ \\
\bottomrule
\end{tabular}
\end{table}

\section{Cross-Patient Generalization}
\label{app:cross_patient}

All CRC evaluations reported in Section~\ref{sec:results} are within-patient: focal cells and donor
neighborhoods are drawn from the same tissue slide (Appendix~\ref{app:donor_sampling}). This design removes patient-level variation from the
comparison, but it does not answer whether the learned relationship between neighborhood and expression
generalizes to unseen patients. We therefore repeat the edge-perturbation evaluation under a leave-one-patient-out split.

\textbf{Protocol.} We run the evaluation on all six CRC slides. We hold out one patient at a time and, on the remaining five, additionally hold out the target
cell type $y$ in the CRC domain exactly as in Section~\ref{sec:evaluation}: cells of type $y$ in the
target domain are excluded from the donor neighborhoods and from the neighborhood
graph $W$ at both training and inference. The
model is fit on the five remaining patients and applied to $\text{REF}_y$ cells of the held-out
patient, whose neighborhoods are rewired following the donor-sampling convention of
Appendix~\ref{app:donor_sampling}; predictions are scored against the observed $\text{CRC}_y$ cells. Consequently
\emph{neither the patient nor the cell-type context was seen during training}, whereas the
within-slide evaluation withholds only the latter. Crossing six held-out patients with five
held-out cell types gives thirty folds per method;
Table~\ref{tab:cross_patient} reports the mean and standard deviation over folds.

The thirty-fold protocol multiplies the cost of an evaluation that is already the most
expensive in this work (Appendix~\ref{app:compute}), which is why it is restricted to three efficient models: the mean shift, which sees only the average REF~$\rightarrow$~CRC
change; CPA, the strongest non-\textit{Cellina} baseline on Tables~\ref{tab:loo_summary_crc}
and~\ref{tab:loo_summary_merfish}, which sees the domain as a label but never a cell's own neighbors;
and \textit{Cellina}, which sees the neighborhood itself.

\begin{table}[htbp]
\centering
\caption{Cross-patient edge perturbation on the CRC cohort (top 50 DEGs). One patient and one cell
type are held out simultaneously, thirty folds per method (six patients and five cell types); mean $\pm$ std over folds. Best per metric in \textbf{bold}; \textit{Cellina} in gray. Margins between methods can be compared with Table~\ref{tab:loo_summary_crc} (same slides and cell types, within-patient split); absolute Pearson is lower here because the split is harder.}
\label{tab:cross_patient}
\begin{tabular}{lcccc}
\toprule
Method & Pearson $\uparrow$ & Precision\textsubscript{signed} $\uparrow$ & E-distance $\downarrow$ & RMSE\textsubscript{LFC} $\downarrow$ \\
\midrule
Mean shift & 0.45 $\pm$ 0.24 & 0.15 $\pm$ 0.16 & 30.98 $\pm$ 10.34 & 4.85 $\pm$ 2.55 \\
CPA & 0.52 $\pm$ 0.18 & 0.12 $\pm$ 0.20 & 15.05 $\pm$ 4.22 & 2.88 $\pm$ 0.76 \\
\rowcolor{gray!10}
Cellina & \textbf{0.79 $\pm$ 0.13} & \textbf{0.30 $\pm$ 0.20} & \textbf{6.55 $\pm$ 1.18} & \textbf{1.37 $\pm$ 0.68} \\
\bottomrule
\end{tabular}
\end{table}

\textbf{Results.} Absolute Pearson is lower here than in Table~\ref{tab:loo_summary_crc} for every
method: \textit{Cellina} $0.79$ against $0.86$ Pearson, CPA $0.52$ against $0.68$, the mean shift
$0.45$ against $0.51$. Nevertheless, \textit{Cellina} leads CPA by $+0.27$ Pearson here ($p = 1.23 \times 10^{-7}$, Wilcoxon signed-rank test with Holm-Bonferroni correction) against $+0.18$ within-slide, and by $+0.18$ Signed Precision ($p = 3.5 \times 10^{-6}$) against $+0.23$ within-slide.
Notably, \textit{Cellina} also leads on E-distance ($6.55$ against CPA's $15.05$), the one metric on which CPA is ahead within-slide
($6.44$ against $7.00$). These results indicate that the
spatial latent $s$ encodes transferable signal driven by neighborhood composition and
expression rather than patient-specific confounding.

\section{Neighborhood Composition and Size}
\label{app:neighborhood}

The subdomain analysis of Section~\ref{sec:subtypes} contrasts how a focal cell could differ
across two cancer neighborhoods. Here, we vary the neighborhoods by i) their composition (cell types) as well as by ii) the number of cells the neighborhood graph admits (each via edge perturbation).

\textbf{Neighbor composition.} Because an edge perturbation can simulate the effect of any neighborhood, it can construct neighbor compositions that never occur in the tissue. Following the leave-one-cell-type-out protocol of Section~\ref{sec:evaluation}, we rewire each held-out cell type to a neighborhood drawn entirely from one CRC cell type at a time (homotypic neighborhood), including its own, in which case the held-out cells themselves serve as neighbors at inference; the main evaluation instead excludes same-type neighbors, so as not to give \textit{Cellina} an unfair advantage. All other elements of the protocol are unchanged: the held-out population is excluded from the neighborhood graph $W$ during training, and predictions are scored against the observed CRC cells of that type.

We see that predictions depend on neighbor identity most strongly for Fibroblast and Epithelial cells (a spread of $0.24$ and $0.22$ across neighbor types), followed by Myeloid cells ($0.12$), and least for T and Endothelial cells ($0.07$ and $0.05$; row-wise read of Table~\ref{tab:homotypic}). Notably, homotypic draws, where the held-out cell type and the neighbor cell type are the same, are not systematically better or worse than heterotypic ones. We see that they are the best draw for
Fibroblasts, the worst for T cells, which suggests that \textit{Cellina} does not just pass the effect of neighbors to the focal cell. If \textit{Cellina} merely passed the neighbors' expression through to the focal cell, the homotypic neighborhood would be the easiest case. This is best contradicted by an Epithelial holdout which scores $0.60$ among Myeloid cells, yet only $0.44$ among Epithelial cells. The model therefore predicts a distinct
response to each neighbor type. Similarly, reading the table column-wise, we see that the results also differ across focal cell types. If the predictions were influenced by the
neighborhood alone, all five focal types placed among Fibroblasts (the Fibroblast column) would score alike. Together, these results suggest that \textit{Cellina}'s predictions depend on both the identity of the focal cell and its neighborhood.

\begin{table}[htbp]
\centering
\caption{Neighborhood composition sweep on the CRC cohort (Pearson $r$ on top-50 DEG log fold changes).
Each held-out cell type (rows) is rewired to a neighborhood drawn entirely from a single CRC cell type
(columns); mean over slides 231, 232 and 242. Homotypic draws lie on the shaded diagonal. Along a row the
focal cell type is fixed and its neighborhood varied, including pairings that do not occur in the tissue;
along a column the neighborhood is fixed and the focal cell type varied.}
\label{tab:homotypic}
\begin{tabular}{lccccc}
\toprule
& \multicolumn{5}{c}{Neighbor cell type} \\
\cmidrule(lr){2-6}
Held-out cell type & Endothelial & Epithelial & Fibroblast & Myeloid & T cell \\
\midrule
Endothelial & \cellcolor{gray!25}0.88 & 0.89 & 0.90 & 0.89 & 0.85 \\
Epithelial  & 0.45 & \cellcolor{gray!25}0.44 & 0.58 & 0.60 & 0.38 \\
Fibroblast  & 0.46 & 0.50 & \cellcolor{gray!25}0.66 & 0.65 & 0.42 \\
Myeloid     & 0.75 & 0.87 & 0.81 & \cellcolor{gray!25}0.78 & 0.75 \\
T cell      & 0.86 & 0.90 & 0.87 & 0.88 & \cellcolor{gray!25}0.83 \\
\bottomrule
\end{tabular}
\end{table}

\textbf{Neighborhood size.} The weighted adjacency matrix $W$ determines how many cells count as neighbors, set via $k$ (nearest neighbors) (Appendix~\ref{app:data}). We therefore refit
\textit{Cellina} at $k \in \{10, 100, 200, 1000, 10000\}$ with the kernel bandwidth fixed to $\infty$, so that
$k$ alone determines the neighborhood, and repeat the leave-one-cell-type-out evaluation at each setting with
the held-out population masked from both the adjacency graph and the spatial aggregation. Because the model is
refit at every setting, the sweep is run on a single slide (CRC 232) with five random seeds per setting.

We observe that sensitivity to $k$ is cell-type-dependent (Table~\ref{tab:k_sweep}). T cells are essentially
unaffected across three orders of magnitude ($0.97$ to $0.94$), suggesting that their state is defined largely
intrinsically. Endothelial and Myeloid cells show a modest degradation, while Fibroblast and Epithelial predictions
strongly depend on $k$ throughout and collapse at $k = 10^{4}$ ($0.10$ and $0.11$), where every neighborhood
converges on the domain average and carries no local signal. 

\begin{table}[htbp]
\centering
\caption{Sensitivity to neighborhood-graph construction (Pearson $r$ on top-50 DEG log fold changes).
The number of nearest neighbors $k$ retained in $W$ is varied with the kernel bandwidth fixed to $\infty$ and
the model refit at each setting; $k = 200$ is the value used for \textit{Cellina} on CRC and MERFISH (Appendix~\ref{app:data}).
CRC slide 232; mean $\pm$ std over five random seeds. Neither the absolute values nor the ordering across cell
types are directly comparable to Table~\ref{tab:loo_summary_crc} and Figure~\ref{fig:crc_barplots}, which
average over all six slides under the default Gaussian kernel ($\sigma = 100$\,\textmu m, $\tau = 0.1$).}
\label{tab:k_sweep}
\begin{tabular}{lccccc}
\toprule
$k$ & Endothelial & Epithelial & Fibroblast & Myeloid & T cell \\
\midrule
$10$      & 0.85 $\pm$ 0.06 & 0.49 $\pm$ 0.12 & 0.52 $\pm$ 0.15 & 0.89 $\pm$ 0.03 & 0.97 $\pm$ 0.01 \\
$100$     & 0.83 $\pm$ 0.04 & 0.48 $\pm$ 0.11 & 0.71 $\pm$ 0.14 & 0.87 $\pm$ 0.02 & 0.96 $\pm$ 0.01 \\
$200$     & 0.83 $\pm$ 0.04 & 0.41 $\pm$ 0.04 & 0.65 $\pm$ 0.16 & 0.87 $\pm$ 0.02 & 0.96 $\pm$ 0.01 \\
$1000$    & 0.79 $\pm$ 0.04 & 0.40 $\pm$ 0.22 & 0.54 $\pm$ 0.18 & 0.85 $\pm$ 0.01 & 0.96 $\pm$ 0.01 \\
$10000$   & 0.68 $\pm$ 0.04 & 0.11 $\pm$ 0.17 & 0.10 $\pm$ 0.24 & 0.77 $\pm$ 0.07 & 0.94 $\pm$ 0.01 \\
\bottomrule
\end{tabular}
\end{table}

\section{Node Perturbation as a Function of the Perturbed Neighbor Fraction}
\label{app:dose}

The gene-set sweep of Appendix~\ref{app:convergence} varies how many genes are shifted in every neighbor. Here we instead vary the proportion of neighbors perturbed by the same $\delta$ shift. We reuse the $k = 200$ checkpoints of the neighborhood-size sweep (Appendix~\ref{app:neighborhood}), one per held-out cell type on CRC slide 232, without any retraining. The perturbation is the cell-type-specific shift of Appendix~\ref{app:convergence} over the top-200 genes per type, with the held-out type $y$ excluded from the pseudobulk and assigned the global shift $\delta_{\setminus y}$ as before. The only change is that the shift is applied to a random fraction $f \in \{0.1, 0.25, 0.5, 0.75, 1\}$ of cells, the remainder keeping their observed expression, so that after aggregation each held-out REF cell has only a fraction $f$ of its neighborhood perturbed. We report the Pearson correlation between observed and predicted log fold changes over the top-50 DEGs, mean $\pm$ std over five seeds.

We see that prediction accuracy rises largely monotonically; from $f = 0.1$, where only a handful of neighbors are shifted, toward each type's ceiling at $f = 1$, for every cell type except T cells, which are already at ceiling at $f = 0.1$ (Table~\ref{tab:dose}). 
The ceiling ordering (T cell $>$ Myeloid $>$ Endothelial $>$ Fibroblast $>$ Epithelial) mirrors the neighborhood-size sweep (Table~\ref{tab:k_sweep}), and the two most neighborhood-dependent types, Fibroblast and Epithelial, are also the ones whose response is most strongly associated with $f$. \textit{Cellina} thus shows a dose-dependent response across most cell types.

\begin{table}[htbp]
\centering
\caption{Node-perturbation dose response (Pearson $r$ on top-50 DEG log fold changes). The cell-type-specific logFC shift is applied to a random fraction $f$ of neighbor cells before neighborhood aggregation, using the $k = 200$ checkpoints of Table~\ref{tab:k_sweep} with each cell type held out. CRC slide 232; mean $\pm$ std over five random seeds.}
\label{tab:dose}
\begin{tabular}{lccccc}
\toprule
$f$ & Endothelial & Epithelial & Fibroblast & Myeloid & T cell \\
\midrule
$0.10$ & 0.65 $\pm$ 0.06 & 0.02 $\pm$ 0.37 & 0.32 $\pm$ 0.23 & 0.78 $\pm$ 0.03 & 0.95 $\pm$ 0.01 \\
$0.25$ & 0.68 $\pm$ 0.05 & 0.05 $\pm$ 0.38 & 0.39 $\pm$ 0.22 & 0.79 $\pm$ 0.03 & 0.95 $\pm$ 0.01 \\
$0.50$ & 0.72 $\pm$ 0.05 & 0.14 $\pm$ 0.35 & 0.50 $\pm$ 0.19 & 0.81 $\pm$ 0.03 & 0.95 $\pm$ 0.01 \\
$0.75$ & 0.77 $\pm$ 0.05 & 0.32 $\pm$ 0.15 & 0.62 $\pm$ 0.15 & 0.84 $\pm$ 0.02 & 0.96 $\pm$ 0.01 \\
$1.00$ & 0.80 $\pm$ 0.05 & 0.43 $\pm$ 0.04 & 0.69 $\pm$ 0.13 & 0.87 $\pm$ 0.02 & 0.96 $\pm$ 0.01 \\
\bottomrule
\end{tabular}
\end{table}
\section{Disentanglement Benchmark}
\label{app:disentanglement}

\begin{figure}[htbp]
    \centering
    \includegraphics[width=\textwidth]{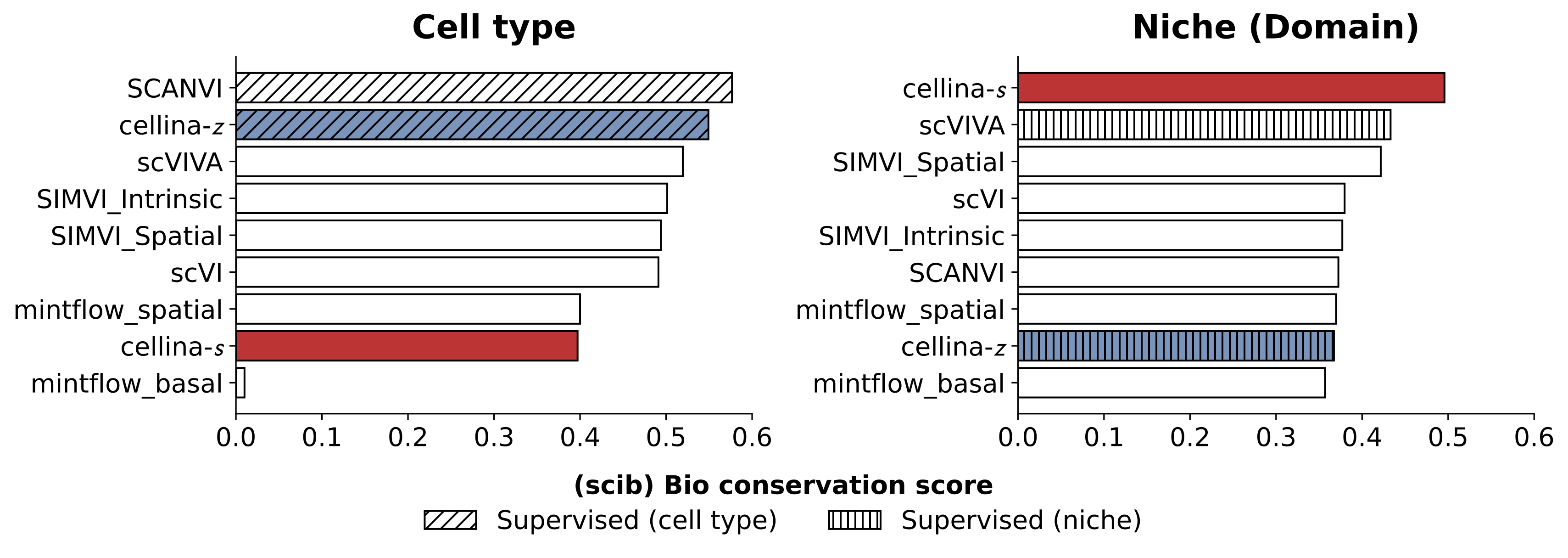}
    \caption{scIB disentanglement benchmark across methods. Bio-conservation score is an aggregate measure of clustering with respect to given label (cell type or spatial domain). \textit{Cellina} shows strong competitive performance across common disentanglement metrics.}
    \label{fig:scib}
\end{figure}

One of our core claims in this work is that supervised disentanglement improves perturbation simulations on graph-structured data. To demonstrate that \textit{Cellina}'s latent factors $z$ and $s$ accordingly absorb cell type and spatial domain information, respectively, we use the single-cell integration benchmark (scIB) package \citep{luecken_benchmarking_2022} -- a comprehensive benchmarking tool for latent factors of single-cell data. Given a categorical label, such as cell type, scIB assigns an aggregate ``Bio Conservation'' score summarizing multiple clustering metrics such as K-means Normalized Mutual Information (NMI), K-means Adjusted Rand Index (ARI) and Silhouette score. We use the CRC cohort to assess both cell type and spatial domain (niche) conservation (Figure~\ref{fig:scib}). As points of comparison, we take standard single-cell methods such as scVI \citep{lopez2018deep}, scANVI \citep{xu2021probabilistic} and scVIVA \citep{levy2025scviva} (details in Appendix~\ref{app:baselines}). Additionally, we compare to MintFlow \citep{akbarnejad2025mapping} and SIMVI \citep{dong2025simvi}, which are designed for disentanglement of spatial and non-spatial variation in single cells. Each model is trained once on the entirety of CRC data (all six slides together) and aggregate scores for each label are reported on the training set, except for SIMVI, which does not scale to millions of cells (see Section~\ref{sec:experiments} and Appendix~\ref{app:baselines}). For SIMVI, we take coherent regions of two slides (231, 242) containing every domain label present in the cohort, for a total of 40K cells. As SIMVI is evaluated on a smaller and distinct subset, this comparison is not directly controlled; we include it as the best available evidence given its scalability constraints. We observe that not only do \textit{Cellina}'s latent $z$ and $s$ encode the desired source of variation, but $s$ also contains less nuisance cell-type information ($0.40$ vs.\ $0.49$ for SIMVI's spatial latent), while $z$'s niche score is on par with non-spatial models such as scVI. The advantage over SIMVI and MintFlow is particularly notable as all three methods are spatially-informed models designed for this task. Our results suggest that \textit{Cellina}'s explicitly supervised disentanglement benefits each latent space as opposed to the implicit or unsupervised disentanglement strategies adopted by MintFlow and SIMVI, respectively.

\begin{figure}[p]
  \centering
  \setlength{\tabcolsep}{2pt}
  \renewcommand{\arraystretch}{1.5}
  \begin{tabular}{@{}cc@{}}
    \raisebox{-0.5\height}{\rotatebox{90}{\scriptsize CRC 231}} &
      \raisebox{-0.5\height}{\includegraphics[width=1\textwidth,height=0.33\textheight,keepaspectratio]{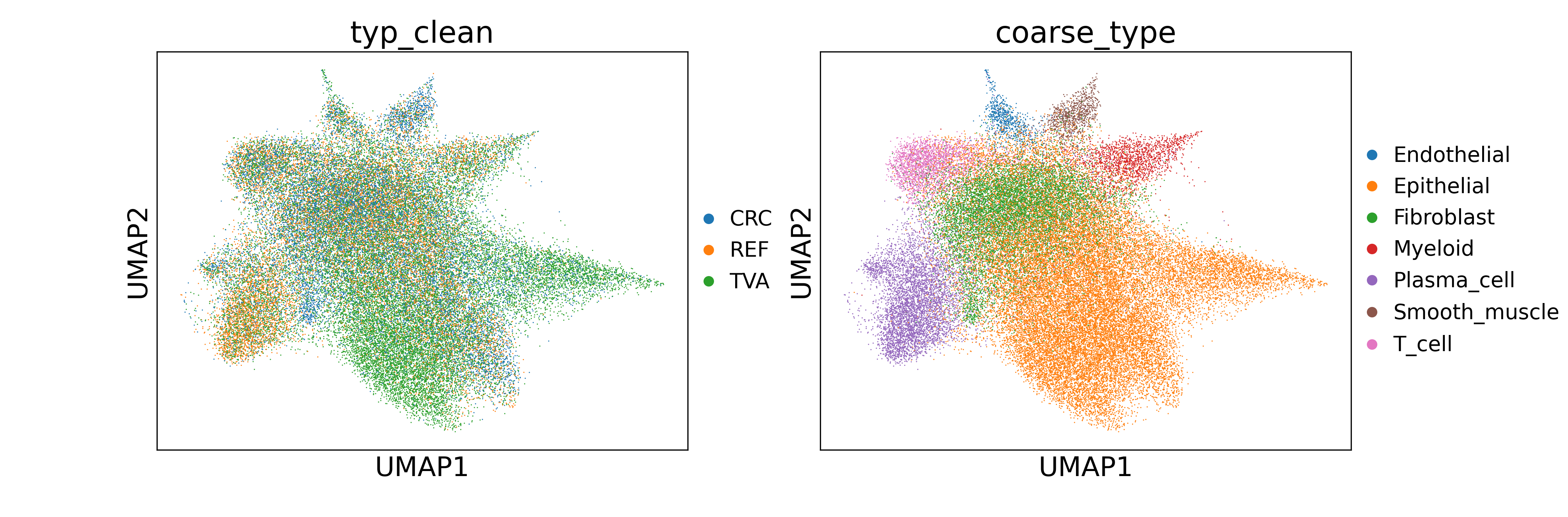}} \\
    \raisebox{-0.5\height}{\rotatebox{90}{\scriptsize CRC 232}} &
      \raisebox{-0.5\height}{\includegraphics[width=1\textwidth,height=0.33\textheight,keepaspectratio]{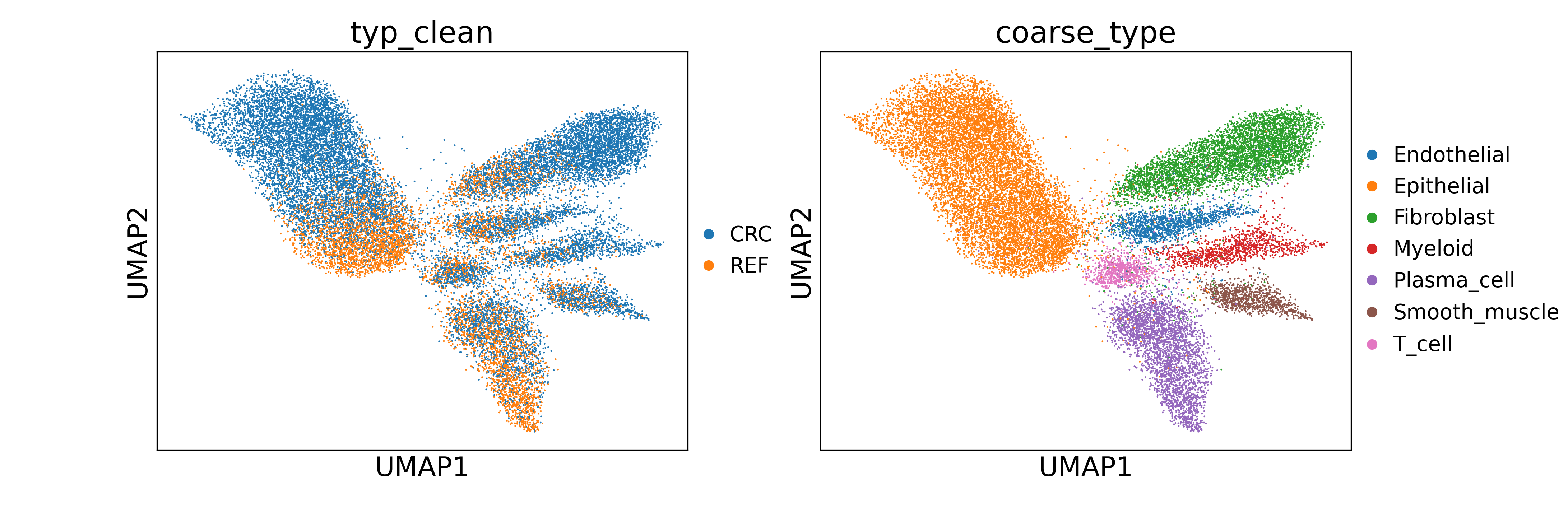}} \\
    \raisebox{-0.5\height}{\rotatebox{90}{\scriptsize CRC 242}} &
      \raisebox{-0.5\height}{\includegraphics[width=1\textwidth,height=0.33\textheight,keepaspectratio]{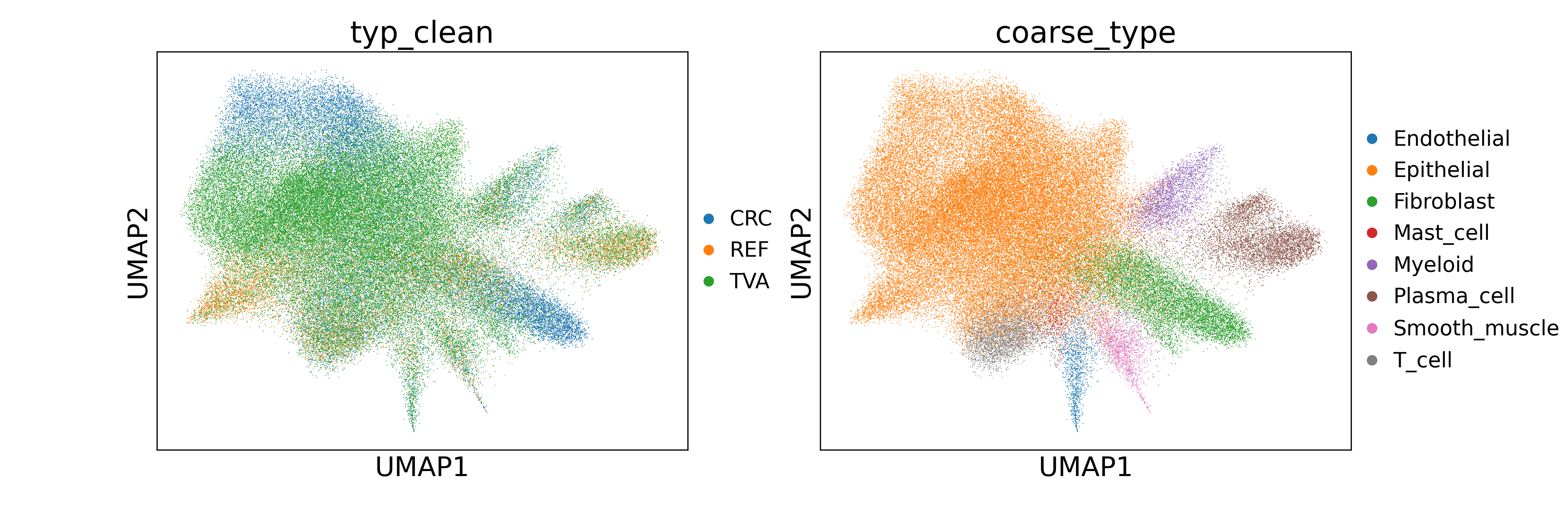}} \\
  \end{tabular}
  \caption{UMAPs of \textit{Cellina} latent space $z$ for each donor used in the scGraph analysis of Appendix~\ref{app:disentanglement}, colored by spatial region (left; TVA: tubulovillous adenoma) and cell type (right).}
  \label{fig:scgraph_umap_z}
\end{figure}

\begin{figure}[p]
  \centering
  \setlength{\tabcolsep}{2pt}
  \renewcommand{\arraystretch}{1.5}
  \begin{tabular}{@{}cc@{}}
    \raisebox{-0.5\height}{\rotatebox{90}{\scriptsize CRC 231}} &
      \raisebox{-0.5\height}{\includegraphics[width=1\textwidth,height=0.33\textheight,keepaspectratio]{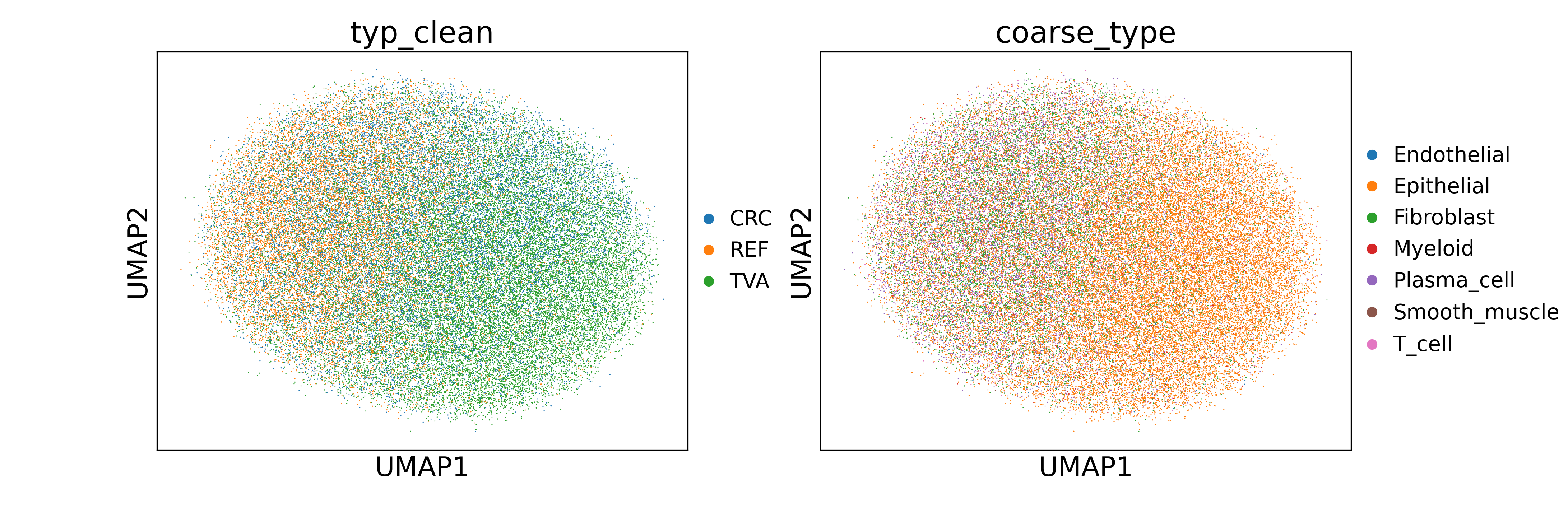}} \\
    \raisebox{-0.5\height}{\rotatebox{90}{\scriptsize CRC 232}} &
      \raisebox{-0.5\height}{\includegraphics[width=1\textwidth,height=0.33\textheight,keepaspectratio]{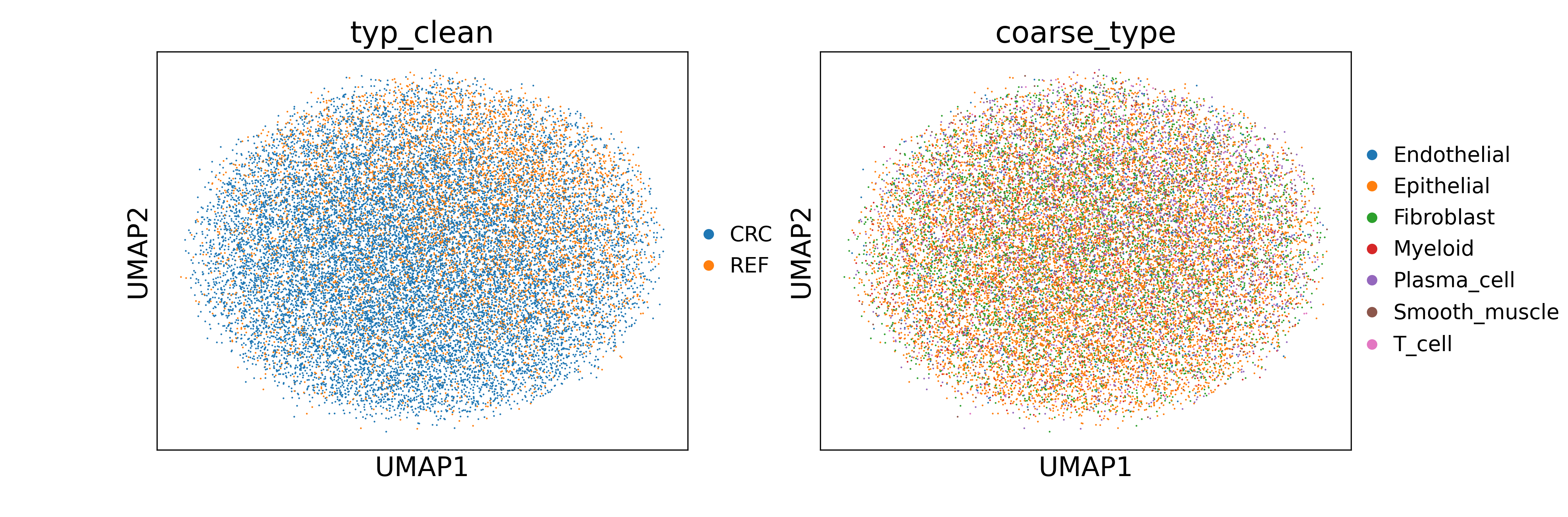}} \\
    \raisebox{-0.5\height}{\rotatebox{90}{\scriptsize CRC 242}} &
      \raisebox{-0.5\height}{\includegraphics[width=1\textwidth,height=0.33\textheight,keepaspectratio]{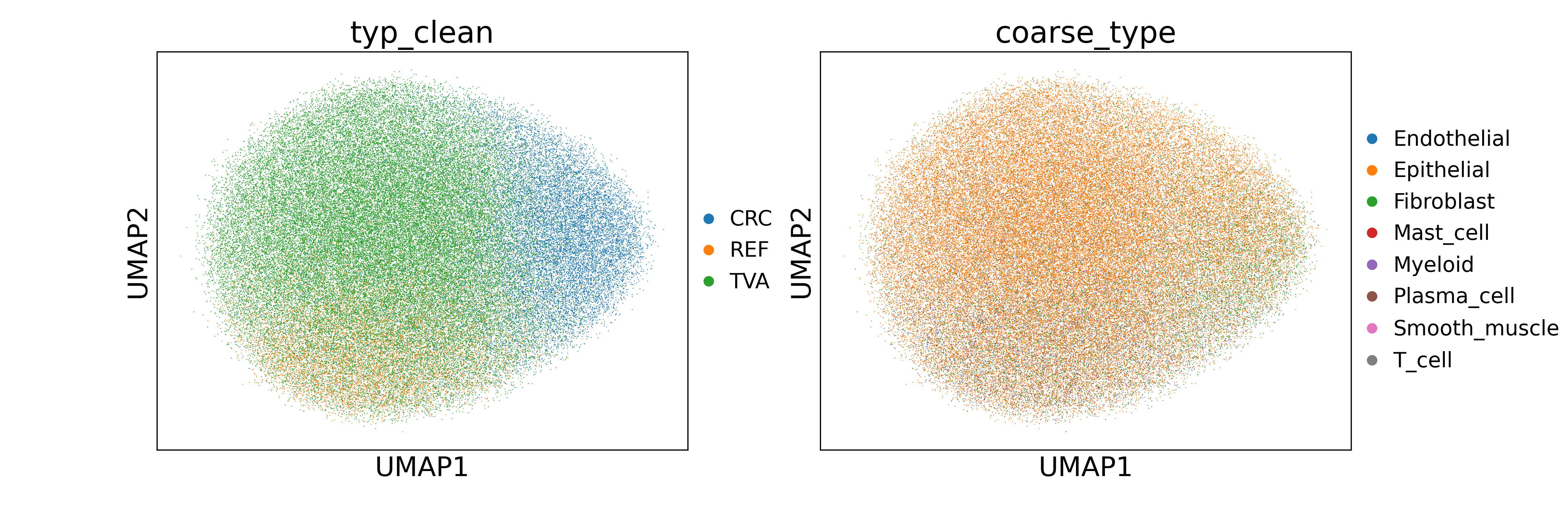}} \\
  \end{tabular}
  \caption{UMAPs of \textit{Cellina} latent space $s$ for each donor used in the scGraph analysis of Appendix~\ref{app:disentanglement}, colored by spatial region (left; TVA: tubulovillous adenoma) and cell type (right).}
  \label{fig:scgraph_umap_s}
\end{figure}

Finally, because \textit{Cellina} anchors $z$ with a cell-type classifier, high cell-type conservation could in principle be achieved by collapsing cells onto their label centroids -- a failure mode that label-based integration metrics reward rather than detect~\citep{wang2026limitations}. As a sanity check, we therefore also scored \textit{Cellina}'s $z$ with the scGraph metrics of \citet{wang2026limitations}, which ask whether an embedding preserves smooth transitions of cell type populations rather than separating them into well-formed but arbitrarily placed clusters or so-called ``floating islands.'' \textit{Cellina}'s $z$ scores far above the CPA latent of Table~\ref{tab:loo_summary_crc} ($0.51 \pm 0.14$ against $0.00 \pm 0.22$ across the CRC cohort), indicating that its intrinsic latent reflects retained biological structure rather than label collapse. This is further supported by plotting UMAPs of three slides (231, 232, 242) as examples in which \textit{Cellina}'s latents show no evidence of floating islands (Figures \ref{fig:scgraph_umap_z}, \ref{fig:scgraph_umap_s}).

\section{Scalability Benchmark}
\label{app:scalability}

\begin{figure}[htbp]
    \centering
    \includegraphics[width=0.7\textwidth]{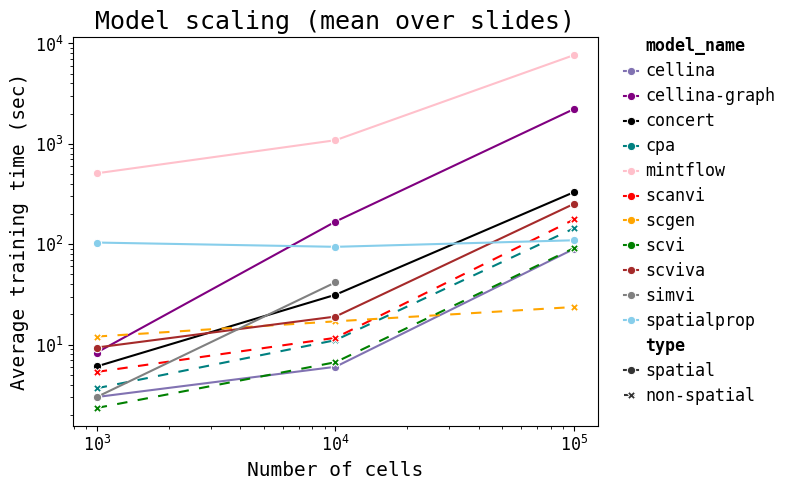}
    \caption{Training time (wall clock, seconds) vs.\ dataset size (number of cells)
    for \textit{Cellina} and related methods. \textit{Cellina} is one of the fastest, remaining competitive with scVI and substantially faster than spatial competitors (SIMVI, scVIVA, MintFlow, Concert). SpatialProp's training time remains stable due to efficient cell-type-based subsampling, while \textit{Cellina-GAT}'s explicit message passing adds a computational penalty over \textit{Cellina}, yet it remains faster than MintFlow. In the legend, cellina-graph denotes \textit{Cellina-GAT}.}
    \label{fig:scalability}
\end{figure}

Training models on emerging spatial datasets which capture millions of cells can incur major costs in user wall-clock time. It is therefore important that models are not only relatively quick to train, but also able to scale as the number of cells increases. We assess the scalability of all models shown in this work and report findings on three CRC slides (221, 242, 232) with subsampling to $10^3$, $10^4$ and $10^5$ cells (Figure~\ref{fig:scalability}). For a fair comparison, we set the same batch size and number of epochs for each model and only compare the train loop (without pre-processing workloads). \textit{Cellina} comes out as one of the fastest-to-train models in the benchmark suite, owing to the efficient pseudobulk-based $\varphi(v)$ computed a priori, omitting the expensive message passing from model training over graphs. Training then proceeds with the same complexity as standard scVI with no graph-structure
overhead at training time. This contrasts with GAT-based methods (including \textit{Cellina-GAT}, MintFlow, and SIMVI) that require subgraph sampling during training.

All methods were trained and evaluated on the same GPU machine with NVIDIA GeForce RTX 4090 GPUs, AMD Ryzen Threadripper PRO 7975WX 32-Cores, and 500GB of RAM. 

\section{Edge vs.\ Node Perturbations}
\label{app:operators}

\textbf{Three properties of neighborhood topology.} A focal cell's neighborhood topology is characterized by three
properties: \emph{proximity} (how close each neighbor is), \emph{cardinality} (how many neighbors there
are), and \emph{composition} (which cells occupy those positions). An edge perturbation (Definition~1)
intervenes on $(\mathcal{E}, W)$ and can therefore act on all three; a node perturbation
(Definition~2) acts on none of them, since it holds the topology fixed and rewrites only the features
$X$ of the neighbors already present. In other words, an edge perturbation modifies the connections between cells, whereas a node perturbation modifies the features of the cells themselves. As such, the two operators are complementary, and coincide only in the narrow case described below.

\textbf{Operations available only to edge perturbation.} Any change to cardinality, proximity, or composition.
For example, our edge perturbation is exactly such an intervention at inference: every
original edge on the focal cell is dropped and replaced by the observed neighborhood of a same-type cell in
the target domain, so both composition and cardinality are set by construction (Appendix~\ref{app:donor_sampling}). Consistent with
this, predictions depend on the neighborhood size used to construct $W$ and degrade
once $k$ reaches the slide scale, where each neighborhood converges on the domain average (Appendix~\ref{app:neighborhood}). Edge perturbation also
permits neighborhoods that never occur in the tissue --- e.g., a fully homotypic one
(Appendix~\ref{app:neighborhood}), or by extension any desired composition and size. None of these are
expressible as node perturbations.
Experimentally, an edge perturbation is the in-silico counterpart of rewiring a cell's surroundings,
as in a co-culture, an organoid transplant, or a xenograft; our REF$\to$CRC evaluation represents one such scenario,
asking what a healthy colon cell would express once its neighbors are replaced by tumor
tissue. The same operator is also implicitly used by MintFlow~\citep{akbarnejad2025mapping}, which the authors use to
deplete regulatory T cells from lesional atopic dermatitis skin, and tertiary-lymphoid-structure
macrophages from clear-cell renal cell carcinoma. Similarly, removing a cell type, e.g., with a drug that kills it, in our formalization, represents an edge perturbation, since the killed cells (nodes) are removed from the graph altogether, and hence disconnected from any focal cell rather than having their expression altered.

\textbf{Operations available only to node perturbation.} On a fixed topology, altering a chosen gene
set $\mathcal{S}$ of the already connected neighbors and nothing else: the pathway shift of
Figure~\ref{fig:application}(d), a single-ligand stimulation, or a double knockout in a neighbor. Edge
perturbation cannot express these, because it can only connect the focal cell to, or disconnect it from, cells
(typically) present in the data. SpatialProp's applications focus on this operator~\citep{sun2025spatialprop}, rescaling a chosen gene set
across the neighbors while leaving the graph untouched. Specifically, their causal benchmark exploits that fixed topology to test response directionality: for a set of mediators,
such as metabolites, lipids, ions, cytokines, and hormones, they perturb the neighbors' genes associated with \emph{producing} the mediator and ask
whether the focal cells' \emph{response} matches known pathways.

\textbf{Where the two operations coincide.} The two operators agree only when (i) the original and target neighborhoods have the same proximity and size, so that nothing needs rewiring, (ii) every neighbor's entire feature vector is overwritten, and (iii) each neighbor is overwritten to match an observed or simulated cell. Our node perturbations do not satisfy (iii), since $\delta$ shifts neighbors toward a population-average profile rather than onto observed cells. Consistently, node perturbation saturates below the edge-perturbation reference for every gene set (Appendix~\ref{app:convergence}), and in the held-out benchmarks it trails edge perturbation for \textit{Cellina}, with a smaller gap for \textit{Cellina-GAT} (Tables~\ref{tab:loo_summary_crc} and~\ref{tab:loo_summary_merfish}).

\textbf{Beyond tissue graphs.} The two operator definitions require only a graph whose nodes carry features; as such,
both operators transfer to other relational data. In a hospital contact network such as that of \citet{vanhems2013estimating}, extended so that
each individual carries a measured readout (e.g., blood screen values or an omics profile) alongside their role,
an edge perturbation asks how that readout would change if their contacts were those of a peer in
another ward. In a labeled collaboration graph, where each person carries a set of skills, a node perturbation
asks how an individual's own skills would develop if their collaborators' skills changed. However, while
the pair of operators together with the intrinsic/extrinsic decomposition are in theory transferable,
\textit{Cellina} as implemented is specifically targeted to spatial omics data. Specifically, cell-type and spatial-domain labels $y_v$ and $d_v$ are required,
by default, its encoder aggregates neighbors through a Gaussian kernel over
spatial coordinates, assuming a diffusion gradient, and its decoder models integer counts with a negative-binomial likelihood.

\begin{table}[htbp]
  \centering
  \caption{Every perturbation experiment in this work and the operator it instantiates.
  Edge perturbations rewire $\mathcal{N}(v)$; node perturbations rewrite the features of a fixed
  $\mathcal{N}(v)$.}
  \label{tab:operator_mapping}
  \begin{tabular}{llc}
    \toprule
    Experiment & Reference & Operator \\
    \midrule
    CRC leave-one-cell-type-out (REF $\rightarrow$ CRC)   & Table~\ref{tab:loo_summary_crc}, edge block     & Edge \\
    Mouse brain leave-one-cell-type-out                   & Table~\ref{tab:loo_summary_merfish}, edge block & Edge \\
    Held-out Fibroblast IGF2 illustration                 & Figure~\ref{fig:example}               & Edge \\
    Subdomain-conditioned rewiring (CRC1/CRC2)            & Figure~\ref{fig:application}(a,b)      & Edge \\
    Cross-patient generalization                          & Table~\ref{tab:cross_patient}          & Edge \\
    Neighborhood composition sweep                        & Table~\ref{tab:homotypic}              & Edge \\
    \midrule
    CRC and mouse brain node perturbation ($\delta_{y,g}$)  & Tables~\ref{tab:loo_summary_crc},~\ref{tab:loo_summary_merfish}, node blocks & Node \\
    Perturbed gene-set sweep                              & Figure~\ref{fig:convergence}           & Node \\
    Perturbed neighbor-fraction sweep                     & Table~\ref{tab:dose}                  & Node \\
    Pathway-guided neighbor perturbation                  & Figure~\ref{fig:application}(d)        & Node \\
    \textit{In vivo} CRISPR xenograft (Perturb-FISH)       & Table~\ref{tab:cellina-invivo}         & Node \\
    \bottomrule
  \end{tabular}
\end{table}

\end{document}